\documentclass[preprint]{aastex62}
\usepackage{color}

\shorttitle{DM and RM from the ejecta of compact binary mergers}
\shortauthors{Zhao et al.}

\usepackage{threeparttable}
\usepackage{hyperref}
\usepackage{natbib}
\usepackage{epsfig}
\usepackage{graphicx}
\usepackage{subfigure}
\usepackage{float}
\usepackage{amsmath}
\usepackage{color}
\usepackage{amssymb}
\usepackage{amsfonts}
\usepackage{units}
\usepackage{mathtools}
\usepackage{bm}
\usepackage{CJKulem}
\newcommand{\ud}{\mathrm{d}}

\begin{document}
\title{Dispersion and Rotation Measures from the Ejecta of Compact Binary Mergers: Clue to the Progenitors of Fast Radio Bursts}

\author[0000-0002-2171-9861]{Z. Y. Zhao}
\affiliation{School of Astronomy and Space Science, Nanjing University, Nanjing 210093, China}

\author[0000-0001-6545-4802]{G. Q. Zhang}
\affiliation{School of Astronomy and Space Science, Nanjing University, Nanjing 210093, China}

\author[0000-0002-3822-0389]{Y. Y. Wang}
\affiliation{School of Astronomy and Space Science, Nanjing University, Nanjing 210093, China}

\author[0000-0001-6606-4347]{Zuo-Lin Tu}
\affil{School of Astronomy and Space Science, Nanjing University, Nanjing 210093, China}

\author[0000-0003-4157-7714]{F. Y. Wang}
\affiliation{School of Astronomy and Space Science, Nanjing University, Nanjing 210093, China}
\affiliation{Key Laboratory of Modern Astronomy and Astrophysics (Nanjing University), Ministry of Education, Nanjing 210093, China}

\correspondingauthor{F. Y. Wang}
\email{fayinwang@nju.edu.cn}

\begin{abstract}
Since the discovery of FRB 200428 associated with the Galactic SGR 1935+2154, magnetars are considered to power fast radio bursts (FRBs). It is widely believed that magnetars could form by core-collapse (CC) explosions and compact binary mergers, such as binary neutron star (BNS), binary white dwarfs (BWD), and neutron star-white dwarf (NSWD) mergers. Therefore, it is important to distinguish the various progenitors. The expansion of the merger ejecta produces a time-evolving dispersion measure (DM) and rotation measure (RM) that can probe the local environments of FRBs. In this paper, we derive the scaling laws for the DM and RM from ejecta with different dynamical structures (the mass and energy distribution) in the uniform ambient medium (merger scenario) and wind environment (CC scenario). We find that the DM and RM will increase in the early phase, while DM will continue to grow slowly but RM will decrease in the later phase in the merger scenario. We fit the DM and RM evolution of FRB 121102 simultaneously for the first time in the BNS merger scenario, and find the source age is $ \sim9-10 $ yr when it was first detected in 2012, and the ambient medium density is $ \sim 2.5-3.1 $ cm$ ^{-3} $. The large offsets of some FRBs are consistent with BNS/NSWD channel. The population synthesis method is used to estimate the rate of compact binary mergers. The rate of BWD mergers is close to the observed FRB rate. Therefore, the progenitors of FRBs may not be unique.
\end{abstract}


\section{Introduction}\label{sec:intro}
Fast radio bursts (FRBs) are bright radio transients, which were first discovered more than a decade ago \citep{2007Sci...318..777L}. The large dispersion measures (DMs) imply their cosmological origins. FRB 121102, the first repeating FRB \citep{2016Natur.531..202S}, has been localized to a star-forming region of a low-metallicity dwarf galaxy at redshift $ z=0.193 $ \citep{2017Natur.541...58C,2017ApJ...843L...8B,2017ApJ...834L...7T}. From several years' observations, its DM was found to increase significantly (an increase of about one percent) \citep{2019ApJ...882L..18J,2019ApJ...876L..23H,2020A&A...635A..61O}. The very high ($\sim 10^5$ rad m$ ^{-2} $) and variable Faraday rotation measure (RM) implies that its source is surrounded by an extreme magnetized environment \citep{2018Natur.553..182M}. FRB 180916.J0158+65 (hereafter FRB 180916) is a repeater discovered by the Canadian Hydrogen Intensity Mapping Experiment (CHIME) telescope \citep{2019ApJ...885L..24C}, and was localized to a nearby massive spiral galaxy at redshift $ z=0.0337 \pm 0.0002 $ \citep{2020Natur.577..190M}.

Among various progenitor models of FRBs, the one relevant to a young magnetar is promising \citep{2013arXiv1307.4924P,2014ApJ...797...70K,2016MNRAS.461.1498M,2017ApJ...841...14M,2017JCAP...03..023W,2017ApJ...843L..26B,2018MNRAS.477.2470L,2018ApJ...868...31Y,2019ApJ...879....4W}. Recently, FRB 200428 originated from the Galactic magnetar SGR 1935+2154 has been discovered \citep{2020Natur.587...54T,2020Natur.587...59B}, which has greatly promoted the study of the relationship between FRBs and activities of magnetars. Previous research suggested that magnetars could be born in the following processes: core-collapses (CC) of massive stars, binary neutron star (BNS) mergers \citep{2003MNRAS.345.1077R,2006Sci...312..719P,2013ApJ...771L..26G,Yamasaki2020}, binary white dwarf (BWD) mergers \citep{2001MNRAS.320L..45K,2007MNRAS.380..933Y,2016MNRAS.463.3461S,Kashiyama2017}, neutron star–white dwarf (NSWD) mergers \citep{2020ApJ...893....9Z} or accretion-induced collapse (AIC) \citep{1991ApJ...367L..19N,2013A&A...558A..39T,2015MNRAS.453.1910S}. Some observational evidence of gamma-ray bursts, including X-ray plateaus \citep{Dai1998,Zhang2001,Zhang2006,Rowlinson2013,Lu2014} and X-ray flares \citep{Burrows2005,Dai2006,Wang2013}, also support that magnetars are produced by CC of massive stars or mergers of compact objects. Therefore, it is important to distinguish between different formation channels.

The evolution of DM and RM can be related to the surrounding environment of FRBs \citep{Piro_2016}. After CC SNe or the merger of two compact stars, the ejecta drives a blast wave into the local environment. The forward and reverse shock wave will generate as a result of the interaction between the energetic ejecta and the surrounding medium. The shocked regions not only have high temperature and density \citep{1982ApJ...258..790C}, but also are the ideal places to amplify magnetic field \citep{2013SSRv..178..201B,2014ApJ...794...46C}. For different formation channels of magnetars, the initial conditions (the mass and energy of the ejecta) vary significantly, which plays an important role in the evolution of DM and RM.

Under the assumption of constant ejecta density, the variations of DM and RM from a supernova remnant (SNR) have been studied \citep{Piro_2016,2017ApJ...847...22Y,Piro_2018}. However, the supernova (SN) ejecta has been found to have an outer region characterized by a steep power-law profile and an inner relatively flat core (\citeauthor{1989ApJ...341..867C} \citeyear{1989ApJ...341..867C}; \citeauthor{1999ApJS..120..299T} \citeyear{1999ApJS..120..299T}\defcitealias{1999ApJS..120..299T}{TM99}, hereafter \citetalias{1999ApJS..120..299T}). The ejecta structure of BWD mergers has been considered by \cite{10.1093/mnras/stz3593} under the assumption of a constant ratio of the forward and reverse shocks radius. Before the reverse shock hits the ejecta core, the constant ratio relation of the forward and reverse shocks radius is accurate \citep{1982ApJ...258..790C}.  However, after the reverse shock hits the unshocked ejecta core, the analytical approximate solutions of the more complex reverse shock wave are given in TM model (TM99 and the methods based on it, i.e. \citetalias{1999ApJS..120..299T}; \citeauthor{2000ApJS..128..403T} \citeyear{2000ApJS..128..403T}; \citeauthor{2012ApJ...746..130H} \citeyear{2012ApJ...746..130H}; \citeauthor{refId0} \citeyear{refId0}). In addition to CC SNe or BWD mergers, the magnetar born in BNS and NSWD mergers channels can also power FRBs. The localized FRBs, whether repeating or not, have large offsets from the galaxies centers \citep{2019Sci...365..565B,2019Sci...366..231P,2019Natur.572..352R,2020arXiv200513158C,2020Natur.577..190M,2020ApJ...903..152H}, which favors the BNS or NSWD channel because NSs receive natal kicks as a result of asymmetric SN explosions \citep{1994A&A...290..496J,1996PhRvL..76..352B}. \cite{Wang_2020} have found that the offsets of FRBs are consistent with the case of BNS mergers, which provides further evidence that FRBs may originate from compact binary mergers (remnants) containing at least one neutron star.

In this paper, we derive the DM and RM evolution in the local environment for compact binary (BNS, NSWD, or BWD) mergers powering FRBs. The evolution of the shocked shell is taken from \cite{2017MNRAS.465.3793T} and \citetalias{1999ApJS..120..299T}, which takes the ejecta dynamical structures and the behavior of the reverse shocks in the later times into account. The initial conditions of BNS, BWD, and NSWD mergers are given by numerical simulations \citep{Bauswein_2013,Radice_2018,2016MNRAS.461.1154M,2007ApJ...669..585D,2009MNRAS.396.1659M,10.1093/mnras/stz316}. As a comparison, taking Cassiopeia A (Cas A) as an example from \cite{refId0}, the DM and RM evolution from the CC remnants is only discussed briefly. The event rate of FRBs has been reported very high, and we also compare it with the rate of compact binary merger estimated by the population synthesis method. Especially, observations show that the DM increases and RM decreases for FRB 121102 \citep{2020arXiv200912135H}, which is hard to explain. We will show that the DM and RM evolution can be well understood in the BNS merger scenario. In section \ref{model}, the approximate solutions of the shocked shell are shown. In section \ref{dm&rm}, we introduce the evolution of DM and RM in the uniform ambient medium (merger scenario) and wind environment (CC scenario). In section \ref{obs}, we explain the DM and RM evolution of FRB 121102, FRB 180916 and FRB 180301 in the merger scenario, and the Markov-chain Monte Carlo (MCMC) method is used to estimate the model parameters of FRB 121102. In section \ref{discu}, we discuss the free-free absorption of the ejecta and shocked shell, and the results of population synthesis are also shown. Finally, a summary is given in section \ref{conlu}.

\section{The Model}\label{model}
After CC explosions or compact binary mergers, the energetic ejecta will sweep up the circumstellar medium (CSM). During the interaction of the ejecta with the CSM, a forward shock with the radius $ R_{\mathrm{b}} $ and a reverse shock with the radius $ R_{\mathrm{r}} $ are generated. The forward shock and reverse shock are separated by the shock contact discontinuity (CD) with the radius $ R_{\mathrm{c}} $. We assume the density $ \rho_{\mathrm{ej}} $ of the ejecta profile and the evolution of the shock is similar to the SNR after compact binary mergers. The ejecta mass $ M_{\mathrm{ej}} $ and the ejecta kinetic energy $ E_{\mathrm{k}} $ are taken as the initial conditions. The outer profile of the expanding ejecta has a power-law density $ \rho_{\mathrm{ej}} \varpropto r^{-n} $, as well as the density profile of the ambient medium density $ \rho_{\mathrm{a}} \varpropto r^{-s} $, where $n$ and $s$ are power-law indices.

At early times, the swept-up mass $ M_{\mathrm{sw}} $ is negligible ($ M_{\mathrm{ej}} \gg M_{\mathrm{sw}} $). If the ejecta has a shallow density profile ($ n<5 $), the early evolution of the shocked shell is characterized by the free expansion (FE) solution $ R_{\mathrm{b}} \varpropto t $ \citep{1963idpbookP,1984ApJ...281..682H}. If the ejecta has a steep density profile ($ n>5 $), the early evolution of the shocked shell is characterized by the the self-similar driven wave (SSDW) solution $ R_{\mathrm{b}} \varpropto t^{(n-3)/(n-s)} $ \citep{1982ApJ...258..790C}. The FE solution has been studied extensively \citep{2017ApJ...847...22Y,Piro_2018}, and we focus on the SSDW solution in this paper. The swept-up mass $ M_{\mathrm{sw}} $ increases with the expanding of blast wave. When $ M_{\mathrm{ej}} \ll M_{\mathrm{sw}} $, the FE or SSDW phase ends and the shocked shell evolves into the self-similar Sedov–Taylor (ST) solution $ R_{\mathrm{b}} \varpropto t^{2/(5-s)} $ \citep{1959sdmm.book.....S,Taylor1946The}. At the end of ST phase, the radiation cannot be ignored any more, and the evolution of the shocked shell will enter the snowplow phase. If the shock expands in the uniform ambient medium, the transition time $ t_{\mathrm{sp}} $ is approximately at \citep{2011piim.book.....D}:
\begin{equation}
 	t_{\mathrm{sp}} \thicksim 5 \times 10^4 \ \left( \frac{E_{\mathrm{k}}}{10^{51} \ \mathrm{erg}}\right)^{0.22} \left( \frac{n_0}{1 \ \mathrm{cm^{-3}}}\right) ^{-0.55} \mathrm{yr},
\end{equation}
where $ n_0 $ is the particle density of the interstellar medium (ISM). The DM evolution in the snowplow phase is beyond the scope of this paper, and the results can be found in \cite{2017ApJ...847...22Y}.

We follow the density profile of the ejecta and the ambient medium in \citetalias{1999ApJS..120..299T}:
\begin{equation}
	\rho(r,t) = \left\{ \begin{array}{ll}
	\rho_{\mathrm{ej}}(r)=\frac{M_{\mathrm{ej}}}{{R_{\mathrm{ej}}}^{3}}f(\frac{r}{R_{\mathrm{ej}}}), &  r \le R_{\mathrm{ej}} \\
$  $	\rho_{\mathrm{a}}(r)=\eta_{\mathrm{s}}r^{-s} & r > R_{\mathrm{ej}},
	\end{array} \right.
	\label{den_prof}
\end{equation}
where $ R_{\mathrm{ej}} $ is the radius of the outer layer of the ejecta, $ \eta_{\mathrm{s}} $ is a constant and $ f(r/R_{\mathrm{ej}}) $ is called the structure function of the ejecta. At the beginning, the ejecta expand freely before encountering the CSM. The structure function can be described in the following power-law profile:
\begin{equation}
	f(w) = \left\{ \begin{array}{ll}
	f_0,                       & 0\le w \le w_{\mathrm{core}}   \\
	f_0(w_{\mathrm{core}}/w)^n & w_{\mathrm{core}}\le w\le 1,   \\
	\end{array} \right.
\end{equation}
where $ w=r/R_{\mathrm{ej}} $ and $ w_{\mathrm{core}}=R_{\mathrm{core}}/R_{\mathrm{ej}} $. Under the assumption of free expansion, $ w_{\mathrm{core}} $ can also be written as $w_{\mathrm{core}} =v_{\mathrm{core}}/v_{\mathrm{ej}} $, where $ v_{\mathrm{core}} $ is the core velocity and $ v_{\mathrm{ej}} $ is the ejecta velocity. The ejecta velocity is several $ 10^4 $ km s$ ^{-1} $ via the observations of SNRs. \cite{2017hsn..book..875C} have found $ v_{\mathrm{core}} \sim 10^3-10^4$ km s$ ^{-1} $. Therefore, the value of $w_{\mathrm{core}}$ is between 0.01 and 0.1. In fact, the difference between the values of $w_{\mathrm{core}}$ is not obvious \citep{2012ApJ...746..130H}, and the approximations of $ w_{\mathrm{core}} \to 0 $ \citep{refId0} and $ w_{\mathrm{core}} \to 1 $ \citep{2003ApJ...597..347L} are both reasonable.
In the FE solution, $ R_{\mathrm{ej}}=R_{\mathrm{c}}=\lambda_{\mathrm{c}}t\sqrt{E_{\mathrm{k}}/M_{\mathrm{ej}}} $, due to mass and energy conservation,
\begin{equation}
	M_{\mathrm{ej}}=\int_{0}^{R_{\mathrm{ej}}}4\pi r^2\rho(r,t)dr,
\end{equation}
\begin{equation}
	E_{\mathrm{k}} = \frac{1}{2}\int_{0}^{R_{\mathrm{ej}}}4\pi r^2 \rho(r,t) (\frac{r}{t})^2dr,
\end{equation}
we obtain
\begin{equation}
	f_0 = \frac{3}{4\pi w_{\mathrm{core}}^n}
	\left[\frac{1-(n/3)}{1-(n/3)w_{\mathrm{core}}^{3-n}}\right] ,
\end{equation}
and
\begin{equation}
	\lambda_{\mathrm{c}}^{2}(n,w_{\mathrm{core}})=2w_{\mathrm{core}}^{-2}
	\left( \frac{5-n}{3-n} \right)
	\left( \frac{w_{\mathrm{core}}^{n-3}-n/3}{w_{\mathrm{core}}^{n-5}-n/5} \right).
\end{equation}
When $ n<3 $, the ejecta core is not necessary and the expression of $ \lambda_{\mathrm{c}} $ can be found in Table \ref{Asymp} under the assumption of $ w_{\mathrm{core}}=0 $.

\subsection{Characteristic scales}\label{chara}
By analogy with a non-radiative SNR, we can define characteristic scales by the following physical variables: the ejecta kinetic energy $ E_{\mathrm{k}} $, ejecta mass $ M_{\mathrm{ej}} $ and the constant $ \eta_{\mathrm{s}} $ describing the ambient medium density $ \rho_{\mathrm{a}}=\eta_{\mathrm{s}}r^{-s} $
\begin{equation}
	M_{\mathrm{ch}}=M_{\mathrm{ej}},
\end{equation}
\begin{equation}
	R_{\mathrm{ch}}={M_{\mathrm{ej}}}^{1/(3-s)}{\eta_{\mathrm{s}}}^{-1/(3-s)},
\end{equation}
\begin{equation}
	t_{\mathrm{ch}}=E_{\mathrm{k}}^{-1/2}{M_{\mathrm{ej}}}^{(5-s)/2(3-s)}{\eta_{\mathrm{s}}}^{-1/(3-s)}.
\end{equation}
Moreover, we can also get the characteristic velocity
\begin{equation}
	v_{\mathrm{ch}}=R_{\mathrm{ch}}/t_{\mathrm{ch}}=(E_{\mathrm{k}}/M_{\mathrm{ej}})^{1/2}.
\end{equation}
For the uniform ambient medium ($ s=0 $), the characteristic radius and time are
\begin{equation}
	R_{\mathrm{ch}}=3.4\ \mathrm{pc}\left( \frac{M_{\mathrm{ej}}}{M_{\odot}}\right) ^{1/3}\left( \frac{m_{{\mathrm{p}}}\ \mathrm{cm^{-3}}}{\eta_{\mathrm{s}}}\right) ^{1/3},
\end{equation}
\begin{equation}
	t_{\mathrm{ch}}=473\ \mathrm{yr}\left( {\frac{10^{51}\ \mathrm{erg}}{E_{\mathrm{k}}}}\right) ^{1/2}\left( \frac{M_{\mathrm{ej}}}{M_{\odot}}\right) ^{5/6}\left( \frac{m_{{\mathrm{p}}}\ \mathrm{cm^{-3}}}{\eta_{\mathrm{s}}}\right) ^{1/3},
\end{equation}
where $ m_{{\mathrm{p}}} $ is the proton mass. For the wind environment ($ s=2 $), $ \eta_{\mathrm{s}}= \dot{M}/4\pi v_{\mathrm{w}} $, where $ \dot{M} $ and $ v_{\mathrm{w}} $ are the mass-loss rate from the progenitor and the wind velocity, respectively. The characteristic scales are
\begin{equation}
	R_{\mathrm{ch}}=12.9\ \mathrm{pc}\ \left( \frac{M_{\mathrm{ej}}}{M_{\odot}}\right) \left( \frac{10^{-5}\ M_{\odot}\ \mathrm{yr^{-1}}}{\dot{M}}\right) \left( \frac{v_{\mathrm{w}}}{10\ \mathrm{km\ s^{-1}}}\right) ,
\end{equation}
\begin{equation}
	t_{\mathrm{ch}}=1772\ \mathrm{yr}\ \left( {\frac{10^{51}\ \mathrm{erg}}{E_{\mathrm{k}}}}\right) ^{1/2}\left( \frac{M_{\mathrm{ej}}}{M_{\odot}}\right) ^{3/2}\left( \frac{10^{-5}\ M_{\odot}\ \mathrm{yr^{-1}}}{\dot{M}}\right) \left( \frac{v_{\mathrm{w}}}{10\ \mathrm{km\ s^{-1}}}\right) .
\end{equation}
In section \ref{appro}, we will express the physical quantity $ X $ in terms of the characteristic scales, i.e. $ X^{*}=X/{M_{\mathrm{ch}}}^{x_{1}}{R_{\mathrm{ch}}}^{x_{2}}{t_{\mathrm{ch}}}^{x_{3}} $, where $ X^{*} $ is the dimensionless quantity and $ x_{1} $,$ x_{2} $ and $ x_{3} $ are constants given by dimensional analysis.

\subsection{Approximate Solutions}\label{appro}
When $ t\to 0 $, if the density envelope of the ejecta is very shallow ($ n<5 $), the forward shock radius can be well described by the FE solution $R_{\mathrm{b}}^*=\lambda_{\mathrm{b}}t^*$ \citep{1963idpbookP,1984ApJ...281..682H}. If the density envelope of the ejecta is much steeper ($ n>5 $), we should use the SSDW solution \citep{1982ApJ...258..790C} $  R_{\mathrm{b}}^*=\zeta_{\mathrm{b}}t^{*{(n-3)/(n-s)}} $, where $ \lambda_{\mathrm{b}} $ and $ \zeta_{\mathrm{b}} $ are dimensionless constants. When $ t\to\infty $, ST solution \citep{1959sdmm.book.....S,Taylor1946The} is a good analytical approximation $ R_{\mathrm{b}}^*= {(\xi t^{*2})}^{1/(5-s)}  $. A detailed description of the asymptotic behavior is listed in Table \ref{Asymp}.

\begin{figure}
	\centering
	\includegraphics[width = 0.5\textwidth]{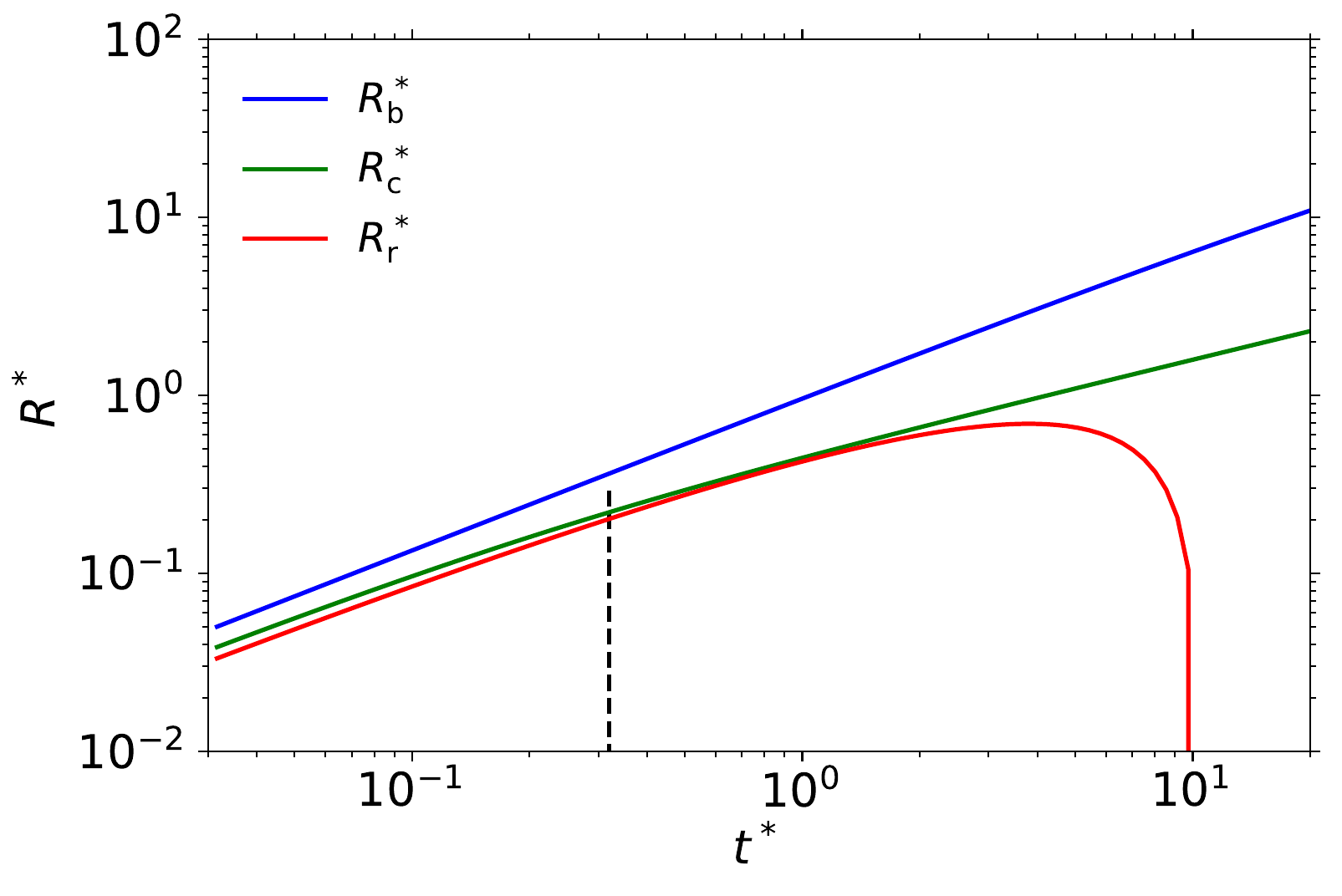}
	\caption{Dimensionless radius of the forward shock $ R_{\mathrm{b}}^* $, CD $ R_{\mathrm{c}}^* $ and reverse shock $ R_{\mathrm{r}}^* $ in the wind environment ($ s=2 $) for $ n=9 $. The transition time is $ t_{\mathrm{tran}}^{*}=0.32 $ in this case.
	}
	\label{fig:Rs2}
\end{figure}

Based on the asymptotic solutions, \cite{2017MNRAS.465.3793T} gave the analytical approximate solution of the evolution of the forward shock radius for $ n>5 $:
\begin{equation}
R_{\mathrm{b}}^*(t^*)=[(\zeta_{\mathrm{b}}t^*)^{-2\alpha}+(\xi t^*)^{-2\alpha/(5-s)}]^{-1/2 \alpha},
\end{equation}
where the values of parameters $ \zeta_{\mathrm{b}} $, $ \xi $ and $ \alpha $ depend on the power-law index $ n $ and $ s $. They can be found in tables 3 and 4 in \cite{2017MNRAS.465.3793T} for a uniform medium and the wind environment, respectively. The transition time $ t_{\mathrm{tran}}^{*} $ from SSDW solution to the ST solution is
\begin{equation}
t_{\mathrm{tran}}^{*} = \left(  \frac{\xi}{\zeta_{\mathrm{b}}^{5-s}}\right) ^{(n-s)/(n-5)(3-s)}.
\end{equation}
The asymptotic behavior of the CD in ST phase is more complex, and \cite{2017MNRAS.465.3793T} have found that a simple power-law ($ ct^{*b} $) provides a good fit to numerical simulations.
Thus, the approximate solutions of the CD radius is
\begin{equation}
R_{\mathrm{c}}^*(t^*)=[(\zeta_{\mathrm{c}}t^{*(n-3)/(n-s)})^{-a}+(c t^{*b})^{-a}]^{-1/a},
\end{equation}
where the values of parameters $ \zeta_{\mathrm{c}} $, $ a $, $ b $ and $ c $ can also be found in tables 5 and 6 in
\cite{2017MNRAS.465.3793T} for different $ n $ and $ s $. The approximate solutions of $ R_{\mathrm{b}}^* $ and $ R_{\mathrm{c}}^* $ only depend on the asymptotic behavior and they are less affected by the different dynamical structures of the ejecta. Because of the uncertainty of the asymptotic behavior of the reverse shock when $ t \to \infty $, we cannot get the analytical approximate solution of $ R_{\mathrm{r}}^* $ in the same way. The approximate solution for the reverse shock can be found in TM model (\citetalias{1999ApJS..120..299T}; \citeauthor{2000ApJS..128..403T} \citeyear{2000ApJS..128..403T}; \citeauthor{2012ApJ...746..130H} \citeyear{2012ApJ...746..130H}; \citeauthor{refId0} \citeyear{refId0}). For example, we use the results of \citetalias{1999ApJS..120..299T}. When the reverse shock wave does not hit the ejecta core ($ t^*<t_{\mathrm{core}}^* $), $R_{\mathrm{r}}^*$ satisfies
\begin{equation}
R_{\mathrm{r}}^*(t^*)=R_{\mathrm{b}}^*(t^*)/l_{\mathrm{ED}},
\end{equation}
where
\begin{equation}
l_{\mathrm{ED}}=1+\frac{8}{n^{2}}+\frac{0.4}{4-s}
\end{equation}
is the lead factor. When $ t^*>t_{\mathrm{core}}^* $, reverse shock radius $R_{\mathrm{r}}^*$ is
\begin{equation}
R_{\mathrm{r}}^*(t^*)=t^{*}\left[ \frac{R_{\mathrm{r,core}}^{*}}{t_{\mathrm{core}}^{*}}-\tilde{a}_{\mathrm{r,core}}(t^{*}-t_{\mathrm{core}}^{*})-(\tilde{v}_{\mathrm{r,core}}-\tilde{a}_{\mathrm{r,core}}t_{\mathrm{core}}^{*})\ln(\frac{t^{*}}{t_{\mathrm{core}}^{*}})\right] ,
\end{equation}
where $ R_{\mathrm{r,core}} $, $ \tilde{v}_{\mathrm{r,core}}$ and $\tilde{a}_{\mathrm{r,core}} $ are the radius, the velocity and acceleration (in the frame of the unshocked ejecta) when the reverse shock reach the ejecta core, respectively. The values of $ t_{\mathrm{core}}^* $, $ R_{\mathrm{r,core}} $, $ \tilde{v}_{\mathrm{r,core}}$ and $\tilde{a}_{\mathrm{r,core}} $ are provided in table 6 of \citetalias{1999ApJS..120..299T} in the uniform medium. The forward shock radius $ R_{\mathrm{b}}^* $, CD radius $ R_{\mathrm{c}}^* $ and reverse shock radius $ R_{\mathrm{r}}^* $ in uniform medium for $ n=10 $ and $ n=6 $ are shown in Figure \ref{fig:Rs0}. For the wind environment, we use the solutions in \cite{refId0}, which are tabulated in Table \ref{Rr}. Follow the study of Cas A from \cite{2003ApJ...597..347L} and \cite{refId0}, $ n=9 $ is considered in this work. The evolution of $ R_{\mathrm{b}}^* $, $ R_{\mathrm{c}}^* $ and $ R_{\mathrm{r}}^* $ in the wind environment for $ n=9 $ are shown in Figure \ref{fig:Rs2}.

\section{DM and RM Evolution}\label{dm&rm}
For cosmological-origin FRBs,  the observed DM contributed by different parts
\begin{equation}
	\mathrm{DM_{obs}} = \mathrm{DM_{MW}} +\mathrm{DM_{IGM}} +\mathrm{DM_{HG}^{obs}} + \mathrm{DM_{local}^{obs}},
\end{equation}
where $ \mathrm{DM_{MW}} $ is from the Milky Way Galaxy,  $\mathrm{DM_{IGM}}$ is the component of the intergalactic medium (IGM),  $ \mathrm{DM_{HG}^{obs}} $ and $ \mathrm{DM_{local}^{obs}} $ are contributed by the host galaxy and the local environment of the source in the frame of observers, respectively. And similarly, RM is given by
\begin{equation}
	\mathrm{RM_{obs}} = \mathrm{RM_{MW}} +\mathrm{RM_{IGM}} +\mathrm{RM_{HG}^{obs}} + \mathrm{RM_{local}^{obs}}.
\end{equation}
In this part, we  derive the DM and RM evolution contributed from the local environment in the frame of sources. The conversions of two frames are: DM$_{\text{source}} $ = DM$_{\text{obs}}(1+z) $, RM$_{\text{source}} $ = RM$_{\text{obs}}(1+z)^2 $ and $t_{\text{source}} $ = $t_{\text{obs}}(1+z)^{-1} $. Henceforth, all values are in source frame and subscripts `source' are omitted unless otherwise specified.

The core radius $R_{\mathrm{core}} $, the reverse shock radius $ R_{\mathrm{r}} $, the CD radius $ R_{\mathrm{c}} $ and the forward shock radius $ R_{\mathrm{b}} $ divide the space around the young magnetar into five regions: unshocked ejecta core ($ r<R_{\mathrm{core}} $), external power-law unshocked ejecta ($ R_{\mathrm{core}}<r<R_{\mathrm{r}} $), shocked ejecta ($ R_{\mathrm{r}}<r<R_{\mathrm{c}} $), shocked ISM ($ R_{\mathrm{c}}<r<R_{\mathrm{b}} $) and unshocked ISM ($ r>R_{\mathrm{b}} $). In this paper, we focus on the DM and RM contributed by the local environment ($ r<R_{\mathrm{b}} $) of the FRB source after the compact binary stars merger or CC SNe.

\subsection{DM from the unshocked regions}
\begin{figure}
	\centering
	\includegraphics[width = 0.5\textwidth]{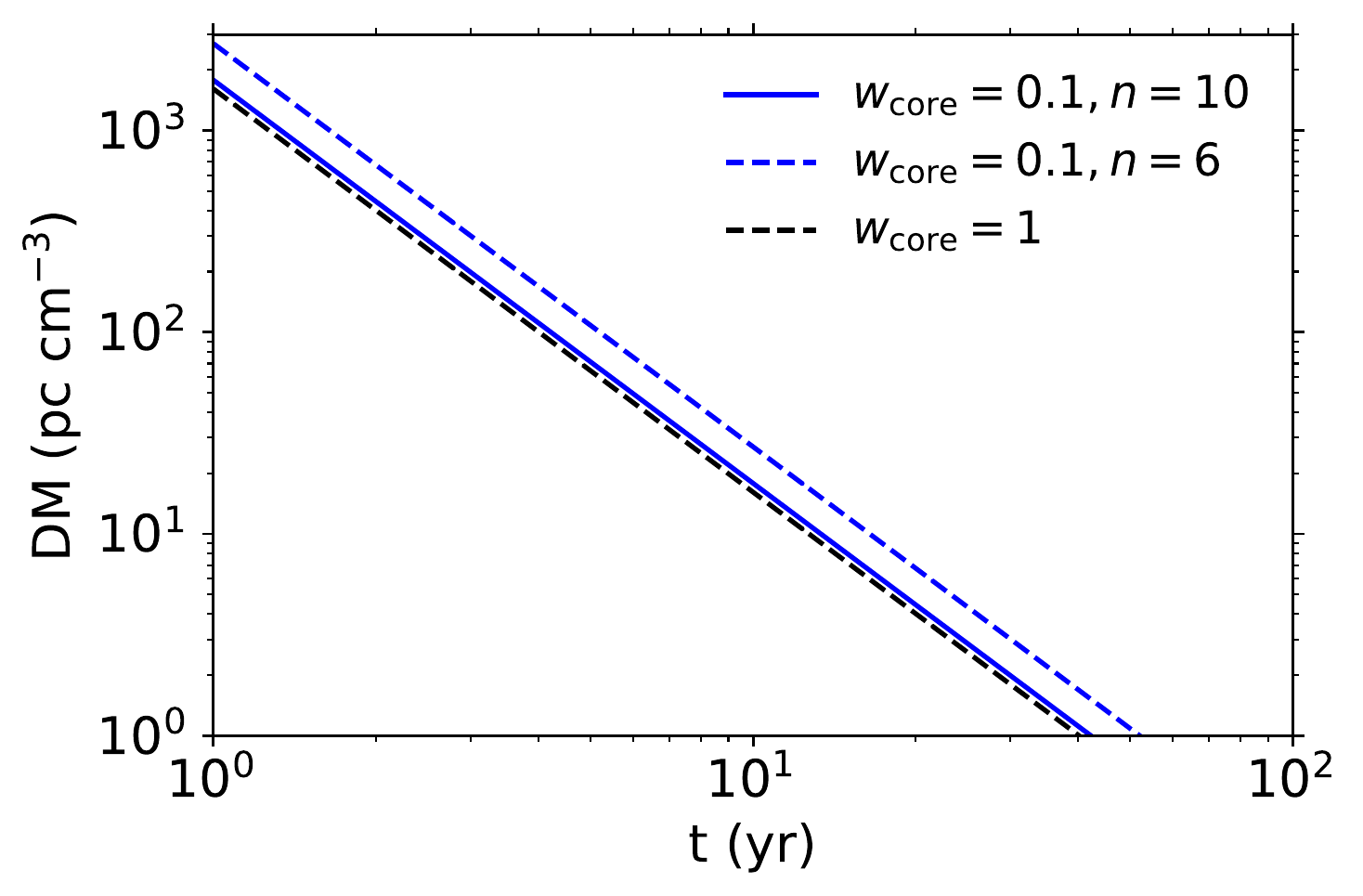}
	\caption{The DM from the unshocked ejecta for $ M \sim M_{\odot} $, $ E \sim 10^{51} $ erg, and $ \eta =0.03 $. The solid and dashed blue lines illustrate the DM for different ejecta structure, $w_{\mathrm{core}}=0.1,n=10$ and $w_{\mathrm{core}}=0.1,n=6$, respectively. The dashed black line represents the case of a constant ejecta density (assuming the ejecta without structure, $  w_{\mathrm{core}} \to 1 $). }
	\label{DMej}
\end{figure}

For unshocked ejecta \citep{2017ApJ...847...22Y,Wang_2020}, the DM will decrease with time (DM$_{\mathrm{unsh,ej}} \varpropto t^{-2}  $ ), so only the early evolution is observable. Here, the initial free expanding ejecta is considered. According to the density profile equation (\ref{den_prof}), the DM from unshocked ejecta , including the contribution of unshocked core $ \mathrm{DM_{core}} $ and the contribution of unshocked external power-law envelope $ \mathrm{DM_{pl}} $, is
\begin{equation}
\mathrm{DM_{unsh,ej}} = \int_{0}^{R_{\mathrm{core}}} \frac{M_{\mathrm{ej}}}{\mu m_pR_{\mathrm{ej}}^3} \eta f_0dr + \int_{R_{\mathrm{core}}}^{R_{\mathrm{r}}} \frac{M_{\mathrm{ej}}}{\mu m_pR_{\mathrm{ej}}^3} \eta f_0 \left( \frac{r}{R_{\mathrm{core}}} \right)^{-n} dr,
\end{equation}
where $ \eta $ is the ionization fraction and $ \mu $ is the mean atomic weight of the ejecta. In the FE solution $ R_{\mathrm{ej}}=R_{\mathrm{c}} $, the DM due to the unshocked core can be written as
\begin{equation}
\mathrm{DM_{core}}  = \frac{M_{\mathrm{ej}}}{\mu m_p} \eta f_0w_{\mathrm{core}}   \lambda_{\mathrm{c}}^{-2}v_{\mathrm{ch}}^{-2}t^{-2}.
\end{equation}
The reverse shock radius $ R_{\mathrm{r}} $ in FE solution is $ R_{\mathrm{r}}=q_{\mathrm{r}}R_{\mathrm{c}} $, where $ q_{\mathrm{r}}=q_{\mathrm{b}}/l_{\mathrm{ED}} $. Therefore, the DM from the external power-law envelope can be written as
\begin{equation}
\mathrm{DM_{pl}}  = \frac{M_{\mathrm{ej}}}{\mu m_p} \eta f_0\frac{w_{\mathrm{core}}-w_{\mathrm{core}}^nq_{\mathrm{r}}^{1-n}}{n-1}\lambda_{\mathrm{c}}^{-2}v_{\mathrm{ch}}^{-2}t^{-2}.
\end{equation}
In the limit $ w_{\mathrm{core}} \to 1 $, the structure function is $ f(n)=f_0 \to 3/4 \pi $, which means that the ejecta density is a constant. Therefore, DM from the ejecta is
\begin{equation}
\mathrm{DM_{unsh,ej}}= \frac{3M_{\mathrm{ej}}}{4 \pi \mu m_p  }\eta q_{\mathrm{r}}\lambda_{\mathrm{c}}^{-2}v_{\mathrm{ch}}^{-2}t^{-2}.
\end{equation}
The ejecta is not magnetized in general, so the RM from the unshocked regions is negligible. In Figure \ref{DMej}, we present the DM from the unshocked ejecta for $ M \sim M_{\odot} $, $ E_{\mathrm{k}} \sim 10^{51} $ erg. The ionization fraction $ \eta \sim 0.03 $ is estimated by \cite{2017hsn..book..875C} for SN 1993J. We also assume the unshocked ejecta of different structures have a similar $ \eta  $. The solid and dashed blue lines illustrate the DM for the ejecta with $w_{\mathrm{core}}=0.1,n=10$ and $w_{\mathrm{core}}=0.1,n=6$, respectively. The solid black line represents the case of a constant ejecta density ($  w_{\mathrm{core}} \to 1 $). The structure of the ejecta density has very little effect on the DM from the unshocked ejecta, and it is detectable only in the first few decades after the explosion. Therefore, in the subsequent discussion of the DM$ _{\mathrm{{uhsh,ej}}} $ in merger scenarios, although there are few observations to constrain $ w_{\mathrm{core}} $, assuming the ejecta density without structure is feasible.

\subsection{DM and RM from the shocked regions}
In the shocked regions, matter is ionized because of the high temperature $ T>10^6$ K \citep{1982ApJ...258..790C,2012A&ARv..20...49V}. Therefore, the DM of surrounding environment is mainly contributed by the shocked shell decades after the merger or CC explosion, including the contribution of the shocked ejecta $ \mathrm{DM_{sh,ej}} $ and the contribution of the shocked ISM $ \mathrm{DM_{sh,ISM}} $:
\begin{equation}
\mathrm{DM_{sh}}=\int_{R_{\mathrm{r}}}^{R_{\mathrm{c}}}n_{e,\mathrm{r}}(r)dr+\int_{R_{\mathrm{c}}}^{R_{\mathrm{b}}}n_{e,\mathrm{b}}(r)dr,
\end{equation}
where $ n_{e,\mathrm{r}} $ is the electron density of the shocked ejecta between $ R_{\mathrm{r}} $ and $ R_{\mathrm{c}} $, and $ n_{e,\mathrm{b}} $ is the electron density of the shocked ISM between $ R_{\mathrm{c}} $ and $ R_{\mathrm{b}} $. For strong shock waves, the shocked matter is compressed 4 times. The electron density of the shocked ISM is
\begin{equation}
	n_{e,\mathrm{b}} =4n_{0}= \frac{4 \eta_{s}r^{-s}}{\mu m_{\mathrm{p}}},
	\label{n_s}
\end{equation}
where $ n_0 $ is the particle density of ISM and $ \mu $ is the mean atomic weight of the ambient medium. In the thin shell approximation of the shocked shell \citep{1982ApJ...258..790C}, $ n_{e,\mathrm{r}} $ is
\begin{equation}
	n_{e,\mathrm{r}} =\frac{(n-3)(n-4)}{(3-s)(4-s)}n_{e,\mathrm{b}}.
	\label{n_r}
\end{equation}
Therefore, the DM from the shocked ejecta is
\begin{equation}\label{DM_sh,ej}
	\mathrm{DM_{sh,ej}}=\frac{4 \eta_{s}}{\mu m_{\mathrm{p}}}\frac{(n-3)(n-4)}{(3-s)(4-s)(1-s)}(R_{\mathrm{c}}^{1-s}-R_{\mathrm{r}}^{1-s}),
\end{equation}
and the DM from the shocked ISM is
\begin{equation}\label{DM_sh,ISM}
	\mathrm{DM_{sh,ISM}}=\frac{4 \eta_{s}}{\mu m_{\mathrm{p}}}\frac{1}{1-s}(R_{\mathrm{b}}^{1-s}-R_{\mathrm{c}}^{1-s}),
\end{equation}
where $ R=R^*R_{\mathrm{ch}} $ is presented in the usual units. The dimensionless radii are given in \ref{appro}, and the characteristic radius is determined by the initial conditions.

The shocked shell will amplify the magnetic field during the expansion \citep{2013SSRv..178..201B,2014ApJ...794...46C}. The energy of the magnetic field in the shocked region is converted from a fraction $ \epsilon_{\mathrm{B}} $ of the shock energy:
\begin{equation}
	\frac{B^2}{8\pi}=\epsilon_{\mathrm{B}}u_{th},
\end{equation}
where $ u_{th}=9 \rho_{\mathrm{a}}v_b^2/8 $ and $ v_{\mathrm{b}}= v_{\mathrm{ch}}dR_{\mathrm{b}}/dt $ is the velocity of the forward shock. We assume the component of magnetic field along the line of sight $ B_{\parallel} \sim B $. Therefore $ B_{\parallel}$ is
\begin{equation}
B_{\parallel} \sim B=\sqrt{9\pi \eta_{s} \epsilon_{\mathrm{B}}}v_{\mathrm{b}}r^{-s/2}.
\end{equation}

The RM from the shocked shell is
\begin{equation}
	\mathrm{RM_{sh}}=K\left(  \int_{R_{\mathrm{r}}}^{R_{\mathrm{c}}}n_{e,\mathrm{r}}(r)B_{\parallel}dr+\int_{R_{\mathrm{c}}}^{R_{\mathrm{b}}}n_{e,\mathrm{b}}(r)B_{\parallel}dr\right) ,
	\label{RM}
\end{equation}
where $ K=\frac{e^3}{2\pi m_e^3c^4} $, $ e $ is the electric charge, $ m_e $ is the mass of an electron and $ c $ is the speed of light in vacuum.
The RM from the shocked ejecta is
\begin{equation}\label{RM_sh,ej}
	\mathrm{RM_{sh,ej}}=K\sqrt{9\pi \eta_{s} \epsilon_{\mathrm{B}}}v_{\mathrm{b}} \frac{4 \eta_{s}}{\mu m_{\mathrm{p}}}\frac{2(n-3)(n-4)}{(3-s)(4-s)(2-3s)}(R_{\mathrm{c}}^{1-3s/2}-R_{\mathrm{r}}^{1-3s/2}),
\end{equation}
and the RM from the shocked ISM is
\begin{equation}\label{RM_sh,ISM}
	\mathrm{RM_{sh,ISM}}=K\sqrt{9\pi \eta_{s} \epsilon_{\mathrm{B}}}v_{\mathrm{b}} \frac{4 \eta_{s}}{\mu m_{\mathrm{p}}}\frac{2}{2-3s}(R_{\mathrm{b}}^{1-3s/2}-R_{\mathrm{c}}^{1-3s/2}).
\end{equation}

\subsubsection{uniform medium}\label{merger}
For different kinds of compact binary (BNS, BWD, and NSWD) mergers that may be associated with FRBs, the different initial conditions determine the evolution of shocked shell, which play an important role in the long-term evolution of $ \mathrm{DM_{sh}} $ and $ \mathrm{RM_{sh}} $.

Due to no good constrains on the nuclear equation of state (EOS), the ejecta mass and the kinetic energy of BNS mergers are highly uncertain. The relativistic hydrodynamical simulations of \cite{Bauswein_2013} investigated 40 representative EOSs, and found the ejecta mass of BNS merger ranges from $ \sim 10^{-3}-10^{-2}M_{\odot} $ with the kinetic energy between $ \sim 5 \times 10^{49} $ erg and $ 10^{51} $ erg. The numerical relativity study of \cite{Radice_2018} found that $ \sim 10^{-3} M_{\odot}  $ material is ejected with the velocity $ v_{\mathrm{ej}} \sim 0.2c $ and energy $ \sim 5 \times 10^{49} $ erg after BNS mergers. A more detailed description of the ejecta can be found in the study of \cite{2017Natur.551...80K}. Two different ejection mechanisms should be considered. When stars get close, the matter with masses $ \sim 10^{-3}-10^{-2} M_{\odot}  $ in the polar regions will be accelerated to $ \sim 0.2-0.3c $ and escape from a neutron star; after the merger, matter in the accretion disk will be blown away, with the velocities of $ 0.05-0.1\ c $ and masses of $ \sim 0.01 -0.1M_{\odot}  $. In this work, we will not consider the detailed ejecta mechanism and assume the ejecta is isotropic for simplicity.

A single NS or magnetar can also be born via the AIC process after the merger of BWD \citep{2016MNRAS.463.3461S,2019MNRAS.484..698R}. The study of \cite{2009MNRAS.396.1659M} has shown that Nickel-rich outflows with the ejecta mass $ M \sim 10^{-2}M_{\odot} $ expand at a typical speed $ v \sim 0.1c $ after BWD merger. MHD simulations of the AIC of a rapidly rotating WD have found that the ejecta mass can be up to $ \sim 0.1 M_{\odot} $ with the kinetic energy $\sim$ 10$^{50} $ erg \citep{2007ApJ...669..585D}.

The NSWD mergers have been less studied because both physical outcomes and observable expectations are not well known at this point. A WD, especially with the small mass $ <0.2 M_{\odot} $ \citep{2017MNRAS.467.3556B} could be tidal disrupted by a NS and the WD debris is sheared into an accretion disc \citep{2016MNRAS.461.1154M}. The simulation of the time-dependent one-dimensional accretion discs model implies a fraction of the initial WD mass $ \sim 0.1 M_{\odot} $ has been ejected at a characteristic velocity of $ \sim 10^{9} $ cm s$ ^{-1} $. The 2D hydrodynamical-thermonuclear simulations of \cite{10.1093/mnras/stz316} have found the explosive transients of NSWD mergers are weak ($ 10^{48}-10^{49} $ erg) with low ejecta mass ($ \sim 0.01-0.1 M_{\odot} $).

In summary, there are some uncertainties about the values of ejecta mass and the kinetic energy during compact binary mergers. In this paper, we choose five kinds of typical initial conditions, with the ejecta mass range of $ 0.001 -0.1 M_{\odot} $ and the kinetic energy range of $ 10^{49}-10^{51} $ erg, and the details are shown in Table \ref{initial}.

The evolution of DM and RM depends not only on the initial conditions, but also on the density profile power-law index $ n $. The characteristic scales and transition time of different models for $ n= 10$ and $ n=6 $ are tabulated in Table \ref{ch}. After the merger, the ejecta electron density with the velocity $ v $ is \citep{Wang_2020}
\begin{equation}
n_{\mathrm{e}} \simeq \frac{\eta Y_{\mathrm{e}}M}{4 \pi m_{\mathrm{p}}v^3t^3}=2.8 \ \mathrm{cm^{-3}}\eta Y_{\mathrm{e,0.2}}M_{-3}v_{0.2}^{-3}t_{\mathrm{yr}}^{-3},
\end{equation}
where $ Y_{\mathrm{e}}=0.2Y_{\mathrm{e,0.2}} $ is the electron fraction, $ M_{-3}=M/10^{-3}\mathrm{M_{\odot}} $, and $ v_{0.2}=v/0.2c $. The DM from the unshocked ejecta is
\begin{equation}
	\mathrm{DM_{unsh}}=n_e \Delta R \simeq 0.17 \ \mathrm{pc \ cm^{-3}} \eta q_{\mathrm{r}} Y_{\mathrm{e,0.2}}M_{-3}v_{0.2}^{-2}t_{\mathrm{yr}}^{-2},
\end{equation}
where $ \Delta R = R_{\mathrm{r}} \simeq q_{\mathrm{r}} vt $ represents the region where the ejecta is not swept by the reverse shock wave. In the case of BNS merger, the DM from unshocked ejecta is negligible ($ <1 \ \mathrm{pc \ cm^{-3}} $) on account of the ejecta expanding at a high velocity. On the contrary, the ${\mathrm{DM_{unsh,ej}}}  $ from BWD or NSWD merger is observable in a few years or decades after the merger.

\begin{figure}
	\centering
	\includegraphics[width = 0.5\textwidth]{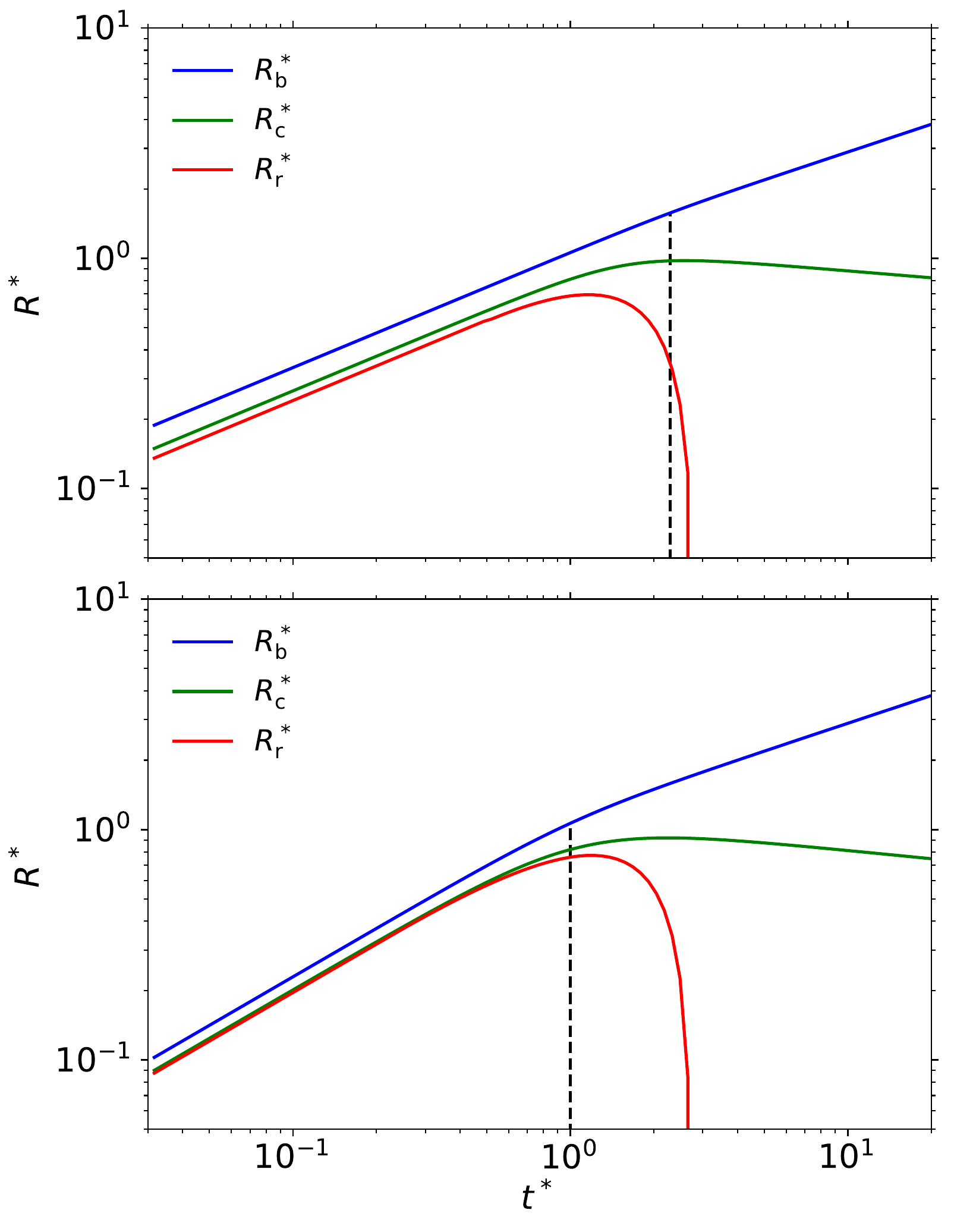}
	\caption{Dimensionless radius of the forward shock $ R_{\mathrm{b}}^* $, CD $ R_{\mathrm{c}}^* $ and reverse shock $ R_{\mathrm{r}}^* $ in a uniform medium ($ s=0 $). The radius evolution is shown for the power-law index $ n=6 $ (top panel) and $ n=10 $ (bottom panel). The $ x- $axis is the dimensionless time $ t^* $ in the range $ 0.03<t^*<20 $. The transition time $ t_{\mathrm{tran}}^{*} $  from SSDW solution to ST solution for $ n=6 $ is $ t_{\mathrm{tran}}^{*}=2.29 $ and for $ n=10 $ is $ t_{\mathrm{tran}}^{*}=1.0 $, which is characterized in the black vertical dashed line.}
	\label{fig:Rs0}
\end{figure}

\begin{figure}
	\centering
	\includegraphics[width = 1\textwidth]{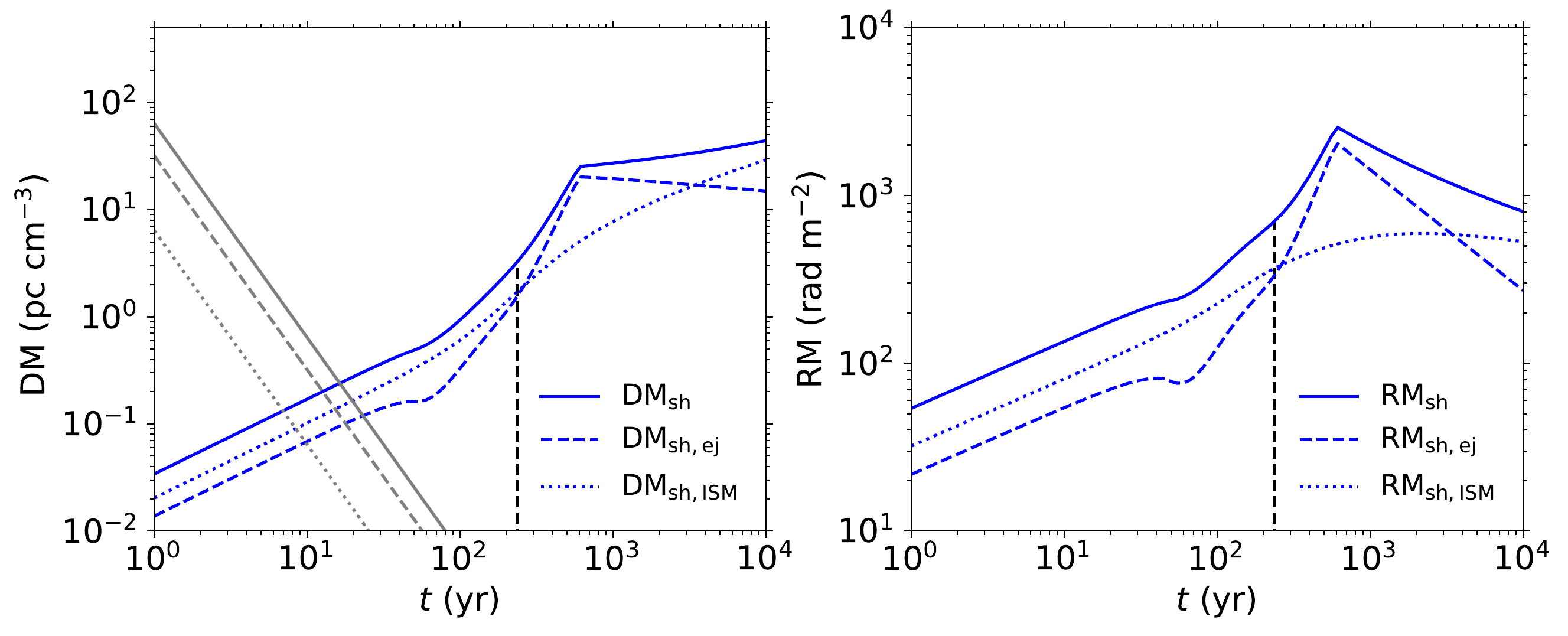}
	\caption{DM (left panel) and RM (right panel) evolution for Case C (BWD/NSWD merger, $ M_{\mathrm{ej}}=0.1M_{\odot}  $, $ E_k = 10^{50} $ erg) , $ n=10 $, $ n_0 = 1 $ cm$ ^{-3} $, $ \mu \sim 1 $ and $ \epsilon_{\mathrm{B}}=0.1 $. The dashed and dotted blue lines illustrate the DM and RM from the shocked ejecta and the shocked ISM, respectively. The solid blue lines show the total contributions from the shocked shell. The solid, dashed and dotted gray lines illustrate the DM from the unshocked ejecta with the ionization fractions of 100\%, 50\% and 10\%, respectively. The black vertical dashed lines represent the transition time $ t_{\mathrm{tran}}^{*} $  from SSDW solution to ST solution. }
	\label{con}
\end{figure}

Taking the limit $ t \to 0 $, the SSDW given in Table \ref{Asymp} is a good approximation, and DM from the shocked ejecta is given by
\begin{equation}
\mathrm{DM_{sh,ej}}=\frac{(n-3)(n-4)}{3}n_0R_{\mathrm{ch}}(1-r_2)\zeta_{\mathrm{c}} \left( \frac{t}{t_{\mathrm{ch}}}\right) ^{(n-3)/n},
\end{equation}
and DM from the shocked ISM is
\begin{equation}
\mathrm{DM_{sh,ISM}}=4n_0R_{\mathrm{ch}}(r_1-1)\zeta_{\mathrm{c}} \left( \frac{t}{t_{\mathrm{ch}}}\right) ^{(n-3)/n},
\end{equation}
where $ r_1=R_1/R_c $ and $ r_2=R_2/R_c $ can be found in \cite{1982ApJ...258..790C}. At this stage, we have $ \mathrm{DM_{sh}} =\mathrm{DM_{sh,ej}}+\mathrm{DM_{sh,ISM}} \varpropto t^{(n-3)/n} (6 \le n \le 14)$, which means for $ n > 5 $, $ \mathrm{DM_{sh}} $ will increase. In SSDW solution, the velocity of the forward shock is
\begin{equation}
v_{\mathrm{b}}=\frac{n-3}{n-s}\frac{R_{\mathrm{b}}}{t}.
\end{equation}
Therefore, RM from the shocked ejecta is
\begin{equation}
\mathrm{RM_{sh,ej}}=K\sqrt{9\pi \eta_{s} \epsilon_{\mathrm{B}}}n_0 \frac{ 2(n-3)^2(n-4)}{3n}r_1(1-r_2) \left( \frac{R_{\mathrm{ch}} \zeta_{\mathrm{c}}}{t_{\mathrm{ch}}^{(n-3)/n}}\right)^2 t^{(n-6)/n} ,
\end{equation}
and the RM from the shocked ISM is
\begin{equation}
\mathrm{RM_{sh,ISM}}=K\sqrt{9\pi \eta_{s} \epsilon_{\mathrm{B}}}n_0 \frac{4 (n-3)}{n}r_1(r_1-1) \left( \frac{R_{\mathrm{ch}} \zeta_{\mathrm{c}}}{t_{\mathrm{ch}}^{(n-3)/n}}\right)^2 t^{(n-6)/n} .
\end{equation}
in the early SSDW phase. When $ t \to 0 $, the RM from the shocked shell is $ \mathrm{RM_{sh}} =\mathrm{RM_{sh,ej}}+\mathrm{RM_{sh,ISM}} \varpropto t^{(n-6)/n} (6 \le n \le 14)$. When $ n $ is equal to 6, $ \mathrm{RM_{sh}}$ is going to be constant at early time. For $ n>6 $,  $ \mathrm{RM_{sh}}$ will go up with time. The asymptotic behavior of the forward shock in the limit $ t \to \infty $ is the ST solution \citep{Taylor1946The,1959sdmm.book.....S}, but the asymptotic behavior of the CD and reverse shock is not clear at this point. Therefore, the analytical approximations of DM and RM cannot be given by similar methods in this work. However, following the assumptions of the CD radius when $ t \to \infty $ in \cite{2017MNRAS.465.3793T} and considering $ R_{\mathrm{r}} \to 0 $ when $ t \to \infty $ (see Figure \ref{fig:Rs0}), we can get the long-term evolution of DM and RM. The evolution of DM and RM from the unshocked ejecta and the shocked shell for Case C (BWD/NSWD merger) with $ n_0 = 1 $ cm$ ^{-3} $, $ \mu \sim 1 $ and $ \epsilon_{\mathrm{B}}=0.1 $ is shown in Figure \ref{con}. The dashed and dotted blue lines illustrate the DM and RM from the shocked ejecta with a power-law index of $ n=10 $ and the shocked ISM, respectively. The solid blue lines show the total contributions from the shocked shell. The solid, dashed and dotted gray lines illustrate the DM from the unshocked ejecta with the ionization fractions of 100\%, 50\% and 10\%, respectively. The discontinuity of DM$_\mathrm{sh,ej}$ occurs at around 600 yr, which is caused by the simplification of $ R_{\mathrm{r}}$. The evolution of $ R_{\mathrm{r}}$ is hard to describe in the later phase because how the reverse shock approaches the center is not clear at this point \citep{2017MNRAS.465.3793T}. To get the long-term evolution of DM$_\mathrm{sh,ej}$, we assume $ R_{\mathrm{r}} \to  $ 0 when $ t$ is very large (see Figure \ref{fig:Rs0}). This assumption will not have a big impact on the value of DM$_\mathrm{sh,ej}$, but the discontinuity is unavoidable. In a uniform medium, the CD radius will decrease at large $ t^* $ (see Figure \ref{fig:Rs0}), and that is why DM$_{\mathrm{sh,ej}}  $ will decrease in the later ST phase. Both RM$_{\mathrm{{sh,ej}}} $ and RM$_{\mathrm{{sh,ISM}}}$ decrease in the later ST phase due to the reduction of forward shock wave velocity. When $ t>t_{\mathrm{core}} $, the forward and reverse shock radius no longer maintain a fixed ratio. As the forward shock wave propagates ahead, the reverse shock wave gradually approaches the remnant center, and the shocked region expands continuously. Thus the DM and RM from the shocked shell in this work is larger than the results of \cite{10.1093/mnras/stz3593}. The DM and RM evolution of different models are shown in Figures \ref{DMRMn10s0} and \ref{DMRMn6s0} for $ n=10 $ and $ n=6 $, respectively. The solid lines illustrate the total DM and RM from the shocked regions. The contributions of DM from the unshocked ejecta are presented in dash-dotted lines with ionization fractions of 50 percent. The top and bottom panels show the ejecta interacting with the ambient medium of $ n_0 =5 $ cm$ ^{-3} $ and $ n_0 =0.1 $ cm$ ^{-3} $, respectively.

\subsubsection{wind environment}\label{wind}
Massive stars ($ >8 \mathrm{M_{\odot}} $) usually end their lives and produce a NS or a magnetar via CC SNe. While sometimes the explosions occur in the uniform medium, in many cases the wind of the progenitor will change the circumstellar environment significantly. Therefore, the CSM density profile can be written as $ \rho_{\mathrm{a}}=\eta_{\mathrm{s}}r^{-2} $, and the value of $ \eta_{\mathrm{s}} = \dot{M}/4 \pi v_{\mathrm{w}} $ is often assumed to be a constant. The SSWD solutions of the wind environment ($ s=2 $) in Table \ref{Asymp} are also applied at an early time, and the DM from the shocked ejecta is
\begin{equation}
	\mathrm{DM_{sh,ej}}= \frac{2 \eta_{\mathrm{s}} (n-3)(n-4)}{\mu m_{\mathrm{p}}} \left(\frac{1}{r_2}-1 \right) \frac{1}{\zeta_{\mathrm{c}} R_{\mathrm{ch}}} \left( \frac{t_{\mathrm{ch}}}{t} \right) ^{(n-3)/(n-2)}.
\end{equation}
Also, the DM from the shocked ISM is
\begin{equation}
	\mathrm{DM_{sh,ISM}}= \frac{4 \eta_{\mathrm{s}}}{\mu m_{\mathrm{p}}} \left(1-\frac{1}{r_1} \right) \frac{1}{\zeta_{\mathrm{c}} R_{\mathrm{ch}}} \left( \frac{t_{\mathrm{ch}}}{t} \right) ^{(n-3)/(n-2)},
\end{equation}
where $ r_1=R_1/R_c $ and $ r_2=R_2/R_c $ is the ratio of the radius for $ s=2 $, and the values can be found in \cite{1982ApJ...258..790C}.

\begin{figure}
	\centering
	\includegraphics[width = 1\textwidth]{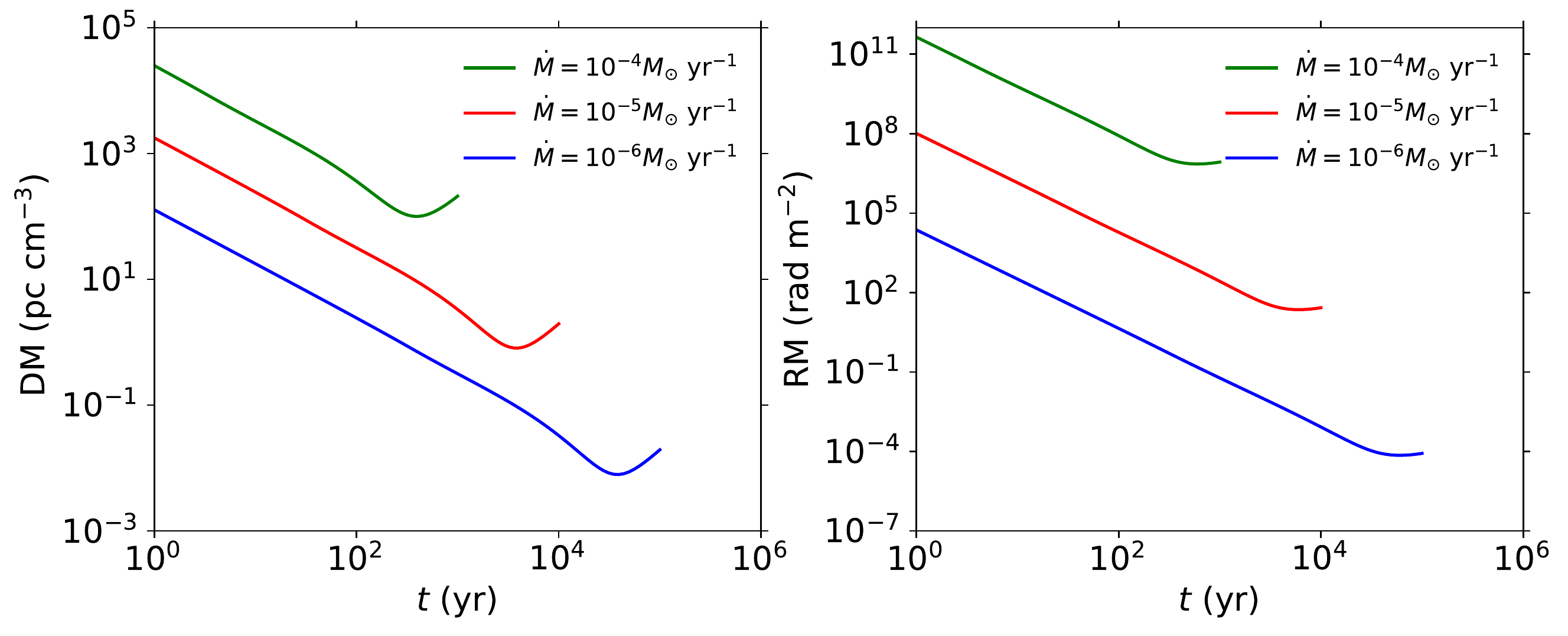}
	\caption{DM (left panel) and RM (right panel) evolution for $ n=9 $, $ M_{\mathrm{ej}}= 2 M_{\odot} $,  $ E=2.2 \times 10^{51} $ erg, $ v_{\mathrm{w}}=10 $ km s$ ^{-1} $ and $ \epsilon_{\mathrm{B}}=0.1 $
	 in the wind environment. The green, red, and blue solid lines represent different mass-loss rate of $ \dot{M} = 1 \times 10^{-4} $, $ \dot{M} = 1 \times 10^{-5} $, and $ \dot{M} = 1 \times 10^{-6} \ M_{\odot }\mathrm{ \ yr^{-1}}$, respectively.}
	\label{DMRMn9s2}
\end{figure}

The evolution of RM from the shocked ejecta is
\begin{equation}
	\mathrm{RM_{sh,ej}}=K\sqrt{9\pi \eta_{s} \epsilon_{\mathrm{B}}} \frac{ \eta_{s}(n-3)^2(n-4)}{\mu m_{\mathrm{p}}(n-2)}\left[ r_1 \left( \frac{1}{r_2^2}-1\right) \right] \frac{t_{\mathrm{ch}}^{(n-3)/(n-2)}}{R_{\mathrm{ch}} \zeta_{\mathrm{c}}}t^{(5-2n)/(n-2)} ,
\end{equation}
and the RM from the shocked ISM is
\begin{equation}
\mathrm{RM_{sh,ISM}}=K\sqrt{9\pi \eta_{s} \epsilon_{\mathrm{B}}} \frac{2 \eta_{s}}{\mu m_{\mathrm{p}}}\frac{n-3}{n-2}\left( r_1- \frac{1}{r_1}\right)\frac{t_{\mathrm{ch}}^{(n-3)/(n-2)}}{R_{\mathrm{ch}} \zeta_{\mathrm{c}}}t^{(5-2n)/(n-2)} .
\end{equation}
For the same reason, we will not give analytic expressions in ST phase anymore. Following the study of Cas A from \cite{refId0}, we consider the ejecta mass $ M_{\mathrm{ej}}=2 M_{\odot} $, the  explosion energy $ E=2.2 \times 10^{51} $ erg, and the power-law index $ n=9 $. The characteristic scales and transition time for different mass-loss rates are tabulated in Table \ref{CCch}. The DM and RM from the shocked shell are shown in Figure \ref{DMRMn9s2}. The green, red, and blue solid lines represent the cases of $ \dot{M} = 1 \times 10^{-4} $, $ \dot{M} = 1 \times 10^{-5} $, and $ \dot{M} = 1 \times 10^{-6} \ M_{\odot }\mathrm{ \ yr^{-1}}$, respectively. From equation (\ref{DM_sh,ej}), we know that DM $ \varpropto R^{-1} $ for the wind environment. When the reverse shock approaching the remnant center, DM increases. Although \cite{Piro_2018} have predicted the increase of the DM in late evolution, the reason is different from our model. In their study, the increase of DM from the shocked ISM in the later period offset the decrease of DM from expanding ejecta in the uniform medium, leading to the increase of overall DM. In this work, the DM from both shocked ejecta and the shocked shell decrease in the wind environment at an early time, but DM$_{\mathrm{sh}}$ will increase when the reverse shock approaches the remnant center. It is worth noting that the subsequent evolution is hard to describe because how the reverse shock approaches the center is not clear \citep{2017MNRAS.465.3793T}.

\section{Comparison to observations}\label{obs}
\subsection{FRB 121102}
\subsubsection{The DM and RM observations}
The DM of FRB 121102 was found to be $557.4 \pm 2.0  $ pc cm$^{-3}  $ in 2012 \citep{2014ApJ...790..101S}, while the recent observations of DM have shown an increase of about one percent over the past several years \citep{2019ApJ...882L..18J,2019ApJ...876L..23H,2020A&A...635A..61O}. The age of the magnetar associated with FRB 121102 is found to be very young ($ < 100 $ yr) \citep{2017ApJ...839L..20C,2017ApJ...839L...3K,2017ApJ...841...14M}, which suggests that it is difficult to observe the increase of DM except in very dense environments in BWD or NSWD merger because of the non-negligible $ \mathrm{DM_{unsh,ej}}$. However, in the case of compact binary mergers, the progenitors are in a very clean environment. For example, through the study of GRB 170817A afterglow, the ambient density is found to be $ \lesssim 10^{-3}-10^{-2} $ cm$ ^{-3} $ \citep{2019ApJ...886L..17H,2020arXiv200602382M}. For the population study of short gamma-ray bursts, the median density of circumburst medium is $ n_0 =0.15$ cm$ ^{-3} $, and most of them are lower than 10 cm$ ^{-3} $ \citep{2014ARA&A..52...43B}. In this work, we set the upper limit of the ISM density around compact binaries $ n_0 = 5$ cm$ ^{-3} $, which is much smaller than the value $ n_0 = 50$ cm$ ^{-3} $ used by \cite{10.1093/mnras/stz3593}.

Furthermore, the RM of FRB 121102 has been found to decay rapidly, $ \sim 10\%$ drop in seven months
\citep{2018Natur.553..182M} and $ \sim 34\%$ in 2.6 years \citep{2020arXiv200912135H}.
The value of RM ($ \sim 10^5 $ rad cm$^{-2} $) is shown in black horizontal lines in Figures \ref{DMRMn10s0} and \ref{DMRMn6s0}. In the case of relatively low ejecta energy (BWD/NSWD), the contribution from the shocked shell is far below $ \sim 10^5 $ rad cm$ ^{-2} $.  The large variation of RM could not be explained simply by the evolution of shock shell, and the central magnetar activity should be considered. \cite{2018ApJ...868L...4M} have proposed that FRB 121102 is embedded in an expanding magnetized electron-ion nebula, which can explain the rapid decay of RM. If the magnetar activity energy input rate is $ \dot{E} \varpropto t^{-\delta} $, RM from the wind nebula (WN) reads
\begin{equation}\label{RM_WN}
\mathrm{RM_{WN}} \approx 1.9\times 10^{37} \left(\frac{E_{\mathrm{B}}}{10^{50}\  \mathrm{erg}}\right) ^{3/2} \left(\frac{v_{\mathrm{n}}}{10^{8}\  \mathrm{cm\ s^{-1}}}\right) ^{-7/2} (\delta-1)^{3/2}t_0^{(\delta-1)/2}t^{-(6+\delta)/2} \ \mathrm{rad \ m^{-2}},
\end{equation}
where $E_{\mathrm{B}} $ is the free magnetic energy of the magnetar, $ v_{\mathrm{n}} $ is the velocity of expanding nebula, $ t_0 $ is the time in seconds describing the nebula becomes active how long after the energy is released, and $ t $ is in seconds since the magnatar was born. The magnetic energy of a magnetar is $ E_{\mathrm{B}} \simeq B_{\star}^2R_{\star}^3/6 \approx 3 \times 10^{49}B_{16}^2 $ erg, where the interior magnetic field $ B_{\star}=B_{16} \times 10^{16} $ G should obey $ B_{\star} \gtrsim 2 \times 10^{16} $ G to satisfy the measured RM \citep{2018ApJ...868L...4M}, and $ R_{\star} \sim 12$ km is the typical radius of a NS. For a FE nebula, the velocity obeys $ v_{\text{n}} \sim (0.1-1)v_{\text{ej}} \lesssim 10^4$ km s$ ^{-1} $. The onset of magnetar's active period is assumed to be $ t_0 \sim 0.2-0.6 $ yr \citep{2018ApJ...868L...4M}. For $t_{\text{age}} \sim 10 $ yr and $ \delta \gtrsim 1 $, the RM$ _{\text{WN}} $ is $ \sim 10^{5} $ rad m$ ^{-2} $.
The time-dependent radio luminosity and the $  \mathrm{RM_{WN}}$ of FRBs with different ages can be well explained in this model \citep{2020ApJ...900L..26W}. The DM due to the nebula \citep{2020arXiv200912135H} is
\begin{equation}\label{DM_WN}
	\mathrm{DM_{WN}}=30  (\delta-1)\left( \frac{E_{\mathrm{B}}}{10^{50}\ \text{erg}}\right) \left( \frac{v_{\text{n}}}{10^{8} \ \mathrm{cm \ s^{-1}}}\right) ^{-2}\left(\frac{t}{10 \text{ yr}} \right) ^{-2} \ \mathrm{pc \ cm^{-3}}.
\end{equation}
In addition, the decrease of RM seems to be not steady. The decline was rapid at first, and then showed a slow downward trend \citep{2020arXiv200912135H}. In the study of \cite{2018ApJ...868L...4M}, the magnetar was produced by the CC of a massive star. However, the RM from both the wind nebula and SN shocked shell show a rapid decrease, which is not consistent with slow decreasing behavior at a later stage. To make matters worse, the growth of DM cannot be well understood from SNR or wind nebula. As mentioned earlier, the magnetar can also be embedded in the uniform medium if the magnetar is the product of the compact binary merger. In this case, especially BNS mergers, the DM and RM observations could be well explained simultaneously.

\subsubsection{Fitting methods}
We use the MCMC method performed through $\mathtt{emcee}$\footnote{ \url{http://emcee.readthedocs.io/}} Python package \citep{2013PASP..125..306F} to estimate the parameters of the model. The MCMC method could efficiently sample from the posterior $N$-dimensional parameter space and provide sampling approximations to the posterior probability of the parameters. For the very large RM of FRB 121102, the RM is mainly contributed by the local environment
\begin{equation}
	\mathrm{RM_{local}}=\mathrm{RM_{WN}}+\mathrm{RM_{sh}} \approx \mathrm{RM_{obs}}(1+z)^2.
\end{equation}
However, the DM from the local environment is only a small component of DM$_{\mathrm{obs}}  $, and the remaining components are highly uncertain. For comparison with observations, we assume that the contribution of other regions DM$_{\mathrm{other}}=  \mathrm{DM_{MW}} +\mathrm{DM_{IGM}} +\mathrm{DM_{HG}}/(1+z)  $ is constant. So the total DM is
\begin{equation}
	\mathrm{DM_{total}} = \mathrm{DM_{other}} + \frac{\mathrm{DM_{sh}}+\mathrm{DM_{WN}}}{1+z}.
\end{equation}
Considering the DM of FRB 121102 was first measured on MJD 56233 \citep{2014ApJ...790..101S}, while RM was first measured on MJD 57747 \citep{2018Natur.553..182M}, we replace $ t'+t_{\mathrm{age}} $ with $ t $ in Equations (\ref{DM_sh,ej}), (\ref{DM_sh,ISM}) and (\ref{DM_WN}), and $ t'+t_{\mathrm{age}}+4.1 $ with $ t $ in Equations (\ref{RM_sh,ej}), (\ref{RM_sh,ISM}) and (\ref{RM_WN}), where $ t_{\mathrm{age}} $ is the age of the FRB 121102 when it was first observed.

\begin{figure}
	\centering
	\includegraphics[width = 1\textwidth]{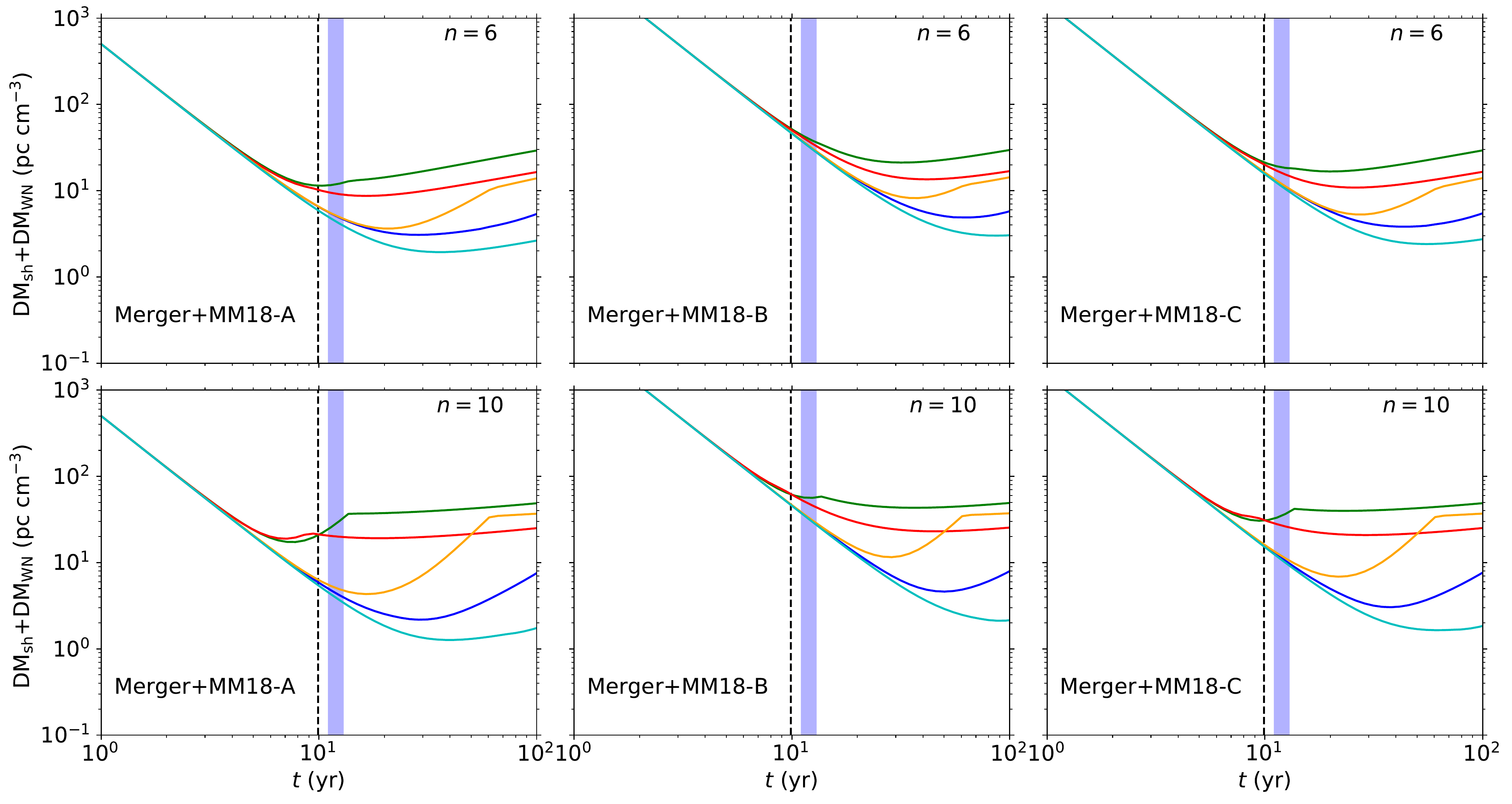}
	\caption{DM evolution of the shocked shell and the wind nebula for different models. The green, red, blue, orange and cyan solid lines represent the merger Case A, B, C, D, and E given in Table \ref{initial}, respectively. The ISM density is $ n_0 $=5 cm$ ^{-3} $. The age of source estimated by \cite{2019ApJ...885..149Y} and \cite{2020arXiv200912135H} is characterized in black vertical dashed lines and purple ranges, respectively. }
	\label{DMshWN}
\end{figure}

\begin{figure}
	\centering
	\includegraphics[width = 1\textwidth]{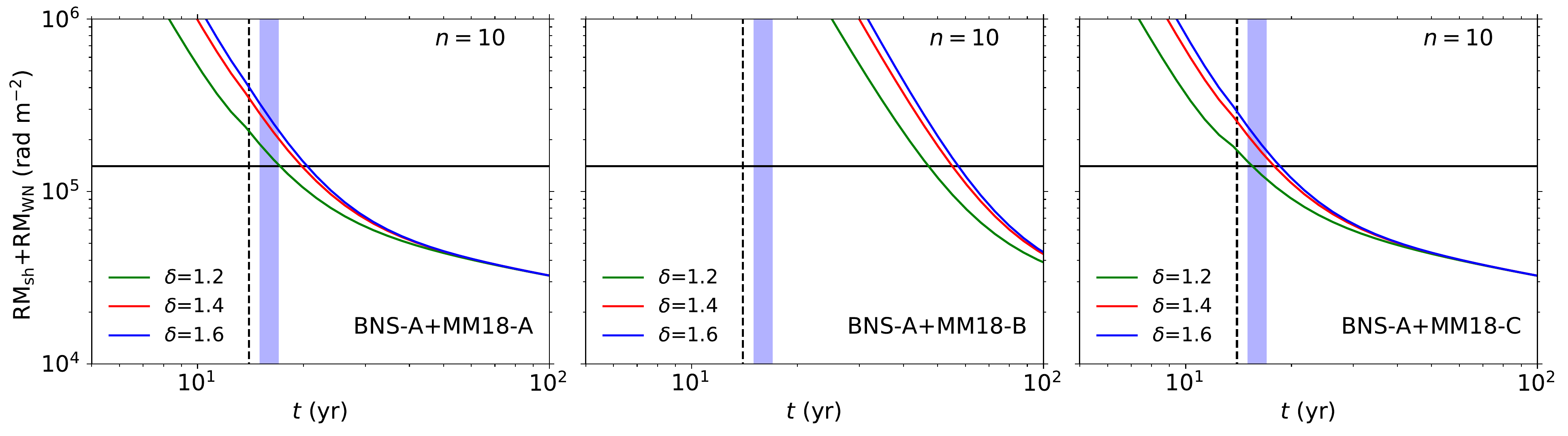}
	\caption{RM evolution of the shocked shell and the wind nebula for BNS-A with $ n=10, n_0=5 $ cm$ ^{-3} $ and $ \epsilon_{\mathrm{B}}=0.01 $. The age of source estimated by \cite{2019ApJ...885..149Y} and \cite{2020arXiv200912135H} is characterized in black vertical dashed lines and purple ranges, respectively. The black horizontal lines represent the value of RM when it was first measured. }
	\label{RMshWN}
\end{figure}

Before fitting, we can restrict some model parameters by the evolution trend of DM and RM. The parameters $E_{\mathrm{B}}$, $ v_{\mathrm{n}} $ and $ t_0 $ are from models A, B, C in \cite{2018ApJ...868L...4M} (hereafter, MM18-A, MM18-B and MM18-C). These models can successfully explain the large value of measured RM and the age was strictly limited, such as $ t_{\text{age}} \sim 14 $ yr from \cite{2019ApJ...885..149Y} and $ t_{\text{age}} \sim 15-17 $ yr from \cite{2020arXiv200912135H} (their $ t $ starts on MJD 57747, the time when RM was first measured). In Figure \ref{DMshWN}, the DM evolution from the shocked shell and wind nebula of different models are shown. The green, red, blue, orange, and cyan solid lines represent the merger cases A, B, C, D, and E given in Table \ref{initial}, respectively. Although the ISM density is dense ($ n_0=5 $ cm$ ^{-3} $), the growth of DM cannot be found in the cases of B, C, D, and E. From observations, DM increases about 1 pc cm$ ^{-3} $ per year. To satisfy both the age constraints and the increasing trend of DM, only BNS merger Case A (hereafter, BNS-A to avoid confusion) with large $ n $ is possible. $ n=10 $ is chosen in our work \citep{2017Natur.551...80K}.
In Figure \ref{RMshWN}, the RM evolution for BNS-A with different wind nebula models is shown. The black horizontal lines represent the value of RM when it was first measured. Considering the constraints of $ t_{\text{age}}$ and RM, only the parameters of MM18-A and MM18-C are reasonable.

The MCMC method is used to estimate the parameters $ \mathrm{DM_{other}} $, $ n_0 $, $t_{\mathrm{age}} $, $ \epsilon_{\mathrm{B}} $, and $ \delta $. The $ \chi_{\mathrm{DM}}^2 $ for the observed DM is given by
\begin{equation}
	\chi_{\mathrm{DM}}^2=\sum_{i=1}^{12}\dfrac{\left(\mathrm{DM^{th}}_i-\mathrm{DM^{obs}}_i\right)^2 }{\sigma_{\mathrm{{DM^{obs}}}_i}^2},
\end{equation}
and the $ \chi_{\mathrm{RM}}^2 $ is
\begin{equation}
	\chi_{\mathrm{RM}}^2=\sum_{i=1}^{20}\dfrac{\left( \mathrm{RM^{th}}_i-\mathrm{RM^{s}}_i \right)^2 }{\sigma_{\mathrm{{RM^{s}}}_i}^2},
\end{equation}
where $ \mathrm{DM^{th}} $ and $ \mathrm{RM^{th}} $ are theoretical values given by our models,  $ \mathrm{DM^{obs}} $ and $ \sigma_{\mathrm{{DM^{obs}}}} $ are the measured values and uncertainties of DM in the frame of observers, and $ \mathrm{RM^{s}} $ and $ \sigma_{\mathrm{{RM^{s}}}} $ are measured values and uncertainties of RM in the frame of the source. We collected 12 observation data points from different telescopes \citep{2014ApJ...790..101S,2016Natur.531..202S,2016ApJ...833..177S,2017ApJ...850...76L,2018ApJ...863....2G,2019ApJ...876L..23H,2019ApJ...877L..19G,2019ApJ...882L..18J,2020A&A...635A..61O}, and we made a weighted average of measured DMs for adjacent time. The RMs of measured 20 bursts are taken from table 1 in \cite{2020arXiv200912135H}. The total likelihood is $ \mathcal{L} \varpropto \exp\left[	-\left( \chi_{\mathrm{DM}}^2+\chi_{\mathrm{RM}}^2\right)/2 \right]  $. We use the $ \mathtt{scipy} $ Python package \citep{2020NatMe..17..261V} to estimate the initial values of parameters by maximizing the likelihood. The uniform priors are used for all the parameters.

\subsubsection{Fitting results}
\begin{figure}
	\centering
	\includegraphics[width = 1\textwidth]{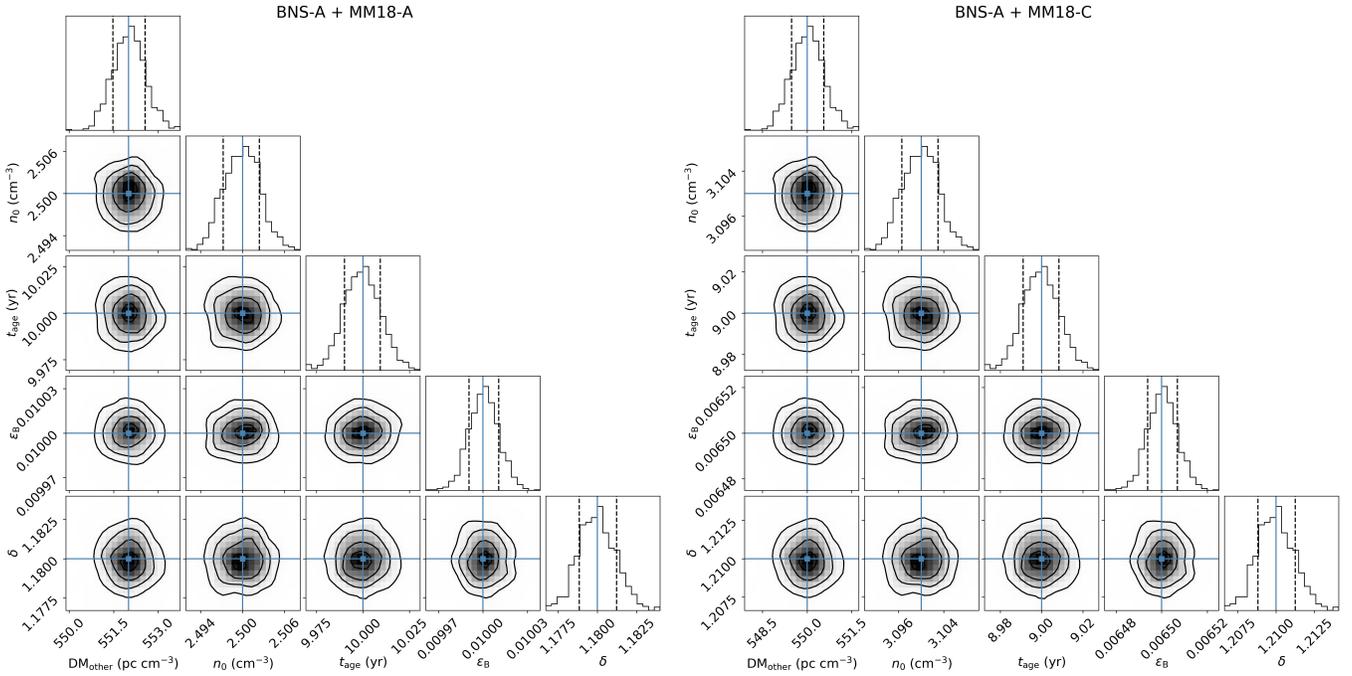}
	\caption{Posterior corner plot for the parameters of BNS-A+MM18-A (left panel) and BNS-A+MM18-C (right panel): $ \mathrm{DM_{other}} $, the ISM density $ n_0 $, the age of the magnetar $t_{\mathrm{age}} $, the ratio of shock kinetic energy to magnetic energy $ \epsilon_{\mathrm{B}} $, and the power-law index of energy released $ \delta $ for BNS-A ( $ M_{\mathrm{ej}} \sim 0.01 M_{\odot}$, $ E_{\mathrm{k}} \sim 10^{51} $ erg)  with $ n=10$. The histograms indicate the posterior probability of each parameter, with the dashed lines denoting the 1$\sigma$ range. The plots show the explored parameter space, with 1, 2, and 3$\sigma$ solid contours, obtained by the maximum likelihood estimates.}
	\label{corner}
\end{figure}

\begin{figure}
	\centering
	\includegraphics[width = 1\textwidth]{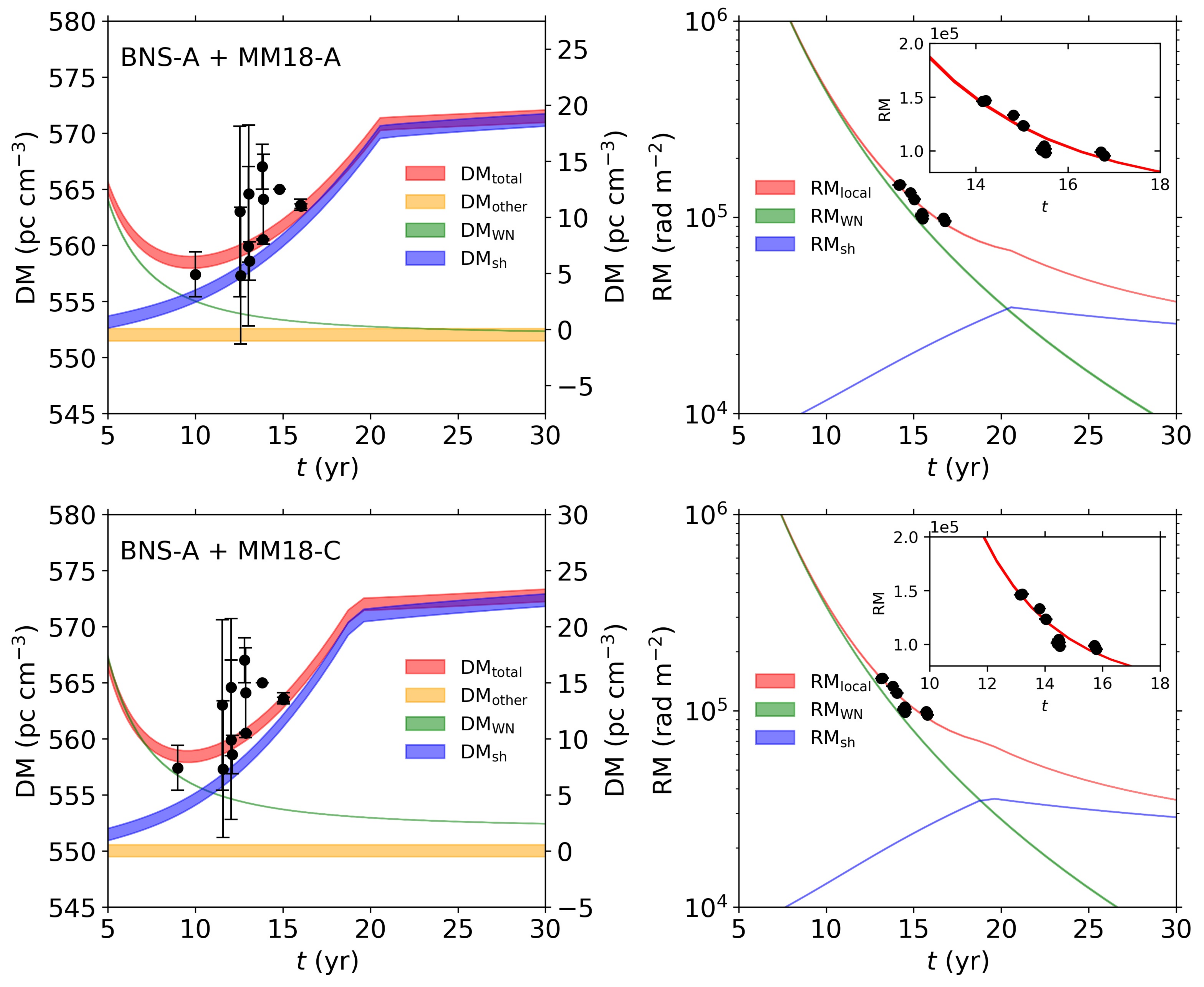}
	\caption{The DM (left panel) and RM (right panel) evolution for the BNS-A ( $ M_{\mathrm{ej}} \sim 0.01 M_{\odot}$, $ E_{\mathrm{k}} \sim 10^{51} $ erg) with $ n=10$. Uncertainties with 1$\sigma$ are shown for DM and RM. The top and bottom panels represent different WN models. The black circles represent measured DMs and RMs of FRB 121102 from \citet{2014ApJ...790..101S,2016Natur.531..202S,2016ApJ...833..177S,2017ApJ...850...76L,2018ApJ...863....2G,2018Natur.553..182M,2019ApJ...876L..23H,2019ApJ...877L..19G,2019ApJ...882L..18J,2020A&A...635A..61O,2020arXiv200912135H}. \emph{left plane}: The red range represents the DM$ _{\mathrm{total}}$. The orange range shows the DM$_{\mathrm{other}} $, which is constant. The left $ y-$axis shows the value of the DM$ _{\mathrm{total}}$ and DM$_{\mathrm{other}} $. The green and blue ranges represent DM$_{\mathrm{WN}} $ and DM$_{\mathrm{sh}} $, respectively.  The evolution of observed DM is mainly caused by the wind nebula and the shocked shell, whose values are shown in the right $ y-$axis. \emph{right plane}: The blue, green and red areas represent the RM$ _{\mathrm{sh}} $ and RM$ _{\mathrm{WN}}$ and RM$ _{\mathrm{local}}$, respectively. The rapid decay of the RM is caused by RM$ _{\mathrm{WN}}$ at first, and the subsequent slow decrease is associated with the behavior of RM$ _{\mathrm{sh}} $. } 	\label{121102}
\end{figure}

The fitting results are listed in Table \ref{results}. The parameters are constrained effectively for the following reasons: the measured structure-optimized DMs and RMs having small errors, and simultaneously fitting of DM and RM. The posterior corner plots are shown in Figures \ref{corner}, and the values of each parameter given by the maximum likelihood estimations are shown by blue solid lines. The parameter ranges with 1-$ \sigma $ are also marked with dashed lines in the histograms. The age $ t_{\mathrm{age}} $ is found to be very young ($ \sim 9-10 $ years), which is consistent with the result $ \sim 14 $ years of \cite{2019ApJ...885..149Y} and $ \sim 15-17 $  years of \cite{2020arXiv200912135H}(their $ t $ starts on MJD 57747, the time when RM was first measured). The ranges of possible DM$_{\mathrm{obs}}$ (in the frame of observers) and RM$ _{\mathrm{local}} $ (in the frame of the source) with 1$ \sigma $ errors are plotted in Figures \ref{121102}. We find DM$_{\mathrm{other}}   \sim 550-552$ pc cm$ ^{-3} $, and the DM from the shocked shell is $ 3$ pc cm$ ^{-3} $ when it was first measured. The DM from the wind nebula is $ \sim3-5$ pc cm$ ^{-3} $, and it decays rapidly over the next few years. The total value of DM has a $ 6-7 $ pc cm$ ^{-3} $ increase in the following several years, which is similar to the observations of \cite{2019ApJ...882L..18J} (having a $\sim 1\% $ growth).

For FRB 121102, the RM from the local environment is dominated by $ \mathrm{RM_{WN}}$ at the beginning, which leads to a rapid decrease in RM. However, after a while, the RM$_{\text{local}} $ will be dominated by $ \mathrm{RM_{sh}}$ because of the fast decay of  $ \mathrm{RM_{WN}}$. At that point, the RM will show a slow change. $ \delta \sim 1.1-1.2 $ is consistent with the values $ \gtrsim 1 $ of \cite{2018ApJ...868L...4M}. The derived $ \epsilon_{\mathrm{B}}$ is $ \sim 0.0065-0.01 $, which implies the RM contributed by the shocked shell is $ (1-4) \times 10 ^4 $ rad m $ ^{-2}$.

We compare our results with previous works. We found the DM and RM evolution of FRB 121102 would be well understood in the assumption of BNS merger progenitors, which is consistent with the predictions of BNS merger (remnants) powering FRBs \citep{2018PASJ...70...39Y,2020arXiv201009214S}. Different from the assumption that the DM and RM evolution is from different regions \citep{2020arXiv200912135H}, the DM increase and RM decrease are all caused by the local environment in this work. In addition, \cite{2020arXiv200912135H} only considered the rapid evolution models to explain the RM evolution, including the SNR model from \cite{Piro_2018} and wind nebula model from \cite{2018ApJ...868L...4M}, which makes it difficult to fit the slow reduction of RM in the later period without adding a large error ($ \sim 10^4 $ rad m$ ^{-2} $) due to instrumental or other kinds of noise processes. In this work, the introduction of slowly changing RM$ _{\mathrm{sh}} $ can avoid the above difficulties, so no additional error is needed. In \cite{10.1093/mnras/stz3593}, they can only explain the DM evolution. A large value of circumstellar density 50 cm$^{-3}$ for BWD mergers are used to explain the large RM, and the rapid decrease is caused by evolving from the free expansion to the ST phase. The variations of the DM and RM can also result from the motion of plasma \citep{Katz_2020} which changes $ n_{\mathrm{e}} $ and $ B_{\parallel}$ of the line of sight.

\subsection{FRB 180916}
\begin{figure}
	\centering
	\includegraphics[width = 1\textwidth]{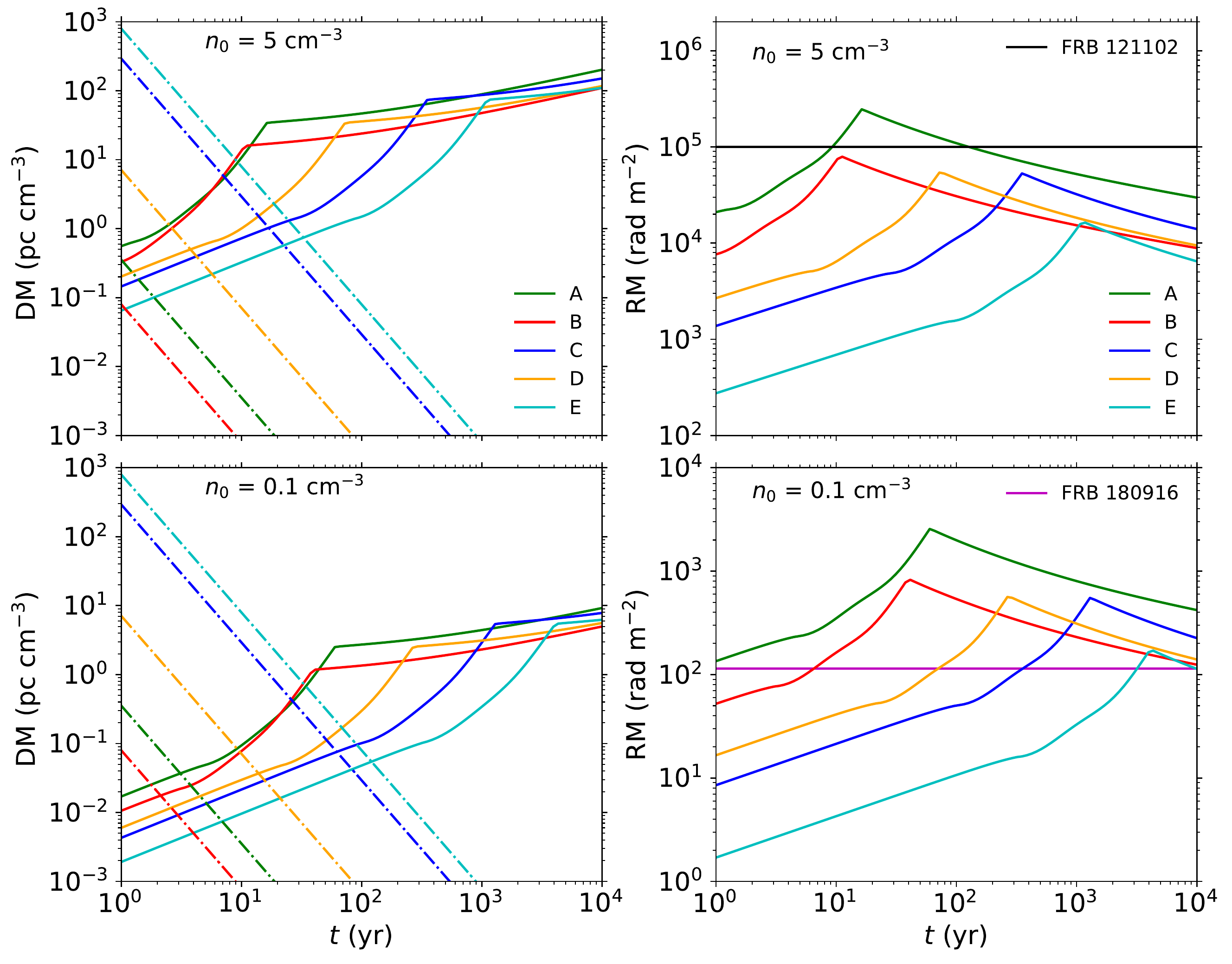}
	\caption{Evolution of DM (left panel) and RM (right panel) for different models of binary compact stars mergers in the uniform medium. The solid lines illustrate the contributions from the shocked shell. The DM from the unshocked ejecta with $ \eta=50 \% $ are shown in dash-dotted lines. Here we assume the ejecta having $ w_{\mathrm{core}} \to 1 $ and $ n=10 $. The top and bottom panels represent the ISM densities of 5 and 0.1 cm$ ^{-3} $, respectively. The RM of FRB 121102/180916 are displayed with black/magenta horizontal lines in the right top/bottom panels.  }
	\label{DMRMn10s0}
\end{figure}

\begin{figure}
	\centering
	\includegraphics[width = 1\textwidth]{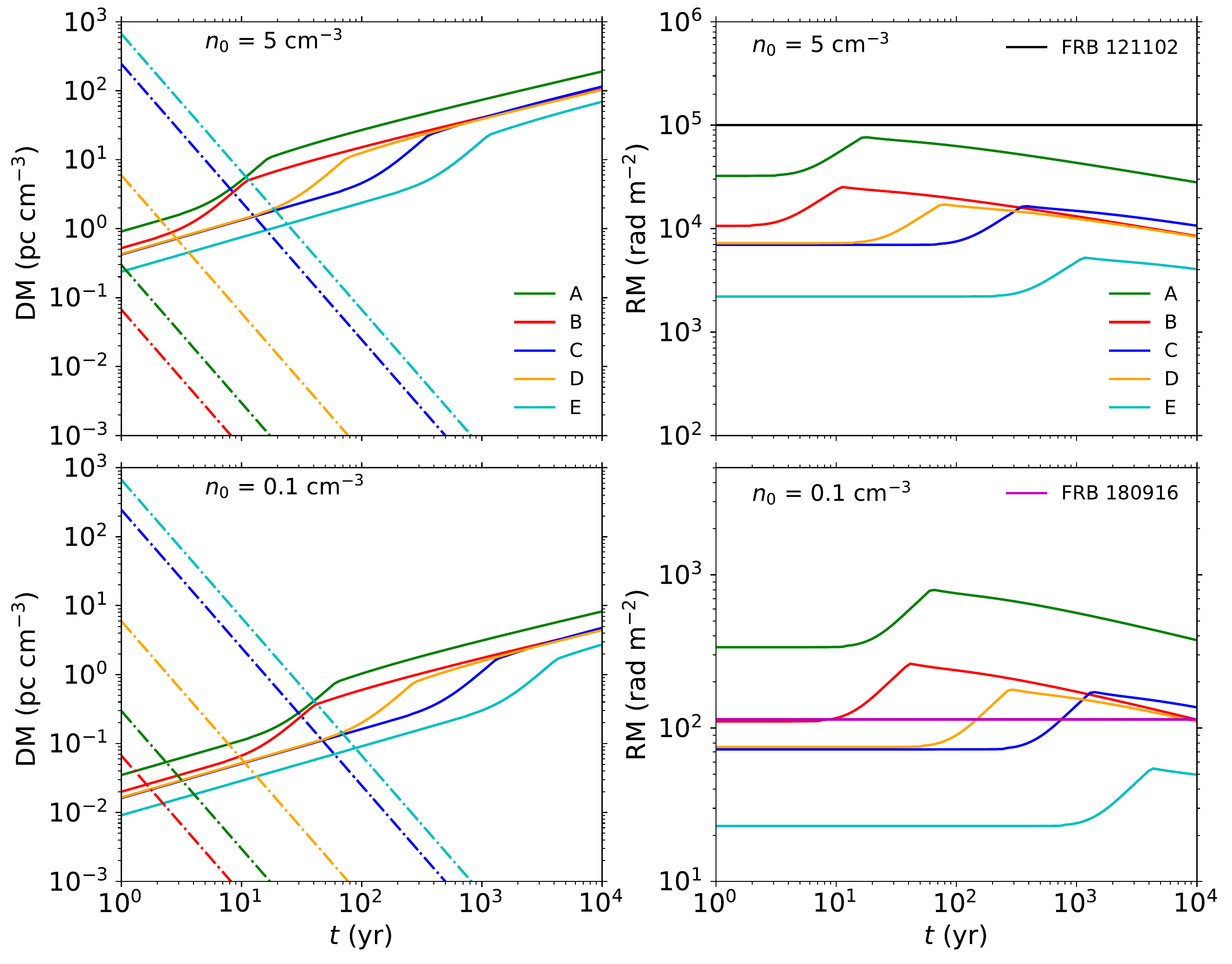}
	\caption{Same as Figure \ref{DMRMn10s0} but the ejecta has a power-law index of 6.}
	\label{DMRMn6s0}
\end{figure}

\begin{figure}
	\centering
	\includegraphics[width = 0.6\textwidth]{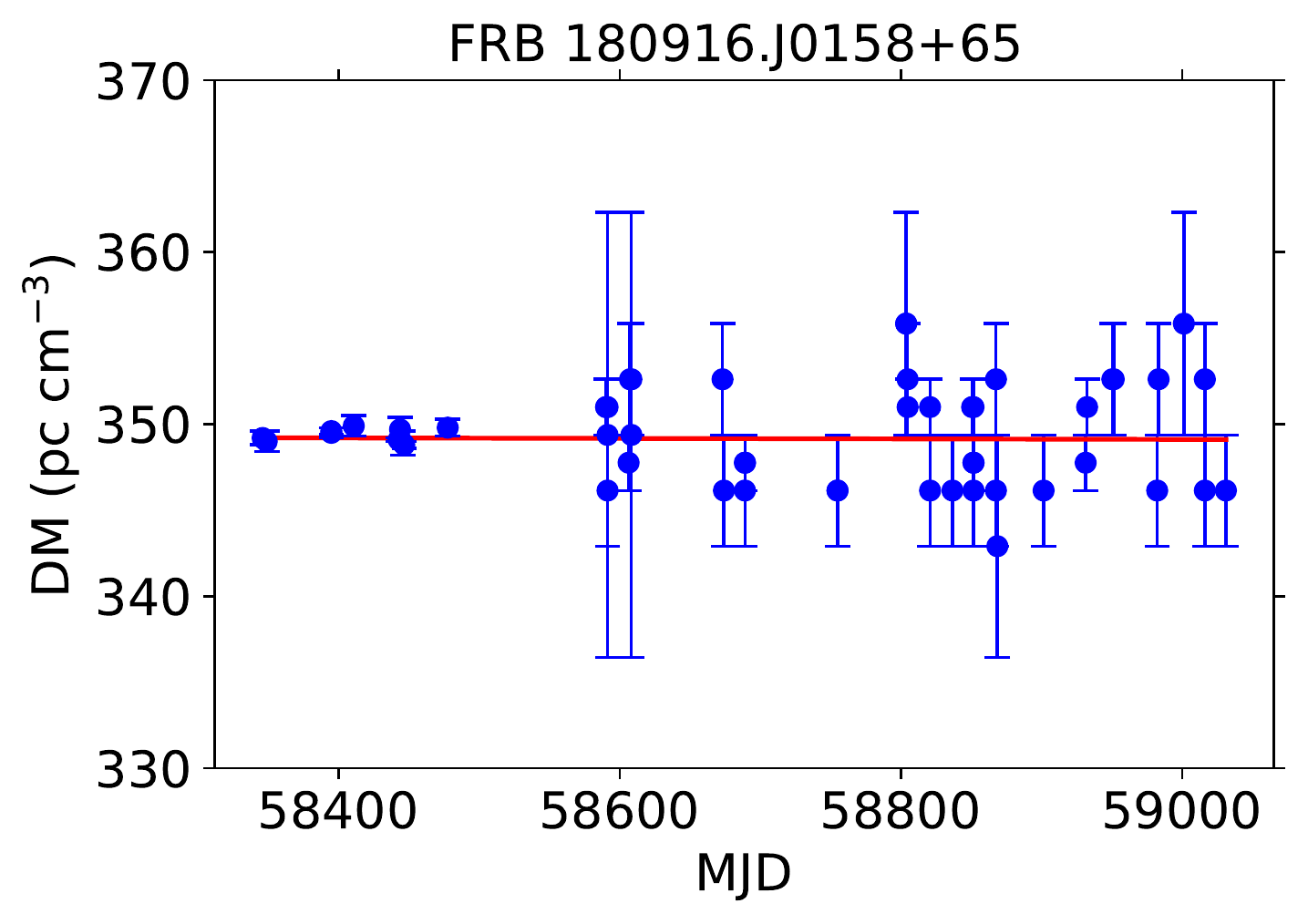}
	\caption{The measured DMs of FRB 180916 are from CHIME/FRB Public Database. The red line is the result of linear fitting with the slope $ |d \mathrm{DM}/dt |= 0.05 $ pc cm$ ^{-3} $ yr$^{-1}$.}
	\label{180916}
\end{figure}

The observed DM of FRB 180916 is $ \sim  $ 349 pc cm$ ^{-3} $ and the RM is $-114 \pm 0.6 $ rad m$^{-2}$ \citep{2019ApJ...885L..24C}. Due to the lack of the evolution of RM, it is difficult to compare the observations to different models. Therefore, the contributions of each region are not considered carefully, and we roughly assume that RM is all from the shocked shell. The RM of FRB 180916 is displayed with magenta horizontal lines in the right bottom panels in Figure \ref{DMRMn10s0} and \ref{DMRMn6s0}, and the corresponding ages of the sources range from a few years to thousands of years. In Figure \ref{180916}, we give the linear fitting of the DM of FRB 180916 using the data from CHIME/FRB Public Database\footnote{The catalog of repeating FRBs detected by CHIME/FRB is available at \url{https://www.chime-frb.ca/}. }, and find that the value of DM is almost constant ($ |d \mathrm{DM}/dt |= 0.05 $ pc cm$ ^{-3} $ yr$^{-1}$). When the density of ISM is very low (see left lower panels in Figures \ref{DMRMn10s0} and \ref{DMRMn6s0}), the DM contribution from the local environment of the source is negligible and therefore does not provide observable changes. Follow-up continuous observation of the RM changes may help us to determine the age and origin of FRB 180916.

\subsection{FRB 180301}
FRB 180301 was first detected by Parkes 64-m radio telescope on MJD 58178 with measured DM $ \sim 522 \pm 5 $ pc cm$ ^{-3} $\citep{2019MNRAS.486.3636P}. The observations of Five-hundred-meter Aperture Spherical radio Telescope (FAST) from July to October in 2019 favor magnetospheric origins of FRBs by measuring the polarization angle \citep{2020Natur.586..693L}. Their observed DM is $ \sim 517 $ pc cm$ ^{-3} $, which seems to be no obvious changes. However, their measured RMs increase rapidly with a slope of 21 rad m$^{-2} $ d$ ^{-1} $.
	
We compare our models with the observations of FRB 180301. As mentioned above, the seemingly unchanging DM may be caused by the ejecta interacting with the low-density ISM.  The rapidly growing RM can be well explained in the merger cases (see Figures \ref{DMRMn10s0} and \ref{DMRMn6s0}).

\section{Discussions}\label{discu}
\subsection{The free-free absorption}
For compact binary (BNS, BWD and NSWD) mergers or CC SNe, the FRB emission may suffer from free-free absorption of young ejecta with a temperature $ T_{\mathrm{ej}} \sim 10^4-10^5  $ K \citep{Bauswein_2013,Margalit_2019,2017hsn..book..875C}. The free-free absorption coefficient of a plasma is
\begin{equation}
\alpha_{\nu}^{\mathrm{ff}} = 0.018T_{\mathrm{ej}}^{-3/2}z_{\mathrm{i}}^{2}n_{\mathrm{e}}n_{\mathrm{i}} \nu^{-2} \bar{g}_{\mathrm{ff}},
\end{equation}
\citep{1986rpa..book.....R}, where $ z_{\mathrm{i}} $ is the atomic number of the ion, $ \bar{g}_{\mathrm{ff}} \sim 1 $ is the Gaunt factor, $ n_{\mathrm{e}} $ and $ n_{\mathrm{i}} $ are the number densities of ions and electrons, respectively. Here we assume $ z_{\mathrm{i}} \sim 1 $, $ n_{\mathrm{e}} \sim n_{\mathrm{i}} $, and just think about simple unshocked free expansion to make an estimate.  For the radio signals $ \nu \sim 1 $ GHz, the free-free optical depth is
\begin{equation}
\tau_{\mathrm{ff,ej}}=\alpha_{\nu}^{\mathrm{ff}} \Delta R \simeq 2.7 \times 10^{-8} \eta^2 Y_{\mathrm{e,0.2}}^2M_{-3}^2T_{\mathrm{ej,4}}^{-3/2}\nu_{9}^{-2} v_{0.2}^{-5}t_{\mathrm{yr}}^{-5},
\end{equation}
where $ T_{\mathrm{ej,4}}=T_{\mathrm{ej}}/10^4 $ K and $  \nu_9= \nu/10^9 $ Hz. The ejecta will become transparent to FRB within a few weeks after BNS merger due to the low ejecta mass ($ \sim 10^{-3}-10^{-2}M_{\odot} $) and high velocity ($ \sim 0.1-0.3c $). In the case of BWD or NSWD merger, the ejecta have higher mass ($ \sim 10^{-2}-10^{-1}M_{\odot} $) and lower velocity ($ \lesssim 0.1c $), and it is opaque to radio signals within several months after the merger. For CC SNe, more material $ \sim M_{\odot} $ with the lower velocity $ \sim 10^4 $ km s$ ^{-1} $ is ejected, so that the ambient medium becomes optically thin around 1-2 years after the explosion.
The free-free optical depth of the shock shell can be estimated from equations (\ref{n_s}) and (\ref{n_r}) in the same way. Here we take $ n_{\mathrm{r}} \sim n_{\mathrm{b}} = 4n_0 $ and $ \Delta R_{\mathrm{sh}} = R_{\mathrm{b}}-R_{\mathrm{c}} \sim R_{\mathrm{c}}-R_{\mathrm{r}}$, then we obtain
\begin{equation}
\tau_{\mathrm{ff,sh}}=1.78 \times 10^{-5} T_{\mathrm{sh},6}^{-3/2}n_{0,100}^{2} \Delta R_{\mathrm{sh,pc}},
\end{equation}
where $ T_{\mathrm{sh},6} = T_{\mathrm{sh}}/10^6$ K is the shock shell temperature and $n_{0,100} =  n_{0}/100 $ cm$ ^{-3}  $. We can see that the shock shell is transparent for radio signals during the evolution.

\subsection{Differentiation of various merger models}
 It is easy to distinguish between the merger and CC SNe scenario via the DM and RM evolution. However, the BNS, BWD and NSWD channel are difficult to distinguish just using the same method because of the great uncertainty of the ejecta mass and energy.  Some of the localized FRBs\footnote{The FRB host galaxy database is available at \url{http://frbhosts.org/}. } have been found to have large offsets from the galaxy centers, such as FRB 180916 ($ 5.46 \pm 0.01 $ kpc) \citep{2020Natur.577..190M}, FRB 180924 (3.43 $\pm  $ 0.64 kpc) \citep{2019Sci...365..565B}, FRB 181112 (1.69 $ \pm$ 2.61  kpc) \citep{2019Sci...366..231P}, FRB 190523 (27.2 $ \pm$ 22.6 kpc) \citep{2019Natur.572..352R},  FRB 190608 (6.60  $\pm $ 0.57 kpc) \citep{2020arXiv200513158C}, FRB 190102 (2.02 $\pm$ 2.01 kpc), FRB 190611 (11.4 $\pm$ 4.1 kpc ), FRB 190711 (3.17 $ \pm$ 2.78 kpc), FRB 190714 (1.88 $ \pm $ 0.62 kpc), FRB 191001 (11.0 $ \pm$ 0.5 kpc) and FRB 200430 (2.97 $ \pm $ 1.62 kpc) \citep{2020ApJ...903..152H}. Large offsets indicate that they may originate from BNS/NSWD mergers because NSs receive natal kicks as a result of asymmetric SN explosions \citep{1994A&A...290..496J,1996PhRvL..76..352B}.

 Besides the difference of offsets, the event rate could also impose constraints on different models. The local volumetric rate of FRBs is estimated as $ \sim 1.1 \times 10^{3}E^{-0.7}_{32} \text{ Gpc}^{-3}\text{ yr}^{-1} $ \citep{2019ApJ...883...40L}, where $ E_{32}=E/10^{32} $ erg Hz$ ^{-1} $ is the energy of FRBs. From 22 Parkes FRBs,  \citet{Cao2018} found that the local event rate is about 3-6$\times 10^4~\text{ Gpc}^{-3}\text{ yr}^{-1}$. The
 volume rate of repeating FRB sources averaged over $0<z<0.5$ is about $500~\text{ Gpc}^{-3}\text{ yr}^{-1}$ \citep{Wang2019}. The local rate of BNS mergers is $ \rho_{\mathrm{BNS}} \sim 1540_{-1220}^{+3200} \text{ Gpc}^{-3}\text{ yr}^{-1}$ according to GW170817 \citep{2017PhRvL.119p1101A}. To estimate the merger rates of NSWD and BWD, we use the $\mathtt{BSE}$ code, a rapid
 binary-evolution algorithm based on a suite of analytical formulae \citep{2002MNRAS.329..897H}, to carry out a population synthesis.
 We created a catalog of 1,000,000 binary systems in which the initial system parameters ($M_1, q, \varepsilon, P$) satisfy the following distributions
 \begin{equation}
 f_{m_{1}}\left(m_{1}\right) \propto\left\{\begin{array}{ll}
 m_{1}^{-1.3}, & \text {for } m_{1} \in[0.08,0.5] M_{\odot} \\
 m_{1}^{-2.2}, & \text {for } m_{1} \in[0.5,1.0] M_{\odot} \\
 m_{1}^{-\beta}, & \text { for } m_{1} \in[1,150] M_{\odot}
 \end{array}\right.
 \end{equation}
 \begin{equation}
 f_{q}(q) \propto q^{\kappa}, \qquad \text { for } q \in[0.1,1]
 \end{equation}
 \begin{equation}
 f_{\varepsilon}(\varepsilon) \propto \varepsilon^{\gamma}, \qquad \text { for } \varepsilon \in[0.0,1.0]
 \end{equation}
 \begin{equation}
 f_{P}(\log_{10} P) \propto(\log_{10} P)^{\pi}, \qquad \text { for }
 \log_{10} P \in[0.15,5.5]
 \end{equation}
 where $M_1$ is the mass of the primary star, $\beta = 2.7$
 \citep{1986FCPh...11....1S}; $q\equiv m_2/m_1$ is the mass ratio of the two
 stars; $P$ and $\varepsilon$ are orbital periods and eccentricity respectively;
 $\kappa = 0$ \citep{1998ApJ...506..780B}, $\gamma= 1$ \citep{1991A&A...248..485D} and $\pi = -0.5$ \citep{2012Sci...337..444S} are used in our simulation. In our population synthesis, primary masses $M_1$ are sampled within the range 4-25 $M_{\odot}$ while the full mass range 0.08-150 $M_{\odot}$ is considered for the normalization of rates. We set the metallicity $Z$ and maximum evolution time $T$ to 0.02 and $15,000$ Myr for all binaries. The constant binary fraction is used in our simulations, and it is assumed to be 75\% \citep{2010ApJS..190....1R,2013ARA&A..51..269D,2014ApJS..215...15S,2018A&A...619A..53T}.

 The merger rate $R_{\mathrm{m}}(z)$ is a convolution of the star
 formation rate history $\rho(z)$ and the probability density
 function (PDF) of delay time
 \begin{equation}
 R_{\mathrm{m}}(z)=\int_{t(z)}^{t(z=\infty)} f \rho(z)\left(t^{\prime}\right) p\left[t(z)-t^{\prime}\right] \ud t^{\prime},
 \end{equation}
 where $\ud t = - H(z)^{-1}(1+z)^{-1}  \ud z$, $H(z)$ is the Hubble parameter at redshift $z$, $f$ is the number of the binaries (NSWD, BWD) evolved from unit mass stellar population, $t(z)$ is the cosmic age at redshift $z$. The cosmic star formation rate (CSFR) is taken from \cite{2014ARA&A..52..415M}
 \begin{equation}
 \rho(z) = 0.015 \frac{(1+z)^{2.7}}{1+[(1+z) / 2.9]^{5.6}} M_{\odot} \text{ Mpc}^{-3}\text{ yr}^{-1}.
 \end{equation}

 From our population synthesis, $f_{\text{NSWD}}\sim 2.1\times 10^{-5} M_{\odot}^{-1}$, $f_{\text{BWD}}\sim 6.5\times 10^{-4} M_{\odot}^{-1}$. Therefore, the expected event rate of NSWD/BWD merger is $\rho_{\mathrm{NSWD}}(0)=3.9\times 10^2\text{ Gpc}^{-3}\text{ yr}^{-1}$ and $\rho_{\mathrm{BWD}}(0)=2.0\times 10^4\text{ Gpc}^{-3}\text{ yr}^{-1}$. Obviously, the merger rate of BNS or NSWD is lower than FRBs ($ E<10^{31} $ erg Hz$^{-1}$). Based on the gravitational-wave events observed by the Advanced LIGO/Virgo gravitational-wave detectors, \cite{Zhang2020} also found that only a small fraction of repeating FRBs are produced by young magnetars from BNS mergers. From the local event rate of compact binary mergers, the value of $ \rho_{\mathrm{BWD}}(0)$ is close to $\rho_{\mathrm{FRB}}(0)$. However, the BWD mergers are hard to explain the large offsets of FRBs. If the different products of BWD mergers are taken into account, such as SNe Ia, a single WD, a single NS or magnetar via the AIC process \citep{2016MNRAS.463.3461S,2019MNRAS.484..698R}, the rate of non-explosion origin cases could be even lower due to no reported SNe Ia observations associated with FRBs. The detection of FRB 200428 from Galactic magnetar SGR 1935+2154 \citep{2020Natur.587...54T,2020Natur.587...59B}, possibly hosted in SNR G57.2+0.8 from the previous research \citep{2014GCN.16533....1G}, suggests that at least some FRBs originated from magnetars hosted in SNR. The above discussions show that the progenitors of FRBs have multiple formation channels, which is consistent with the study of FRB host galaxy properties \citep{2020ApJ...903..152H}.

\section{Summary}\label{conlu}
Motivated by the hypothesis that FRBs arise from magnetars born in CC SNe or compact binary mergers, we have investigated the impact of the local environment in different scenarios. The ejecta and shocked shell after the CC explosion of massive stars or compact binary mergers will have a significant impact on the DM and RM evolution of FRBs. We consider the ejecta of compact binary mergers encountering the uniform medium, while SN ejecta expanding in the wind environment. Our conclusions are as follows.

\begin{itemize}
	\item The contribution of DM$ _{\mathrm{local}} $ mainly consists of three parts: the unshocked ejecta, the wind nebula, and the shocked shell. Since the unshocked ejecta is not magnetized in general, the contribution of RM$ _{\mathrm{local}} $ is only from the wind nebula and shocked shell.
	\item The DM from the ejecta of the BNS merger is negligible, while the DM is dominated by the ejecta in BWD, NSWD, and CC SNe scenarios at a very early time. The DM from the unshocked ejecta decreases rapidly over time, so it can only be detected in the first few years or decades.
	\item For compact binary mergers, both DM$_{\mathrm{sh}} $ and RM$_{\mathrm{sh}}  $ increase over time in the early SSDW phase. In ST phase, DM$_{\mathrm{sh}}  $ continues to grow but RM$_{\mathrm{sh}} $ goes down. For CC SNe, both DM$_{\mathrm{sh}}  $ and RM$_{\mathrm{sh}}  $ decrease over time in the early SSDW phase. The evolution of DM and RM become complex due to the uncertainty of the behavior of the reverse shock radius in the ST phase. When the reverse shock radius approaching the center, DM is likely to increase.
	\item The DM and RM evolution is associated with the local environment of FRBs. For FRB 121102, we consider the contributions from the wind nebula and the shocked shell in the case of the magnetar born in BNS mergers ($ M_{\mathrm{ej}} \sim 0.01 M_{\odot}$, $ E_{\mathrm{k}} \sim 10^{51} $ erg with the ejecta power-law index $ n=10 $), and the evolution of DM and RM can be explained simultaneously. In our fitting, the ISM density is $ \sim 2.5-3.1$ cm$^{-3}  $ and the age of the FRB source is very young ($ \sim 9-10  $ yr). The DM from local environment is $ \sim5-7 $ pc cm$ ^{-3} $ when it was first observed, and the total contributions from DM$_{\mathrm{local}} $ have an increase of $ \Delta $DM $ \sim  6$ pc cm$ ^{-3} $ in six years. The rapid decaying RM is from the wind nebula in 2.6 years, and subsequent slow decrease is due to the shocked shell (in the result of our fitting, RM$_{\mathrm{sh}}=(1-4) \times 10^{4} $ rad m$ ^{-2} $). For FRB 180916 and FRB 180301, the DM is almost constant, which can be caused by the ejecta interacting with the very small ISM density. For FRB 180301, the observed growing RM is consistent with our model predictions.
	\item From the observations of the large offsets, FRBs are more likely to arise from the remnants of BNS or NSWD mergers. The merger rate of BWD is approximately consistent with the event rate of FRBs from the result of population synthesis. Therefore, the progenitors of FRBs may not be unique.
\end{itemize}

\section*{acknowledgments} We thank the anonymous referee for constructive and helpful comments. This work was supported by the National Natural Science
Foundation of China (grant No. U1831207). We also thank Z. G. Dai,Y. P. Yang, Q. C. Li, J. P. Yuan, L. C. Oostrum, A. Bobrick, N. Sridhar and  S. Yamasaki for helpful discussions. We acknowledge use of the CHIME/FRB Public Database, provided at https://www.chime-frb.ca/ by the CHIME/FRB Collaboration.

\bibliographystyle{aasjournal}
\bibliography{evolution}

\begin{thebibliography}{}
\expandafter\ifx\csname natexlab\endcsname\relax\def\natexlab#1{#1}\fi
\providecommand{\url}[1]{\href{#1}{#1}}

\bibitem[{{Abbott} {et~al.}(2017){Abbott}, {Abbott}, {Abbott}, {Acernese},
  {Ackley}, {Adams}, {Adams}, {Addesso}, {Adhikari}, {Adya}, {Affeldt},
  {Afrough}, {Agarwal}, {Agathos}, {Agatsuma}, {Aggarwal}, {Aguiar}, {Aiello},
  {Ain}, {Ajith}, {Allen}, {Allen}, {Allocca}, {Altin}, {Amato}, {Ananyeva},
  {Anderson}, {Anderson}, {Angelova}, {Antier}, {Appert}, {Arai}, {Araya},
  {Areeda}, {Arnaud}, {Arun}, {Ascenzi}, {Ashton}, {Ast}, {Aston}, {Astone},
  {Atallah}, {Aufmuth}, {Aulbert}, {AultONeal}, {Austin}, {Avila-Alvarez},
  {Babak}, {Bacon}, {Bader}, {Bae}, {Bailes}, {Baker}, {Baldaccini},
  {Ballardin}, {Ballmer}, {Banagiri}, {Barayoga}, {Barclay}, {Barish},
  {Barker}, {Barkett}, {Barone}, {Barr}, {Barsotti}, {Barsuglia}, {Barta},
  {Barthelmy}, {Bartlett}, {Bartos}, {Bassiri}, {Basti}, {Batch}, {Bawaj},
  {Bayley}, {Bazzan}, {B{\'e}csy}, {Beer}, {Bejger}, {Belahcene}, {Bell},
  {Berger}, {Bergmann}, {Bernuzzi}, {Bero}, {Berry}, {Bersanetti}, {Bertolini},
  {Betzwieser}, {Bhagwat}, {Bhandare}, {Bilenko}, {Billingsley}, {Billman},
  {Birch}, {Birney}, {Birnholtz}, {Biscans}, {Biscoveanu}, {Bisht}, {Bitossi},
  {Biwer}, {Bizouard}, {Blackburn}, {Blackman}, {Blair}, {Blair}, {Blair},
  {Bloemen}, {Bock}, {Bode}, {Boer}, {Bogaert}, {Bohe}, {Bondu}, {Bonilla},
  {Bonnand}, {Boom}, {Bork}, {Boschi}, {Bose}, {Bossie}, {Bouffanais}, {Bozzi},
  {Bradaschia}, {Brady}, {Branchesi}, {Brau}, {Briant}, {Brillet}, {Brinkmann},
  {Brisson}, {Brockill}, {Broida}, {Brooks}, {Brown}, {Brown}, {Brunett},
  {Buchanan}, {Buikema}, {Bulik}, {Bulten}, {Buonanno}, {Buskulic}, {Buy},
  {Byer}, {Cabero}, {Cadonati}, {Cagnoli}, {Cahillane}, {Calder{\'o}n
  Bustillo}, {Callister}, {Calloni}, {Camp}, {Canepa}, {Canizares}, {Cannon},
  {Cao}, {Cao}, {Capano}, {Capocasa}, {Carbognani}, {Caride}, {Carney},
  {Carullo}, {Casanueva Diaz}, {Casentini}, {Caudill}, {Cavagli{\`a}},
  {Cavalier}, {Cavalieri}, {Cella}, {Cepeda}, {Cerd{\'a}-Dur{\'a}n},
  {Cerretani}, {Cesarini}, {Chamberlin}, {Chan}, {Chao}, {Charlton}, {Chase},
  {Chassande-Mottin}, {Chatterjee}, {Chatziioannou}, {Cheeseboro}, {Chen},
  {Chen}, {Chen}, {Cheng}, {Chia}, {Chincarini}, {Chiummo}, {Chmiel}, {Cho},
  {Cho}, {Chow}, {Christensen}, {Chu}, {Chua}, {Chua}, {Chung}, {Chung},
  {Ciani}, {Ciolfi}, {Cirelli}, {Cirone}, {Clara}, {Clark}, {Clearwater},
  {Cleva}, {Cocchieri}, {Coccia}, {Cohadon}, {Cohen}, {Colla}, {Collette},
  {Cominsky}, {Constancio}, {Conti}, {Cooper}, {Corban}, {Corbitt},
  {Cordero-Carri{\'o}n}, {Corley}, {Cornish}, {Corsi}, {Cortese}, {Costa},
  {Coughlin}, {Coughlin}, {Coulon}, {Countryman}, {Couvares}, {Covas}, {Cowan},
  {Coward}, {Cowart}, {Coyne}, {Coyne}, {Creighton}, {Creighton}, {Cripe},
  {Crowder}, {Cullen}, {Cumming}, {Cunningham}, {Cuoco}, {Dal Canton},
  {D{\'a}lya}, {Danilishin}, {D'Antonio}, {Danzmann}, {Dasgupta}, {Da Silva
  Costa}, {Dattilo}, {Dave}, {Davier}, {Davis}, {Daw}, {Day}, {De}, {DeBra},
  {Degallaix}, {De Laurentis}, {Del{\'e}glise}, {Del Pozzo}, {Demos}, {Denker},
  {Dent}, {De Pietri}, {Dergachev}, {De Rosa}, {DeRosa}, {De Rossi}, {DeSalvo},
  {de Varona}, {Devenson}, {Dhurandhar}, {D{\'\i}az}, {Dietrich}, {Di Fiore},
  {Di Giovanni}, {Di Girolamo}, {Di Lieto}, {Di Pace}, {Di Palma}, {Di Renzo},
  {Doctor}, {Dolique}, {Donovan}, {Dooley}, {Doravari}, {Dorrington},
  {Douglas}, {Dovale {\'A}lvarez}, {Downes}, {Drago}, {Dreissigacker},
  {Driggers}, {Du}, {Ducrot}, {Dudi}, {Dupej}, {Dwyer}, {Edo}, {Edwards},
  {Effler}, {Eggenstein}, {Ehrens}, {Eichholz}, {Eikenberry}, {Eisenstein},
  {Essick}, {Estevez}, {Etienne}, {Etzel}, {Evans}, {Evans}, {Factourovich},
  {Fafone}, {Fair}, {Fairhurst}, {Fan}, {Farinon}, {Farr}, {Farr},
  {Fauchon-Jones}, {Favata}, {Fays}, {Fee}, {Fehrmann}, {Feicht}, {Fejer},
  {Fernandez-Galiana}, {Ferrante}, {Ferreira}, {Ferrini}, {Fidecaro},
  {Finstad}, {Fiori}, {Fiorucci}, {Fishbach}, {Fisher}, {Fitz-Axen},
  {Flaminio}, {Fletcher}, {Fong}, {Font}, {Forsyth}, {Forsyth}, {Fournier},
  {Frasca}, {Frasconi}, {Frei}, {Freise}, {Frey}, {Frey}, {Fries}, {Fritschel},
  {Frolov}, {Fulda}, {Fyffe}, {Gabbard}, {Gadre}, {Gaebel}, {Gair},
  {Gammaitoni}, {Ganija}, {Gaonkar}, {Garcia-Quiros}, {Garufi}, {Gateley},
  {Gaudio}, {Gaur}, {Gayathri}, {Gehrels}, {Gemme}, {Genin}, {Gennai},
  {George}, {George}, {Gergely}, {Germain}, {Ghonge}, {Ghosh}, {Ghosh},
  {Ghosh}, {Giaime}, {Giardina}, {Giazotto}, {Gill}, {Glover}, {Goetz},
  {Goetz}, {Gomes}, {Goncharov}, {Gonz{\'a}lez}, {Gonzalez Castro},
  {Gopakumar}, {Gorodetsky}, {Gossan}, {Gosselin}, {Gouaty}, {Grado}, {Graef},
  {Granata}, {Grant}, {Gras}, {Gray}, {Greco}, {Green}, {Gretarsson}, {Groot},
  {Grote}, {Grunewald}, {Gruning}, {Guidi}, {Guo}, {Gupta}, {Gupta}, {Gushwa},
  {Gustafson}, {Gustafson}, {Halim}, {Hall}, {Hall}, {Hamilton}, {Hammond},
  {Haney}, {Hanke}, {Hanks}, {Hanna}, {Hannam}, {Hannuksela}, {Hanson},
  {Hardwick}, {Harms}, {Harry}, {Harry}, {Hart}, {Haster}, {Haughian}, {Healy},
  {Heidmann}, {Heintze}, {Heitmann}, {Hello}, {Hemming}, {Hendry}, {Heng},
  {Hennig}, {Heptonstall}, {Heurs}, {Hild}, {Hinderer}, {Ho}, {Hoak}, {Hofman},
  {Holt}, {Holz}, {Hopkins}, {Horst}, {Hough}, {Houston}, {Howell}, {Hreibi},
  {Hu}, {Huerta}, {Huet}, {Hughey}, {Husa}, {Huttner}, {Huynh-Dinh}, {Indik},
  {Inta}, {Intini}, {Isa}, {Isac}, {Isi}, {Iyer}, {Izumi}, {Jacqmin}, {Jani},
  {Jaranowski}, {Jawahar}, {Jim{\'e}nez-Forteza}, {Johnson},
  {Johnson-McDaniel}, {Jones}, {Jones}, {Jonker}, {Ju}, {Junker}, {Kalaghatgi},
  {Kalogera}, {Kamai}, {Kandhasamy}, {Kang}, {Kanner}, {Kapadia}, {Karki},
  {Karvinen}, {Kasprzack}, {Kastaun}, {Katolik}, {Katsavounidis}, {Katzman},
  {Kaufer}, {Kawabe}, {K{\'e}f{\'e}lian}, {Keitel}, {Kemball}, {Kennedy},
  {Kent}, {Key}, {Khalili}, {Khan}, {Khan}, {Khan}, {Khazanov}, {Kijbunchoo},
  {Kim}, {Kim}, {Kim}, {Kim}, {Kim}, {Kim}, {Kimbrell}, {King}, {King},
  {Kinley-Hanlon}, {Kirchhoff}, {Kissel}, {Kleybolte}, {Klimenko}, {Knowles},
  {Koch}, {Koehlenbeck}, {Koley}, {Kondrashov}, {Kontos}, {Korobko}, {Korth},
  {Kowalska}, {Kozak}, {Kr{\"a}mer}, {Kringel}, {Krishnan}, {Kr{\'o}lak},
  {Kuehn}, {Kumar}, {Kumar}, {Kumar}, {Kuo}, {Kutynia}, {Kwang}, {Lackey},
  {Lai}, {Landry}, {Lang}, {Lange}, {Lantz}, {Lanza}, {Larson},
  {Lartaux-Vollard}, {Lasky}, {Laxen}, {Lazzarini}, {Lazzaro}, {Leaci},
  {Leavey}, {Lee}, {Lee}, {Lee}, {Lee}, {Lee}, {Lehmann}, {Lenon}, {Leon},
  {Leonardi}, {Leroy}, {Letendre}, {Levin}, {Li}, {Linker}, {Littenberg},
  {Liu}, {Liu}, {Lo}, {Lockerbie}, {London}, {Lord}, {Lorenzini}, {Loriette},
  {Lormand}, {Losurdo}, {Lough}, {Lousto}, {Lovelace}, {L{\"u}ck}, {Lumaca},
  {Lundgren}, {Lynch}, {Ma}, {Macas}, {Macfoy}, {Machenschalk}, {MacInnis},
  {Macleod}, {Maga{\~n}a Hernandez}, {Maga{\~n}a-Sandoval}, {Maga{\~n}a
  Zertuche}, {Magee}, {Majorana}, {Maksimovic}, {Man}, {Mandic}, {Mangano},
  {Mansell}, {Manske}, {Mantovani}, {Marchesoni}, {Marion}, {M{\'a}rka},
  {M{\'a}rka}, {Markakis}, {Markosyan}, {Markowitz}, {Maros}, {Marquina},
  {Marsh}, {Martelli}, {Martellini}, {Martin}, {Martin}, {Martynov}, {Marx},
  {Mason}, {Massera}, {Masserot}, {Massinger}, {Masso-Reid}, {Mastrogiovanni},
  {Matas}, {Matichard}, {Matone}, {Mavalvala}, {Mazumder}, {McCarthy},
  {McClelland}, {McCormick}, {McCuller}, {McGuire}, {McIntyre}, {McIver},
  {McManus}, {McNeill}, {McRae}, {McWilliams}, {Meacher}, {Meadors}, {Mehmet},
  {Meidam}, {Mejuto-Villa}, {Melatos}, {Mendell}, {Mercer}, {Merilh},
  {Merzougui}, {Meshkov}, {Messenger}, {Messick}, {Metzdorff}, {Meyers},
  {Miao}, {Michel}, {Middleton}, {Mikhailov}, {Milano}, {Miller}, {Miller},
  {Miller}, {Millhouse}, {Milovich-Goff}, {Minazzoli}, {Minenkov}, {Ming},
  {Mishra}, {Mitra}, {Mitrofanov}, {Mitselmakher}, {Mittleman}, {Moffa},
  {Moggi}, {Mogushi}, {Mohan}, {Mohapatra}, {Molina}, {Montani}, {Moore},
  {Moraru}, {Moreno}, {Morisaki}, {Morriss}, {Mours}, {Mow-Lowry}, {Mueller},
  {Muir}, {Mukherjee}, {Mukherjee}, {Mukherjee}, {Mukund}, {Mullavey}, {Munch},
  {Mu{\~n}iz}, {Muratore}, {Murray}, {Nagar}, {Napier}, {Nardecchia},
  {Naticchioni}, {Nayak}, {Neilson}, {Nelemans}, {Nelson}, {Nery}, {Neunzert},
  {Nevin}, {Newport}, {Newton}, {Ng}, {Nguyen}, {Nguyen}, {Nichols}, {Nielsen},
  {Nissanke}, {Nitz}, {Noack}, {Nocera}, {Nolting}, {North}, {Nuttall},
  {Oberling}, {O'Dea}, {Ogin}, {Oh}, {Oh}, {Ohme}, {Okada}, {Oliver},
  {Oppermann}, {Oram}, {O'Reilly}, {Ormiston}, {Ortega}, {O'Shaughnessy},
  {Ossokine}, {Ottaway}, {Overmier}, {Owen}, {Pace}, {Page}, {Page}, {Pai},
  {Pai}, {Palamos}, {Palashov}, {Palomba}, {Pal-Singh}, {Pan}, {Pan}, {Pang},
  {Pang}, {Pankow}, {Pannarale}, {Pant}, {Paoletti}, {Paoli}, {Papa}, {Parida},
  {Parker}, {Pascucci}, {Pasqualetti}, {Passaquieti}, {Passuello}, {Patil},
  {Patricelli}, {Pearlstone}, {Pedraza}, {Pedurand}, {Pekowsky}, {Pele},
  {Penn}, {Perez}, {Perreca}, {Perri}, {Pfeiffer}, {Phelps}, {Piccinni},
  {Pichot}, {Piergiovanni}, {Pierro}, {Pillant}, {Pinard}, {Pinto}, {Pirello},
  {Pitkin}, {Poe}, {Poggiani}, {Popolizio}, {Porter}, {Post}, {Powell},
  {Prasad}, {Pratt}, {Pratten}, {Predoi}, {Prestegard}, {Prijatelj},
  {Principe}, {Privitera}, {Prix}, {Prodi}, {Prokhorov}, {Puncken}, {Punturo},
  {Puppo}, {P{\"u}rrer}, {Qi}, {Quetschke}, {Quintero}, {Quitzow-James},
  {Raab}, {Rabeling}, {Radkins}, {Raffai}, {Raja}, {Rajan}, {Rajbhandari},
  {Rakhmanov}, {Ramirez}, {Ramos-Buades}, {Rapagnani}, {Raymond}, {Razzano},
  {Read}, {Regimbau}, {Rei}, {Reid}, {Reitze}, {Ren}, {Reyes}, {Ricci},
  {Ricker}, {Rieger}, {Riles}, {Rizzo}, {Robertson}, {Robie}, {Robinet},
  {Rocchi}, {Rolland}, {Rollins}, {Roma}, {Romano}, {Romano}, {Romel}, {Romie},
  {Rosi{\'n}ska}, {Ross}, {Rowan}, {R{\"u}diger}, {Ruggi}, {Rutins}, {Ryan},
  {Sachdev}, {Sadecki}, {Sadeghian}, {Sakellariadou}, {Salconi}, {Saleem},
  {Salemi}, {Samajdar}, {Sammut}, {Sampson}, {Sanchez}, {Sanchez},
  {Sanchis-Gual}, {Sandberg}, {Sanders}, {Sassolas}, {Sathyaprakash},
  {Saulson}, {Sauter}, {Savage}, {Sawadsky}, {Schale}, {Scheel}, {Scheuer},
  {Schmidt}, {Schmidt}, {Schnabel}, {Schofield}, {Sch{\"o}nbeck}, {Schreiber},
  {Schuette}, {Schulte}, {Schutz}, {Schwalbe}, {Scott}, {Scott}, {Seidel},
  {Sellers}, {Sengupta}, {Sentenac}, {Sequino}, {Sergeev}, {Shaddock},
  {Shaffer}, {Shah}, {Shahriar}, {Shaner}, {Shao}, {Shapiro}, {Shawhan},
  {Sheperd}, {Shoemaker}, {Shoemaker}, {Siellez}, {Siemens}, {Sieniawska},
  {Sigg}, {Silva}, {Singer}, {Singh}, {Singhal}, {Sintes}, {Slagmolen},
  {Smith}, {Smith}, {Smith}, {Somala}, {Son}, {Sonnenberg}, {Sorazu},
  {Sorrentino}, {Souradeep}, {Spencer}, {Srivastava}, {Staats}, {Staley},
  {Steinke}, {Steinlechner}, {Steinlechner}, {Steinmeyer}, {Stevenson},
  {Stone}, {Stops}, {Strain}, {Stratta}, {Strigin}, {Strunk}, {Sturani},
  {Stuver}, {Summerscales}, {Sun}, {Sunil}, {Suresh}, {Sutton}, {Swinkels},
  {Szczepa{\'n}czyk}, {Tacca}, {Tait}, {Talbot}, {Talukder}, {Tanner},
  {T{\'a}pai}, {Taracchini}, {Tasson}, {Taylor}, {Taylor}, {Tewari}, {Theeg},
  {Thies}, {Thomas}, {Thomas}, {Thomas}, {Thorne}, {Thorne}, {Thrane},
  {Tiwari}, {Tiwari}, {Tokmakov}, {Toland}, {Tonelli}, {Tornasi},
  {Torres-Forn{\'e}}, {Torrie}, {T{\"o}yr{\"a}}, {Travasso}, {Traylor},
  {Trinastic}, {Tringali}, {Trozzo}, {Tsang}, {Tse}, {Tso}, {Tsukada}, {Tsuna},
  {Tuyenbayev}, {Ueno}, {Ugolini}, {Unnikrishnan}, {Urban}, {Usman},
  {Vahlbruch}, {Vajente}, {Valdes}, {Vallisneri}, {van Bakel}, {van Beuzekom},
  {van den Brand}, {Van Den Broeck}, {Vander-Hyde}, {van der Schaaf}, {van
  Heijningen}, {van Veggel}, {Vardaro}, {Varma}, {Vass}, {Vas{\'u}th},
  {Vecchio}, {Vedovato}, {Veitch}, {Veitch}, {Venkateswara}, {Venugopalan},
  {Verkindt}, {Vetrano}, {Vicer{\'e}}, {Viets}, {Vinciguerra}, {Vine}, {Vinet},
  {Vitale}, {Vo}, {Vocca}, {Vorvick}, {Vyatchanin}, {Wade}, {Wade}, {Wade},
  {Walet}, {Walker}, {Wallace}, {Walsh}, {Wang}, {Wang}, {Wang}, {Wang},
  {Wang}, {Ward}, {Warner}, {Was}, {Watchi}, {Weaver}, {Wei}, {Weinert},
  {Weinstein}, {Weiss}, {Wen}, {Wessel}, {We{\ss}els}, {Westerweck},
  {Westphal}, {Wette}, {Whelan}, {Whitcomb}, {Whiting}, {Whittle}, {Wilken},
  {Williams}, {Williams}, {Williamson}, {Willis}, {Willke}, {Wimmer},
  {Winkler}, {Wipf}, {Wittel}, {Woan}, {Woehler}, {Wofford}, {Wong}, {Worden},
  {Wright}, {Wu}, {Wysocki}, {Xiao}, {Yamamoto}, {Yancey}, {Yang}, {Yap},
  {Yazback}, {Yu}, {Yu}, {Yvert}, {Zadro{\.Z}ny}, {Zanolin}, {Zelenova},
  {Zendri}, {Zevin}, {Zhang}, {Zhang}, {Zhang}, {Zhang}, {Zhao}, {Zhou},
  {Zhou}, {Zhu}, {Zhu}, {Zimmerman}, {Zucker}, {Zweizig}, {LIGO Scientific
  Collaboration}, \& {Virgo Collaboration}}]{2017PhRvL.119p1101A}
{Abbott}, B.~P., {Abbott}, R., {Abbott}, T.~D., {et~al.} 2017, \prl, 119,
  161101

\bibitem[{{Bannister} {et~al.}(2019){Bannister}, {Deller}, {Phillips},
  {Macquart}, {Prochaska}, {Tejos}, {Ryder}, {Sadler}, {Shannon}, {Simha},
  {Day}, {McQuinn}, {North-Hickey}, {Bhandari}, {Arcus}, {Bennert}, {Burchett},
  {Bouwhuis}, {Dodson}, {Ekers}, {Farah}, {Flynn}, {James}, {Kerr}, {Lenc},
  {Mahony}, {O'Meara}, {Os{\l}owski}, {Qiu}, {Treu}, {U}, {Bateman}, {Bock},
  {Bolton}, {Brown}, {Bunton}, {Chippendale}, {Cooray}, {Cornwell}, {Gupta},
  {Hayman}, {Kesteven}, {Koribalski}, {MacLeod}, {McClure-Griffiths},
  {Neuhold}, {Norris}, {Pilawa}, {Qiao}, {Reynolds}, {Roxby}, {Shimwell},
  {Voronkov}, \& {Wilson}}]{2019Sci...365..565B}
{Bannister}, K.~W., {Deller}, A.~T., {Phillips}, C., {et~al.} 2019, Science,
  365, 565

\bibitem[{{Bassa} {et~al.}(2017){Bassa}, {Tendulkar}, {Adams}, {Maddox},
  {Bogdanov}, {Bower}, {Burke-Spolaor}, {Butler}, {Chatterjee}, {Cordes},
  {Hessels}, {Kaspi}, {Law}, {Marcote}, {Paragi}, {Ransom}, {Scholz},
  {Spitler}, \& {van Langevelde}}]{2017ApJ...843L...8B}
{Bassa}, C.~G., {Tendulkar}, S.~P., {Adams}, E.~A.~K., {et~al.} 2017, \apjl,
  843, L8

\bibitem[{{Bauswein} {et~al.}(2013){Bauswein}, {Goriely}, \&
  {Janka}}]{Bauswein_2013}
{Bauswein}, A., {Goriely}, S., \& {Janka}, H.~T. 2013, \apj, 773, 78

\bibitem[{{Beloborodov}(2017)}]{2017ApJ...843L..26B}
{Beloborodov}, A.~M. 2017, \apjl, 843, L26

\bibitem[{{Berger}(2014)}]{2014ARA&A..52...43B}
{Berger}, E. 2014, \araa, 52, 43

\bibitem[{{Bethe} \& {Brown}(1998)}]{1998ApJ...506..780B}
{Bethe}, H.~A., \& {Brown}, G.~E. 1998, \apj, 506, 780

\bibitem[{{Bobrick} {et~al.}(2017){Bobrick}, {Davies}, \&
  {Church}}]{2017MNRAS.467.3556B}
{Bobrick}, A., {Davies}, M.~B., \& {Church}, R.~P. 2017, \mnras, 467, 3556

\bibitem[{{Bochenek} {et~al.}(2020){Bochenek}, {Ravi}, {Belov}, {Hallinan},
  {Kocz}, {Kulkarni}, \& {McKenna}}]{2020Natur.587...59B}
{Bochenek}, C.~D., {Ravi}, V., {Belov}, K.~V., {et~al.} 2020, \nat, 587, 59

\bibitem[{{Burrows} \& {Hayes}(1996)}]{1996PhRvL..76..352B}
{Burrows}, A., \& {Hayes}, J. 1996, \prl, 76, 352

\bibitem[{{Burrows} {et~al.}(2005){Burrows}, {Romano}, {Falcone}, {Kobayashi},
  {Zhang}, {Moretti}, {O'Brien}, {Goad}, {Campana}, {Page}, {Angelini},
  {Barthelmy}, {Beardmore}, {Capalbi}, {Chincarini}, {Cummings}, {Cusumano},
  {Fox}, {Giommi}, {Hill}, {Kennea}, {Krimm}, {Mangano}, {Marshall},
  {M{\'e}sz{\'a}ros}, {Morris}, {Nousek}, {Osborne}, {Pagani}, {Perri},
  {Tagliaferri}, {Wells}, {Woosley}, \& {Gehrels}}]{Burrows2005}
{Burrows}, D.~N., {Romano}, P., {Falcone}, A., {et~al.} 2005, Science, 309,
  1833

\bibitem[{Bykov {et~al.}(2013)Bykov, Brandenburg, Malkov, \&
  Osipov}]{2013SSRv..178..201B}
Bykov, A.~M., Brandenburg, A., Malkov, M.~A., \& Osipov, S.~M. 2013, \ssr, 178,
  201

\bibitem[{{Cao} {et~al.}(2017){Cao}, {Yu}, \& {Dai}}]{2017ApJ...839L..20C}
{Cao}, X.-F., {Yu}, Y.-W., \& {Dai}, Z.-G. 2017, \apjl, 839, L20

\bibitem[{{Cao} {et~al.}(2018){Cao}, {Yu}, \& {Zhou}}]{Cao2018}
{Cao}, X.-F., {Yu}, Y.-W., \& {Zhou}, X. 2018, \apj, 858, 89

\bibitem[{Caprioli \& Spitkovsky(2014)}]{2014ApJ...794...46C}
Caprioli, D., \& Spitkovsky, A. 2014, \apj, 794, 46

\bibitem[{{Chatterjee} {et~al.}(2017){Chatterjee}, {Law}, {Wharton},
  {Burke-Spolaor}, {Hessels}, {Bower}, {Cordes}, {Tendulkar}, {Bassa},
  {Demorest}, {Butler}, {Seymour}, {Scholz}, {Abruzzo}, {Bogdanov}, {Kaspi},
  {Keimpema}, {Lazio}, {Marcote}, {McLaughlin}, {Paragi}, {Ransom}, {Rupen},
  {Spitler}, \& {van Langevelde}}]{2017Natur.541...58C}
{Chatterjee}, S., {Law}, C.~J., {Wharton}, R.~S., {et~al.} 2017, \nat, 541, 58

\bibitem[{Chevalier(1982)}]{1982ApJ...258..790C}
Chevalier, R.~A. 1982, \apj, 258, 790

\bibitem[{{Chevalier} \& {Fransson}(2017)}]{2017hsn..book..875C}
{Chevalier}, R.~A., \& {Fransson}, C. 2017, {Thermal and Non-thermal Emission
  from Circumstellar Interaction}, ed. A.~W. {Alsabti} \& P.~{Murdin}, 875

\bibitem[{{Chevalier} \& {Soker}(1989)}]{1989ApJ...341..867C}
{Chevalier}, R.~A., \& {Soker}, N. 1989, \apj, 341, 867

\bibitem[{{CHIME/FRB Collaboration} {et~al.}(2019){CHIME/FRB Collaboration},
  {Andersen}, {Bandura}, {Bhardwaj}, {Boubel}, {Boyce}, {Boyle}, {Brar},
  {Cassanelli}, {Chawla}, {Cubranic}, {Deng}, {Dobbs}, {Fandino}, {Fonseca},
  {Gaensler}, {Gilbert}, {Giri}, {Good}, {Halpern}, {Hill}, {Hinshaw},
  {H{\"o}fer}, {Josephy}, {Kaspi}, {Kothes}, {Landecker}, {Lang}, {Li}, {Lin},
  {Masui}, {Mena-Parra}, {Merryfield}, {Mckinven}, {Michilli}, {Milutinovic},
  {Naidu}, {Newburgh}, {Ng}, {Patel}, {Pen}, {Pinsonneault-Marotte}, {Pleunis},
  {Rafiei-Ravandi}, {Rahman}, {Ransom}, {Renard}, {Scholz}, {Siegel}, {Singh},
  {Smith}, {Stairs}, {Tendulkar}, {Tretyakov}, {Vanderlinde}, {Yadav}, \&
  {Zwaniga}}]{2019ApJ...885L..24C}
{CHIME/FRB Collaboration}, {Andersen}, B.~C., {Bandura}, K., {et~al.} 2019,
  \apjl, 885, L24

\bibitem[{{Chittidi} {et~al.}(2020){Chittidi}, {Simha}, {Mannings},
  {Prochaska}, {Rafelski}, {Neeleman}, {Macquart}, {Tejos}, {Jorgenson},
  {Ryder}, {Day}, {Marnoch}, {Bhandari}, {Deller}, {Qiu}, {Bannister},
  {Shannon}, \& {Heintz}}]{2020arXiv200513158C}
{Chittidi}, J.~S., {Simha}, S., {Mannings}, A., {et~al.} 2020, arXiv e-prints,
  arXiv:2005.13158

\bibitem[{{Dai} \& {Lu}(1998)}]{Dai1998}
{Dai}, Z.~G., \& {Lu}, T. 1998, \aap, 333, L87

\bibitem[{{Dai} {et~al.}(2006){Dai}, {Wang}, {Wu}, \& {Zhang}}]{Dai2006}
{Dai}, Z.~G., {Wang}, X.~Y., {Wu}, X.~F., \& {Zhang}, B. 2006, Science, 311,
  1127

\bibitem[{{Dessart} {et~al.}(2007){Dessart}, {Burrows}, {Livne}, \&
  {Ott}}]{2007ApJ...669..585D}
{Dessart}, L., {Burrows}, A., {Livne}, E., \& {Ott}, C.~D. 2007, \apj, 669, 585

\bibitem[{Draine(2011)}]{2011piim.book.....D}
Draine, B.~T. 2011, {Physics of the Interstellar and Intergalactic Medium}

\bibitem[{{Duch{\^e}ne} \& {Kraus}(2013)}]{2013ARA&A..51..269D}
{Duch{\^e}ne}, G., \& {Kraus}, A. 2013, \araa, 51, 269

\bibitem[{{Duquennoy} \& {Mayor}(1991)}]{1991A&A...248..485D}
{Duquennoy}, A., \& {Mayor}, M. 1991, \aap, 500, 337

\bibitem[{{Foreman-Mackey} {et~al.}(2013){Foreman-Mackey}, {Hogg}, {Lang}, \&
  {Goodman}}]{2013PASP..125..306F}
{Foreman-Mackey}, D., {Hogg}, D.~W., {Lang}, D., \& {Goodman}, J. 2013, \pasp,
  125, 306

\bibitem[{{Gaensler}(2014)}]{2014GCN.16533....1G}
{Gaensler}, B.~M. 2014, GRB Coordinates Network, 16533, 1

\bibitem[{{Gajjar} {et~al.}(2018){Gajjar}, {Siemion}, {Price}, {Law},
  {Michilli}, {Hessels}, {Chatterjee}, {Archibald}, {Bower}, {Brinkman},
  {Burke-Spolaor}, {Cordes}, {Croft}, {Enriquez}, {Foster}, {Gizani},
  {Hellbourg}, {Isaacson}, {Kaspi}, {Lazio}, {Lebofsky}, {Lynch}, {MacMahon},
  {McLaughlin}, {Ransom}, {Scholz}, {Seymour}, {Spitler}, {Tendulkar},
  {Werthimer}, \& {Zhang}}]{2018ApJ...863....2G}
{Gajjar}, V., {Siemion}, A.~P.~V., {Price}, D.~C., {et~al.} 2018, \apj, 863, 2

\bibitem[{{Giacomazzo} \& {Perna}(2013)}]{2013ApJ...771L..26G}
{Giacomazzo}, B., \& {Perna}, R. 2013, \apjl, 771, L26

\bibitem[{{Gourdji} {et~al.}(2019){Gourdji}, {Michilli}, {Spitler}, {Hessels},
  {Seymour}, {Cordes}, \& {Chatterjee}}]{2019ApJ...877L..19G}
{Gourdji}, K., {Michilli}, D., {Spitler}, L.~G., {et~al.} 2019, \apjl, 877, L19

\bibitem[{{Hajela} {et~al.}(2019){Hajela}, {Margutti}, {Alexander},
  {Kathirgamaraju}, {Baldeschi}, {Guidorzi}, {Giannios}, {Fong}, {Wu},
  {MacFadyen}, {Paggi}, {Berger}, {Blanchard}, {Chornock}, {Coppejans},
  {Cowperthwaite}, {Eftekhari}, {Gomez}, {Hosseinzadeh}, {Laskar}, {Metzger},
  {Nicholl}, {Paterson}, {Radice}, {Sironi}, {Terreran}, {Villar}, {Williams},
  {Xie}, \& {Zrake}}]{2019ApJ...886L..17H}
{Hajela}, A., {Margutti}, R., {Alexander}, K.~D., {et~al.} 2019, \apjl, 886,
  L17

\bibitem[{Hamilton \& Sarazin(1984)}]{1984ApJ...281..682H}
Hamilton, A.~J.~S., \& Sarazin, C.~L. 1984, \apj, 281, 682

\bibitem[{{Heintz} {et~al.}(2020){Heintz}, {Prochaska}, {Simha}, {Platts},
  {Fong}, {Tejos}, {Ryder}, {Aggerwal}, {Bhandari}, {Day}, {Deller},
  {Kilpatrick}, {Law}, {Macquart}, {Mannings}, {Marnoch}, {Sadler}, \&
  {Shannon}}]{2020ApJ...903..152H}
{Heintz}, K.~E., {Prochaska}, J.~X., {Simha}, S., {et~al.} 2020, \apj, 903, 152

\bibitem[{{Hessels} {et~al.}(2019){Hessels}, {Spitler}, {Seymour}, {Cordes},
  {Michilli}, {Lynch}, {Gourdji}, {Archibald}, {Bassa}, {Bower}, {Chatterjee},
  {Connor}, {Crawford}, {Deneva}, {Gajjar}, {Kaspi}, {Keimpema}, {Law},
  {Marcote}, {McLaughlin}, {Paragi}, {Petroff}, {Ransom}, {Scholz}, {Stappers},
  \& {Tendulkar}}]{2019ApJ...876L..23H}
{Hessels}, J.~W.~T., {Spitler}, L.~G., {Seymour}, A.~D., {et~al.} 2019, \apjl,
  876, L23

\bibitem[{{Hilmarsson} {et~al.}(2020){Hilmarsson}, {Michilli}, {Spitler},
  {Wharton}, {Demorest}, {Desvignes}, {Gourdji}, {Hackstein}, {Hessels},
  {Nimmo}, {Seymour}, {Kramer}, \& {McKinven}}]{2020arXiv200912135H}
{Hilmarsson}, G.~H., {Michilli}, D., {Spitler}, L.~G., {et~al.} 2020, arXiv
  e-prints, arXiv:2009.12135

\bibitem[{{Hurley} {et~al.}(2002){Hurley}, {Tout}, \&
  {Pols}}]{2002MNRAS.329..897H}
{Hurley}, J.~R., {Tout}, C.~A., \& {Pols}, O.~R. 2002, \mnras, 329, 897

\bibitem[{{Hwang} \& {Laming}(2012)}]{2012ApJ...746..130H}
{Hwang}, U., \& {Laming}, J.~M. 2012, \apj, 746, 130

\bibitem[{{Janka} \& {Mueller}(1994)}]{1994A&A...290..496J}
{Janka}, H.~T., \& {Mueller}, E. 1994, \aap, 290, 496

\bibitem[{{Josephy} {et~al.}(2019){Josephy}, {Chawla}, {Fonseca}, {Ng},
  {Patel}, {Pleunis}, {Scholz}, {Andersen}, {Bandura}, {Bhardwaj}, {Boyce},
  {Boyle}, {Brar}, {Cubranic}, {Dobbs}, {Gaensler}, {Gill}, {Giri}, {Good},
  {Halpern}, {Hinshaw}, {Kaspi}, {Landecker}, {Lang}, {Lin}, {Masui},
  {Mckinven}, {Mena-Parra}, {Merryfield}, {Michilli}, {Milutinovic}, {Naidu},
  {Pen}, {Rafiei-Ravand i}, {Rahman}, {Ransom}, {Renard}, {Siegel}, {Smith},
  {Stairs}, {Tendulkar}, {Vanderlinde}, {Yadav}, \&
  {Zwaniga}}]{2019ApJ...882L..18J}
{Josephy}, A., {Chawla}, P., {Fonseca}, E., {et~al.} 2019, \apjl, 882, L18

\bibitem[{Kasen {et~al.}(2017)Kasen, Metzger, Barnes, Quataert, \&
  Ramirez-Ruiz}]{2017Natur.551...80K}
Kasen, D., Metzger, B., Barnes, J., Quataert, E., \& Ramirez-Ruiz, E. 2017,
  \nat, 551, 80

\bibitem[{{Kashiyama} \& {Murase}(2017{\natexlab{a}})}]{Kashiyama2017}
{Kashiyama}, K., \& {Murase}, K. 2017{\natexlab{a}}, \apjl, 839, L3

\bibitem[{{Kashiyama} \& {Murase}(2017{\natexlab{b}})}]{2017ApJ...839L...3K}
---. 2017{\natexlab{b}}, \apjl, 839, L3

\bibitem[{{Katz}(2020)}]{Katz_2020}
{Katz}, J.~I. 2020, arXiv e-prints, arXiv:2011.11666

\bibitem[{{King} {et~al.}(2001){King}, {Pringle}, \&
  {Wickramasinghe}}]{2001MNRAS.320L..45K}
{King}, A.~R., {Pringle}, J.~E., \& {Wickramasinghe}, D.~T. 2001, \mnras, 320,
  L45

\bibitem[{{Kulkarni} {et~al.}(2014){Kulkarni}, {Ofek}, {Neill}, {Zheng}, \&
  {Juric}}]{2014ApJ...797...70K}
{Kulkarni}, S.~R., {Ofek}, E.~O., {Neill}, J.~D., {Zheng}, Z., \& {Juric}, M.
  2014, \apj, 797, 70

\bibitem[{{Kundu} \& {Ferrario}(2020)}]{10.1093/mnras/stz3593}
{Kundu}, E., \& {Ferrario}, L. 2020, \mnras, 492, 3753

\bibitem[{{Laming} \& {Hwang}(2003)}]{2003ApJ...597..347L}
{Laming}, J.~M., \& {Hwang}, U. 2003, \apj, 597, 347

\bibitem[{{Law} {et~al.}(2017){Law}, {Abruzzo}, {Bassa}, {Bower},
  {Burke-Spolaor}, {Butler}, {Cantwell}, {Carey}, {Chatterjee}, {Cordes},
  {Demorest}, {Dowell}, {Fender}, {Gourdji}, {Grainge}, {Hessels}, {Hickish},
  {Kaspi}, {Lazio}, {McLaughlin}, {Michilli}, {Mooley}, {Perrott}, {Ransom},
  {Razavi-Ghods}, {Rupen}, {Scaife}, {Scott}, {Scholz}, {Seymour}, {Spitler},
  {Stovall}, {Tendulkar}, {Titterington}, {Wharton}, \&
  {Williams}}]{2017ApJ...850...76L}
{Law}, C.~J., {Abruzzo}, M.~W., {Bassa}, C.~G., {et~al.} 2017, \apj, 850, 76

\bibitem[{{Lorimer} {et~al.}(2007){Lorimer}, {Bailes}, {McLaughlin},
  {Narkevic}, \& {Crawford}}]{2007Sci...318..777L}
{Lorimer}, D.~R., {Bailes}, M., {McLaughlin}, M.~A., {Narkevic}, D.~J., \&
  {Crawford}, F. 2007, Science, 318, 777

\bibitem[{{L{\"u}} \& {Zhang}(2014)}]{Lu2014}
{L{\"u}}, H.-J., \& {Zhang}, B. 2014, \apj, 785, 74

\bibitem[{{Lu} \& {Kumar}(2018)}]{2018MNRAS.477.2470L}
{Lu}, W., \& {Kumar}, P. 2018, \mnras, 477, 2470

\bibitem[{{Lu} \& {Piro}(2019)}]{2019ApJ...883...40L}
{Lu}, W., \& {Piro}, A.~L. 2019, \apj, 883, 40

\bibitem[{{Luo} {et~al.}(2020){Luo}, {Wang}, {Men}, {Zhang}, {Jiang}, {Xu},
  {Wang}, {Lee}, {Han}, {Zhang}, {Caballero}, {Chen}, {Chen}, {Gan}, {Guo},
  {Hao}, {Huang}, {Jiang}, {Li}, {Li}, {Li}, {Luo}, {Pan}, {Pei}, {Qian},
  {Sun}, {Wang}, {Wang}, {Wen}, {Xu}, {Xu}, {Yan}, {Yan}, {Yu}, {Yuan},
  {Zhang}, \& {Zhu}}]{2020Natur.586..693L}
{Luo}, R., {Wang}, B.~J., {Men}, Y.~P., {et~al.} 2020, \nat, 586, 693

\bibitem[{{Madau} \& {Dickinson}(2014)}]{2014ARA&A..52..415M}
{Madau}, P., \& {Dickinson}, M. 2014, \araa, 52, 415

\bibitem[{{Makhathini} {et~al.}(2020){Makhathini}, {Mooley}, {Brightman},
  {Hotokezaka}, {Nayana}, {Intema}, {Dobie}, {Lenc}, {Perley}, {Fremling},
  {Moldon}, {Lazzati}, {Kaplan}, {Balasubramanian}, {Brown}, {Carbone},
  {Chandra}, {Corsi}, {Camilo}, {Deller}, {Frail}, {Murphy}, {Murphy}, {Nakar},
  {Smirnov}, {Beswick}, {Fender}, {Hallinan}, {Heywood}, {Kasliwal}, {Lee},
  {Lu}, {Rana}, {Perkins}, {White}, {Jozsa}, {Hugo}, \&
  {Kamphuis}}]{2020arXiv200602382M}
{Makhathini}, S., {Mooley}, K.~P., {Brightman}, M., {et~al.} 2020, arXiv
  e-prints, arXiv:2006.02382

\bibitem[{{Marcote} {et~al.}(2020){Marcote}, {Nimmo}, {Hessels}, {Tendulkar},
  {Bassa}, {Paragi}, {Keimpema}, {Bhardwaj}, {Karuppusamy}, {Kaspi}, {Law},
  {Michilli}, {Aggarwal}, {Andersen}, {Archibald}, {Bandura}, {Bower}, {Boyle},
  {Brar}, {Burke-Spolaor}, {Butler}, {Cassanelli}, {Chawla}, {Demorest},
  {Dobbs}, {Fonseca}, {Giri}, {Good}, {Gourdji}, {Josephy}, {Kirichenko},
  {Kirsten}, {Landecker}, {Lang}, {Lazio}, {Li}, {Lin}, {Linford}, {Masui},
  {Mena-Parra}, {Naidu}, {Ng}, {Patel}, {Pen}, {Pleunis}, {Rafiei-Ravandi},
  {Rahman}, {Renard}, {Scholz}, {Siegel}, {Smith}, {Stairs}, {Vanderlinde}, \&
  {Zwaniga}}]{2020Natur.577..190M}
{Marcote}, B., {Nimmo}, K., {Hessels}, J.~W.~T., {et~al.} 2020, \nat, 577, 190

\bibitem[{{Margalit} {et~al.}(2019){Margalit}, {Berger}, \&
  {Metzger}}]{Margalit_2019}
{Margalit}, B., {Berger}, E., \& {Metzger}, B.~D. 2019, \apj, 886, 110

\bibitem[{{Margalit} \& {Metzger}(2016)}]{2016MNRAS.461.1154M}
{Margalit}, B., \& {Metzger}, B.~D. 2016, \mnras, 461, 1154

\bibitem[{{Margalit} \& {Metzger}(2018)}]{2018ApJ...868L...4M}
---. 2018, \apjl, 868, L4

\bibitem[{{Metzger} {et~al.}(2017){Metzger}, {Berger}, \&
  {Margalit}}]{2017ApJ...841...14M}
{Metzger}, B.~D., {Berger}, E., \& {Margalit}, B. 2017, \apj, 841, 14

\bibitem[{Metzger {et~al.}(2009)Metzger, Piro, \&
  Quataert}]{2009MNRAS.396.1659M}
Metzger, B.~D., Piro, A.~L., \& Quataert, E. 2009, \mnras, 396, 1659

\bibitem[{{Micelotta} {et~al.}(2016){Micelotta}, {Dwek, Eli}, \& {Slavin,
  Jonathan D.}}]{refId0}
{Micelotta}, E.~R., {Dwek, Eli}, \& {Slavin, Jonathan D.} 2016, A\&A, 590, A65

\bibitem[{{Michilli} {et~al.}(2018){Michilli}, {Seymour}, {Hessels}, {Spitler},
  {Gajjar}, {Archibald}, {Bower}, {Chatterjee}, {Cordes}, {Gourdji}, {Heald},
  {Kaspi}, {Law}, {Sobey}, {Adams}, {Bassa}, {Bogdanov}, {Brinkman},
  {Demorest}, {Fernand ez}, {Hellbourg}, {Lazio}, {Lynch}, {Maddox}, {Marcote},
  {McLaughlin}, {Paragi}, {Ransom}, {Scholz}, {Siemion}, {Tendulkar}, {van
  Rooy}, {Wharton}, \& {Whitlow}}]{2018Natur.553..182M}
{Michilli}, D., {Seymour}, A., {Hessels}, J.~W.~T., {et~al.} 2018, \nat, 553,
  182

\bibitem[{{Murase} {et~al.}(2016){Murase}, {Kashiyama}, \&
  {M{\'e}sz{\'a}ros}}]{2016MNRAS.461.1498M}
{Murase}, K., {Kashiyama}, K., \& {M{\'e}sz{\'a}ros}, P. 2016, \mnras, 461,
  1498

\bibitem[{{Nomoto} \& {Kondo}(1991)}]{1991ApJ...367L..19N}
{Nomoto}, K., \& {Kondo}, Y. 1991, \apjl, 367, L19

\bibitem[{{Oostrum} {et~al.}(2020){Oostrum}, {Maan}, {van Leeuwen}, {Connor},
  {Petroff}, {Attema}, {Bast}, {Gardenier}, {Hargreaves}, {Kooistra}, {van der
  Schuur}, {Sclocco}, {Smits}, {Straal}, {ter Veen}, {Vohl}, {Adams},
  {Adebahr}, {de Blok}, {van den Brink}, {van Cappellen}, {Coolen}, {Damstra},
  {van Diepen}, {Frank}, {Hess}, {van der Hulst}, {Hut}, {Ivashina}, {Loose},
  {Lucero}, {Mika}, {Morganti}, {Moss}, {Mulder}, {Norden}, {Oosterloo},
  {Orr{\'u}}, {de Reijer}, {Ruiter}, {Vermaas}, {Wijnholds}, \&
  {Ziemke}}]{2020A&A...635A..61O}
{Oostrum}, L.~C., {Maan}, Y., {van Leeuwen}, J., {et~al.} 2020, \aap, 635, A61

\bibitem[{Parker(1963)}]{1963idpbookP}
Parker, E.~N. 1963, {Interplanetary dynamical processes.}

\bibitem[{{Piro}(2016)}]{Piro_2016}
{Piro}, A.~L. 2016, \apjl, 824, L32

\bibitem[{{Piro} \& {Gaensler}(2018)}]{Piro_2018}
{Piro}, A.~L., \& {Gaensler}, B.~M. 2018, \apj, 861, 150

\bibitem[{{Popov} \& {Postnov}(2013)}]{2013arXiv1307.4924P}
{Popov}, S.~B., \& {Postnov}, K.~A. 2013, arXiv e-prints, arXiv:1307.4924

\bibitem[{{Price} {et~al.}(2019){Price}, {Foster}, {Geyer}, {van Straten},
  {Gajjar}, {Hellbourg}, {Karastergiou}, {Keane}, {Siemion}, {Arcavi}, {Bhat},
  {Caleb}, {Chang}, {Croft}, {DeBoer}, {de Pater}, {Drew}, {Enriquez}, {Farah},
  {Gizani}, {Green}, {Isaacson}, {Hickish}, {Jameson}, {Lebofsky}, {MacMahon},
  {M{\"o}ller}, {Onken}, {Petroff}, {Werthimer}, {Wolf}, {Worden}, \&
  {Zhang}}]{2019MNRAS.486.3636P}
{Price}, D.~C., {Foster}, G., {Geyer}, M., {et~al.} 2019, \mnras, 486, 3636

\bibitem[{{Price} \& {Rosswog}(2006)}]{2006Sci...312..719P}
{Price}, D.~J., \& {Rosswog}, S. 2006, Science, 312, 719

\bibitem[{{Prochaska} {et~al.}(2019){Prochaska}, {Macquart}, {McQuinn},
  {Simha}, {Shannon}, {Day}, {Marnoch}, {Ryder}, {Deller}, {Bannister},
  {Bhandari}, {Bordoloi}, {Bunton}, {Cho}, {Flynn}, {Mahony}, {Phillips},
  {Qiu}, \& {Tejos}}]{2019Sci...366..231P}
{Prochaska}, J.~X., {Macquart}, J.-P., {McQuinn}, M., {et~al.} 2019, Science,
  366, 231

\bibitem[{{Radice} {et~al.}(2018){Radice}, {Perego}, {Hotokezaka}, {Fromm},
  {Bernuzzi}, \& {Roberts}}]{Radice_2018}
{Radice}, D., {Perego}, A., {Hotokezaka}, K., {et~al.} 2018, \apj, 869, 130

\bibitem[{{Raghavan} {et~al.}(2010){Raghavan}, {McAlister}, {Henry}, {Latham},
  {Marcy}, {Mason}, {Gies}, {White}, \& {ten Brummelaar}}]{2010ApJS..190....1R}
{Raghavan}, D., {McAlister}, H.~A., {Henry}, T.~J., {et~al.} 2010, \apjs, 190,
  1

\bibitem[{{Ravi} {et~al.}(2019){Ravi}, {Catha}, {D'Addario}, {Djorgovski},
  {Hallinan}, {Hobbs}, {Kocz}, {Kulkarni}, {Shi}, {Vedantham}, {Weinreb}, \&
  {Woody}}]{2019Natur.572..352R}
{Ravi}, V., {Catha}, M., {D'Addario}, L., {et~al.} 2019, \nat, 572, 352

\bibitem[{{Rosswog} {et~al.}(2003){Rosswog}, {Ramirez-Ruiz}, \&
  {Davies}}]{2003MNRAS.345.1077R}
{Rosswog}, S., {Ramirez-Ruiz}, E., \& {Davies}, M.~B. 2003, \mnras, 345, 1077

\bibitem[{{Rowlinson} {et~al.}(2013){Rowlinson}, {O'Brien}, {Metzger},
  {Tanvir}, \& {Levan}}]{Rowlinson2013}
{Rowlinson}, A., {O'Brien}, P.~T., {Metzger}, B.~D., {Tanvir}, N.~R., \&
  {Levan}, A.~J. 2013, \mnras, 430, 1061

\bibitem[{{Ruiter} {et~al.}(2019){Ruiter}, {Ferrario}, {Belczynski},
  {Seitenzahl}, {Crocker}, \& {Karakas}}]{2019MNRAS.484..698R}
{Ruiter}, A.~J., {Ferrario}, L., {Belczynski}, K., {et~al.} 2019, \mnras, 484,
  698

\bibitem[{{Rybicki} \& {Lightman}(1986)}]{1986rpa..book.....R}
{Rybicki}, G.~B., \& {Lightman}, A.~P. 1986, {Radiative Processes in
  Astrophysics}

\bibitem[{{Sana} {et~al.}(2012){Sana}, {de Mink}, {de Koter}, {Langer},
  {Evans}, {Gieles}, {Gosset}, {Izzard}, {Le Bouquin}, \&
  {Schneider}}]{2012Sci...337..444S}
{Sana}, H., {de Mink}, S.~E., {de Koter}, A., {et~al.} 2012, Science, 337, 444

\bibitem[{{Sana} {et~al.}(2014){Sana}, {Le Bouquin}, {Lacour}, {Berger},
  {Duvert}, {Gauchet}, {Norris}, {Olofsson}, {Pickel}, {Zins}, {Absil}, {de
  Koter}, {Kratter}, {Schnurr}, \& {Zinnecker}}]{2014ApJS..215...15S}
{Sana}, H., {Le Bouquin}, J.~B., {Lacour}, S., {et~al.} 2014, \apjs, 215, 15

\bibitem[{{Scalo}(1986)}]{1986FCPh...11....1S}
{Scalo}, J.~M. 1986, \fcp, 11, 1

\bibitem[{{Scholz} {et~al.}(2016){Scholz}, {Spitler}, {Hessels}, {Chatterjee},
  {Cordes}, {Kaspi}, {Wharton}, {Bassa}, {Bogdanov}, {Camilo}, {Crawford},
  {Deneva}, {van Leeuwen}, {Lynch}, {Madsen}, {McLaughlin}, {Mickaliger},
  {Parent}, {Patel}, {Ransom}, {Seymour}, {Stairs}, {Stappers}, \&
  {Tendulkar}}]{2016ApJ...833..177S}
{Scholz}, P., {Spitler}, L.~G., {Hessels}, J.~W.~T., {et~al.} 2016, \apj, 833,
  177

\bibitem[{{Schwab} {et~al.}(2015){Schwab}, {Quataert}, \&
  {Bildsten}}]{2015MNRAS.453.1910S}
{Schwab}, J., {Quataert}, E., \& {Bildsten}, L. 2015, \mnras, 453, 1910

\bibitem[{{Schwab} {et~al.}(2016){Schwab}, {Quataert}, \&
  {Kasen}}]{2016MNRAS.463.3461S}
{Schwab}, J., {Quataert}, E., \& {Kasen}, D. 2016, \mnras, 463, 3461

\bibitem[{Sedov(1959)}]{1959sdmm.book.....S}
Sedov, L.~I. 1959, {Similarity and Dimensional Methods in Mechanics}

\bibitem[{{Spitler} {et~al.}(2014){Spitler}, {Cordes}, {Hessels}, {Lorimer},
  {McLaughlin}, {Chatterjee}, {Crawford}, {Deneva}, {Kaspi}, {Wharton},
  {Allen}, {Bogdanov}, {Brazier}, {Camilo}, {Freire}, {Jenet},
  {Karako-Argaman}, {Knispel}, {Lazarus}, {Lee}, {van Leeuwen}, {Lynch},
  {Ransom}, {Scholz}, {Siemens}, {Stairs}, {Stovall}, {Swiggum},
  {Venkataraman}, {Zhu}, {Aulbert}, \& {Fehrmann}}]{2014ApJ...790..101S}
{Spitler}, L.~G., {Cordes}, J.~M., {Hessels}, J.~W.~T., {et~al.} 2014, \apj,
  790, 101

\bibitem[{{Spitler} {et~al.}(2016){Spitler}, {Scholz}, {Hessels}, {Bogdanov},
  {Brazier}, {Camilo}, {Chatterjee}, {Cordes}, {Crawford}, {Deneva}, {Ferdman},
  {Freire}, {Kaspi}, {Lazarus}, {Lynch}, {Madsen}, {McLaughlin}, {Patel},
  {Ransom}, {Seymour}, {Stairs}, {Stappers}, {van Leeuwen}, \&
  {Zhu}}]{2016Natur.531..202S}
{Spitler}, L.~G., {Scholz}, P., {Hessels}, J.~W.~T., {et~al.} 2016, \nat, 531,
  202

\bibitem[{{Sridhar} {et~al.}(2020){Sridhar}, {Zrake}, {Metzger}, {Sironi}, \&
  {Giannios}}]{2020arXiv201009214S}
{Sridhar}, N., {Zrake}, J., {Metzger}, B.~D., {Sironi}, L., \& {Giannios}, D.
  2020, arXiv e-prints, arXiv:2010.09214

\bibitem[{Tang \& Chevalier(2017)}]{2017MNRAS.465.3793T}
Tang, X., \& Chevalier, R.~A. 2017, \mnras, 465, 3793

\bibitem[{{Tauris} {et~al.}(2013){Tauris}, {Sanyal}, {Yoon}, \&
  {Langer}}]{2013A&A...558A..39T}
{Tauris}, T.~M., {Sanyal}, D., {Yoon}, S.~C., \& {Langer}, N. 2013, \aap, 558,
  A39

\bibitem[{Taylor(1946)}]{Taylor1946The}
Taylor, G.~I. 1946, Proceedings of the Royal Society of London, 186, 273

\bibitem[{{Tendulkar} {et~al.}(2017){Tendulkar}, {Bassa}, {Cordes}, {Bower},
  {Law}, {Chatterjee}, {Adams}, {Bogdanov}, {Burke-Spolaor}, {Butler},
  {Demorest}, {Hessels}, {Kaspi}, {Lazio}, {Maddox}, {Marcote}, {McLaughlin},
  {Paragi}, {Ransom}, {Scholz}, {Seymour}, {Spitler}, {van Langevelde}, \&
  {Wharton}}]{2017ApJ...834L...7T}
{Tendulkar}, S.~P., {Bassa}, C.~G., {Cordes}, J.~M., {et~al.} 2017, \apjl, 834,
  L7

\bibitem[{{The Chime/Frb Collaboration} {et~al.}(2020){The Chime/Frb
  Collaboration}, {Bandura}, {Bhardwaj}, {Bij}, {Boyce}, {Boyle}, {Brar},
  {Cassanelli}, {Chawla}, {Chen}, {Cliche}, {Cook}, {Cubranic}, {Curtin},
  {Denman}, {Dobbs}, {Dong}, {Fandino}, {Fonseca}, {Gaensler}, {Giri}, {Good},
  {Halpern}, {Hill}, {Hinshaw}, {H{\"o}fer}, {Josephy}, {Kania}, {Kaspi},
  {Landecker}, {Leung}, {Li}, {Lin}, {Masui}, {McKinven}, {Mena-Parra},
  {Merryfield}, {Meyers}, {Michilli}, {Milutinovic}, {Mirhosseini},
  {M{\"u}nchmeyer}, {Naidu}, {Newburgh}, {Ng}, {Patel}, {Pen},
  {Pinsonneault-Marotte}, {Pleunis}, {Quine}, {Rafiei-Ravandi}, {Rahman},
  {Ransom}, {Renard}, {Sanghavi}, {Scholz}, {Shaw}, {Shin}, {Siegel}, {Singh},
  {Smegal}, {Smith}, {Stairs}, {Tan}, {Tendulkar}, {Tretyakov}, {Vanderlinde},
  {Wang}, {Wulf}, \& {Zwaniga}}]{2020Natur.587...54T}
{The Chime/Frb Collaboration}, Andersen, B.~C., {Bandura}, K.~M., {Bhardwaj},
  M., {et~al.} 2020, \nat, 587, 54

\bibitem[{{Toonen} {et~al.}(2018){Toonen}, {Perets}, {Igoshev}, {Michaely}, \&
  {Zenati}}]{2018A&A...619A..53T}
{Toonen}, S., {Perets}, H.~B., {Igoshev}, A.~P., {Michaely}, E., \& {Zenati},
  Y. 2018, \aap, 619, A53

\bibitem[{Truelove \& McKee(1999)}]{1999ApJS..120..299T}
Truelove, J.~K., \& McKee, C.~F. 1999, \apjs, 120, 299

\bibitem[{{Truelove} \& {McKee}(2000)}]{2000ApJS..128..403T}
{Truelove}, J.~K., \& {McKee}, C.~F. 2000, \apjs, 128, 403

\bibitem[{{Vink}(2012)}]{2012A&ARv..20...49V}
{Vink}, J. 2012, \aapr, 20, 49

\bibitem[{{Virtanen} {et~al.}(2020){Virtanen}, {Gommers}, {Oliphant},
  {Haberland}, {Reddy}, {Cournapeau}, {Burovski}, {Peterson}, {Weckesser},
  {Bright}, {van der Walt}, {Brett}, {Wilson}, {Millman}, {Mayorov}, {Nelson},
  {Jones}, {Kern}, {Larson}, {Carey}, {Polat}, {Feng}, {Moore}, {Vand erPlas},
  {Laxalde}, {Perktold}, {Cimrman}, {Henriksen}, {Quintero}, {Harris},
  {Archibald}, {Ribeiro}, {Pedregosa}, {van Mulbregt}, \& {SciPy 1. 0
  Contributors}}]{2020NatMe..17..261V}
{Virtanen}, P., {Gommers}, R., {Oliphant}, T.~E., {et~al.} 2020, Nature
  Methods, 17, 261

\bibitem[{{Wadiasingh} \& {Timokhin}(2019)}]{2019ApJ...879....4W}
{Wadiasingh}, Z., \& {Timokhin}, A. 2019, \apj, 879, 4

\bibitem[{{Wang} \& {Dai}(2013)}]{Wang2013}
{Wang}, F.~Y., \& {Dai}, Z.~G. 2013, Nature Physics, 9, 465

\bibitem[{{Wang} {et~al.}(2020){Wang}, {Wang}, {Yang}, {Yu}, {Zuo}, \&
  {Dai}}]{Wang_2020}
{Wang}, F.~Y., {Wang}, Y.~Y., {Yang}, Y.-P., {et~al.} 2020, \apj, 891, 72

\bibitem[{{Wang} \& {Yu}(2017)}]{2017JCAP...03..023W}
{Wang}, F.~Y., \& {Yu}, H. 2017, \jcap, 2017, 023

\bibitem[{{Wang} \& {Zhang}(2019)}]{Wang2019}
{Wang}, F.~Y., \& {Zhang}, G.~Q. 2019, \apj, 882, 108

\bibitem[{{Wu} {et~al.}(2020){Wu}, {Zhang}, {Wang}, \&
  {Dai}}]{2020ApJ...900L..26W}
{Wu}, Q., {Zhang}, G.~Q., {Wang}, F.~Y., \& {Dai}, Z.~G. 2020, \apjl, 900, L26

\bibitem[{{Yamasaki} {et~al.}(2018){Yamasaki}, {Totani}, \&
  {Kiuchi}}]{2018PASJ...70...39Y}
{Yamasaki}, S., {Totani}, T., \& {Kiuchi}, K. 2018, \pasj, 70, 39

\bibitem[{{Yamasaki} {et~al.}(2020){Yamasaki}, {Totani}, \&
  {Kiuchi}}]{Yamasaki2020}
---. 2020, arXiv e-prints, arXiv:2010.07796

\bibitem[{{Yang} \& {Dai}(2019)}]{2019ApJ...885..149Y}
{Yang}, Y.-H., \& {Dai}, Z.-G. 2019, \apj, 885, 149

\bibitem[{Yang \& Zhang(2017)}]{2017ApJ...847...22Y}
Yang, Y.-P., \& Zhang, B. 2017, \apj, 847, 22

\bibitem[{{Yang} \& {Zhang}(2018)}]{2018ApJ...868...31Y}
{Yang}, Y.-P., \& {Zhang}, B. 2018, \apj, 868, 31

\bibitem[{{Yoon} {et~al.}(2007){Yoon}, {Podsiadlowski}, \&
  {Rosswog}}]{2007MNRAS.380..933Y}
{Yoon}, S.~C., {Podsiadlowski}, P., \& {Rosswog}, S. 2007, \mnras, 380, 933

\bibitem[{{Zenati} {et~al.}(2019){Zenati}, {Perets}, \&
  {Toonen}}]{10.1093/mnras/stz316}
{Zenati}, Y., {Perets}, H.~B., \& {Toonen}, S. 2019, \mnras, 486, 1805

\bibitem[{{Zhang} {et~al.}(2006){Zhang}, {Fan}, {Dyks}, {Kobayashi},
  {M{\'e}sz{\'a}ros}, {Burrows}, {Nousek}, \& {Gehrels}}]{Zhang2006}
{Zhang}, B., {Fan}, Y.~Z., {Dyks}, J., {et~al.} 2006, \apj, 642, 354

\bibitem[{{Zhang} \& {M{\'e}sz{\'a}ros}(2001)}]{Zhang2001}
{Zhang}, B., \& {M{\'e}sz{\'a}ros}, P. 2001, \apjl, 552, L35

\bibitem[{{Zhang} {et~al.}(2020){Zhang}, {Yi}, \& {Wang}}]{Zhang2020}
{Zhang}, G.~Q., {Yi}, S.~X., \& {Wang}, F.~Y. 2020, \apj, 893, 44

\bibitem[{{Zhong} \& {Dai}(2020)}]{2020ApJ...893....9Z}
{Zhong}, S.-Q., \& {Dai}, Z.-G. 2020, \apj, 893, 9

\end{thebibliography}

\clearpage

\begin{table}
	\centering
	\caption{Asymptotic behavior of the shock.}
	\begin{threeparttable}
		\begin{tabular}{cl}
			\hline
			{$ t \to 0 $} & FE soultion $ (n<5) $   \\
			& $ R_{\mathrm{c}}^* = \lambda_{\mathrm{c}} t^* $  \\
			& $ R_{\mathrm{b}}^* = \lambda_{\mathrm{b}} t^* $   \\
			& \tnote{a} $ R_{\mathrm{r}}^*(t^*)=R_{\mathrm{b}}^*(t^*)/l_{\mathrm{ED}} $   \\
			& \tnote{b} $ \lambda_{\mathrm{c}}^2(n<3) = 2(\frac{5-n}{3-n}) $ \\
			& \tnote{b} $ \lambda_{\mathrm{c}}^2(n>3) =2w_{\mathrm{core}}^{-2}(\frac{5-n}{3-n})(\frac{w_{\mathrm{core}}^{n-3}-n/3}{w_{\mathrm{core}}^{n-5}-n/5}) $ \\
			& \tnote{c} $ \lambda_{\mathrm{b}} = q_b \lambda_{\mathrm{c}} $ \\
			& $ l_{\mathrm{ED}}=1+\frac{8}{n^{2}}+\frac{0.4}{4-s} $\\
			& \tnote{d} \ SSDW solution $ (n>5) $ \\
			& $ R_{\mathrm{c}}^*=\zeta_{\mathrm{c}}t^{*{(n-3)/(n-s)}} $ \\
			& $ R_{\mathrm{b}}^*=\zeta_{\mathrm{b}}t^{*{(n-3)/(n-s)}} $ \\
			& $ R_{\mathrm{r}}^*=\zeta_{\mathrm{r}}t^{*{(n-3)/(n-s)}} $ \\
			& \tnote{d} $ \zeta_{\mathrm{c}}=(Af_0w_{\mathrm{core}}^n \lambda_{\mathrm{c}}^{n-3})^{1/(n-s)} $ \\
			& \tnote{d} $ \zeta_{\mathrm{b}}= (\frac{R_1}{R_c})\zeta_{\mathrm{c}} \ $ \\
			& \tnote{d} $ \zeta_{\mathrm{r}}= (\frac{R_2}{R_c})\zeta_{\mathrm{c}} $ \\
			
			\hline
			$ t \to \infty $ & \tnote{e} $ R_{\mathrm{b}}^*= [\xi(s) t^*]^{2/(5-s)} $ \\
			& $ \xi(0) = 2.026 $ and $ \xi(2) = 3/2 \pi $ \\
			\hline		
		\end{tabular}
		\begin{tablenotes}
			\footnotesize
			\item[a] Reference \citetalias{1999ApJS..120..299T}.
			\item[b] Reference \cite{2017MNRAS.465.3793T}, where $ w_{\mathrm{core}}=0 \ (n<3) $ and $ w_{\mathrm{core}}=0.05 \ (n>3) $ for $ s=0 $, and $ w_{\mathrm{core}}=0 \ (n<3) $ and $ w_{\mathrm{core}}=0.1 \ (n>3) $ for $ s=2 $.
			\item[c] Reference \cite{1963idpbookP} and \cite{1984ApJ...281..682H}, where $ q_b \ (s=0)=1.1 $ and $ q_b \ (s=2)=1.19 $.
			\item[d] The value of $ A $ , $ R_1/R_c $ and $ R_2/R_c $ can be found in \cite{1982ApJ...258..790C}.
			\item[e] Reference \cite{Taylor1946The} and \cite{1959sdmm.book.....S}.
		\end{tablenotes}
	\end{threeparttable}
	
	\label{Asymp}
\end{table}

\begin{table}
	\centering
	\caption{The reverse shock radius $ R_{\mathrm{r}}^* $ from \cite{refId0}.}
	\begin{tabular}{cl}
		\hline
		$ t^*<t_{\mathrm{core}}^* $      & $ R_{\mathrm{r}}^* =\frac{1}{l_{\mathrm{ED}}}\left\lbrace v_{\mathrm{core}}^{*n-3}\frac{(3-s)^2}{n(n-3)}\frac{3}{4 \pi}\frac{l_{\mathrm{ED}}^{n-2}}{\phi_{\mathrm{ED}}}\right\rbrace^{1/(n-s)} t^{*\frac{n-3}{n-s}} $, \\
		                                & $ l_{\mathrm{ED}}=1+\frac{8}{n^{2}}+\frac{0.4}{4-s} $ \\
		                                & $ \phi_{\mathrm{ED}}=[0.65-\mathrm{exp}(-n/4)]\sqrt{1-\frac{s}{3}} $\\
		                                & $ v_{\mathrm{core}}^*=\left[ \frac{10(n-5)}{3(n-3)} \right]^{1/2}  $ \\
		$ t^* \ge t_{\mathrm{core}}^* $ & $ R_{\mathrm{r}}^* =\left[ \frac{R_{\mathrm{b}}^*(t^*=t_{\mathrm{core}}^*)}{l_{\mathrm{ED}}t_{\mathrm{core}}^*}-\frac{3-s}{n-3}\frac{v_{\mathrm{b}}^*(t^*=t_{\mathrm{core}}^*)}{l_{\mathrm{ED}}} \ln\frac{t^*}{t_{\mathrm{core}}} \right]t^*  $ \\
		                                & $t_{\mathrm{core}}^*= \left[ \frac{l_{\mathrm{ED}}^{s-2}}{\phi_{\mathrm{ED}}} \frac{3}{4 \pi} \frac{(3-s)^2}{n(n-3)}
		                                \right] ^{1/(3-s)}\frac{1}{v_{\mathrm{core}}^*} $\\
		\hline
	\end{tabular}
	\label{Rr}
\end{table}

\begin{table}
	\centering
	\caption{Ejecta mass and the kinetic energy of different kinds of binary compact star mergers.}
		\begin{tabular}{ccccc}
			\hline
			  & Model    & $M_{\mathrm{ej}}$ ($M_{\odot}$) & $E_{k}$ (erg) & Reference \\
			\hline
			A & BNS      & $\sim$ 0.01  & $\sim$ 10$^{51} $         & \cite{Bauswein_2013} \\
			B & BNS      & $\sim$ 0.001 & $\sim$ 5$\times 10^{49} $ & \cite{Bauswein_2013,Radice_2018}   \\
			C & BWD/NSWD & $\sim$ 0.1   & $\sim$ 10$^{50} $         & \cite{2016MNRAS.461.1154M,2007ApJ...669..585D}  \\
			D & BWD      & $\sim$ 0.01  & $\sim$ 5$\times 10^{49} $ & \cite{2009MNRAS.396.1659M}      \\
			E & NSWD     & $\sim$ 0.01  & $\sim$ 10$ ^{49} $        & \cite{10.1093/mnras/stz316}   \\
			\hline
		\end{tabular}
		\label{initial}
\end{table}

\begin{table}
	\centering
	\caption{Characteristic scales and transition times for uniform medium.}
		\begin{tabular}{cccccc}
			\hline
			 &  $ n_0 (\mathrm{cm^{-3}}) $ & $ t_{\mathrm{ch}} $ (yr) & $ R_{\mathrm{ch}} $ (pc) & n & $ t_{\mathrm{tran}} $ (yr) \\
			 \hline
			 A  & 5  & 6.0  & 0.4 & 10 & 6.0 \\
			    &     &      &          &  6 & 13.6 \\
			    & 0.1 & 22   & 1.6  & 10 & 22  \\
			    &       &         &      &  6 & 50  \\
			 B  & 5  & 3.9  & 0.2 & 10 & 3.9 \\
			      &     &      &      & 6  & 9.0 \\
			    & 0.1 & 14   & 0.73 & 10 & 14  \\
			    &     &      &      & 6  & 33  \\
			 C  & 5  & 128.4  & 0.9 & 10 & 128.4 \\
			    &     &      &      & 6  & 294.0 \\
			    & 0.1 & 473  & 3.4  & 10 & 473  \\
			    &     &      &      & 6  & 1083 \\
			 D  & 5  & 26.7   & 0.4 & 10 & 26.7  \\
			    &     &      &      & 6  & 61.0  \\
			    & 0.1 & 98   & 1.58 & 10 & 98   \\
			    &     &      &      & 6  & 225  \\
			 E  & 5  & 406.0  & 0.9 & 10 & 406.0  \\
			    &     &      &      & 6  & 929.8  \\
			    & 0.1 & 1496 & 3.4  & 10 & 1496 \\
			    &     &      &      & 6  & 3425 \\
			 \hline
		\end{tabular}
		\label{ch}
\end{table}

\begin{table}
	\centering
	\caption{Characteristic scales and transition times for CC SNe with the wind velocity $ v_{\mathrm{w}}=10 $ km s$ ^{-1} $. }
	\begin{tabular}{cccc}
		\hline
		$ \dot{M} $ ($ \mathrm{M_{\odot} \ yr^{-1}} $) & $ t_{\mathrm{ch}} $ (yr) & $ R_{\mathrm{ch}} $ (pc) &  $ t_{\mathrm{tran}} $ (yr) \\
		\hline
		$ 10^{-4} $ & 337.9  & 2.58 & 108.1 \\
		$ 10^{-5} $ & 3379   & 25.8 & 1081  \\
		$ 10^{-6} $ & 33790  & 258  & 10812 \\
		\hline
	\end{tabular}
	\label{CCch}
\end{table}

\begin{table}
	\centering
	\caption{Model parameters.}
	\begin{threeparttable}
		\begin{tabular}{ccccccc}
			\hline
			Merger Model & WN Model & DM$_{\text{other}} $ (pc cm$^{-3} $) & $ n_0 $ (cm$^{-3}  $) & $ t_{\text{age}} $ (yr) & $ \epsilon_{\mathrm{B}} $ & $ \delta $ \\
			\hline
			BNS-A\tnote{a} & MM18-A\tnote{b} & 552.03$^{+0.53}_{-0.55}$ & 2.5$^{+0.002}_{-0.003}$ & 10$ ^{+0.01}_{-0.01} $& 0.01$^{+0.00001}_{-0.00001} $ & 1.18$ ^{+0.001}_{-0.001} $ \\
			BNS-A\tnote{a} & MM18-C\tnote{c} & 550.02$^{+0.53}_{-0.55}$ & 3$^{+0.003}_{-0.003}$ & 9$ ^{+0.01}_{-0.01} $& 0.0065$^{+0.00001}_{-0.00001} $ & 1.21$ ^{+0.001}_{-0.001} $ \\
			\hline
		\end{tabular}
		\begin{tablenotes}
			\footnotesize
			\item[a] The ejecta mass and kinetic energy is 0.01$ M_{\odot} $ and 10$ ^{51} $ erg for BNS mergers from \cite{Bauswein_2013}, respectively.
			\item[b] From Model A in \cite{2018ApJ...868L...4M}. The parameters are: $E_{\mathrm{B}}  \sim 5 \times 10^{50}$ erg , $ v_{\mathrm{n}} \sim 3 \times$ 10$ ^8 $ cm s$ ^{-1} $  and $ t_0 =0.2$ yr.	
			\item[c] From Model C in \cite{2018ApJ...868L...4M}. The parameters are: $E_{\mathrm{B}}  \sim 4.9 \times 10^{51}$ erg , $ v_{\mathrm{n}} \sim 9 \times$ 10$ ^8 $ cm s$ ^{-1} $  and $ t_0 =0.2$ yr.		
		\end{tablenotes}
	\end{threeparttable}
	\label{results}
\end{table}

\end{document}